\documentclass[11pt]{article}

\usepackage[T1]{fontenc}
\usepackage[a4paper,margin=1in]{geometry}
\usepackage{amsmath,amssymb,amsfonts,amsthm,mathtools}
\usepackage{mathrsfs}
\usepackage[all]{xy}
\usepackage{tensor}
\usepackage{comment}
\usepackage{stmaryrd}
\usepackage{bm}
\usepackage[normalem]{ulem}
\usepackage[dvipsnames]{xcolor}
\usepackage{ragged2e}
\usepackage{tikz}
\usetikzlibrary{calc,patterns,decorations.pathmorphing,arrows.meta,positioning,decorations.pathreplacing,intersections}
\usepackage{physics}
\usepackage{diagbox}
\usepackage{makecell}
\usepackage{hhline}
\usepackage{stackengine}
\usepackage{simplewick}
\usepackage[labelfont={small, bf, sf}, textfont={small,sf}]{caption}
\usepackage{authblk}
\usepackage{cite}
\usepackage{microtype}
\usepackage{enumitem}
\usepackage{booktabs}
\usepackage{array}
\usepackage[colorlinks=true, citecolor=Blue, linkcolor=Blue]{hyperref}
\usepackage{cleveref}
\usepackage{quantikz}

\hypersetup{
	breaklinks=true,
	colorlinks=true,
}

\numberwithin{equation}{section}
\newtheorem*{theorem-non}{Theorem}

\creflabelformat{equation}{#2#1#3}
\crefrangelabelformat{equation}{(#3#1#4)–(#5#2#6)}

\crefname{equation}{Eq.}{Eqs.}
\crefname{section}{Section}{Sections}
\crefname{appendix}{Appendix}{Appendices}
\crefname{figure}{Fig.}{Figs.}
\crefname{definition}{Def.}{Defs.}
\crefname{prop}{Prop.}{Props.}
\crefname{lemma}{Lemma}{Lemmas}
\crefname{corollary}{Cor.}{Cors.}
\crefname{thm}{Theorem}{Theorems}
\crefname{remark}{Remark}{Remarks}

\newcommand{\Dd}{\mathrm{d}}
\newcommand{\ii}{\mathrm{i}}
\newcommand{\ee}{\mathrm{e}}

\newcommand{\R}{\mathbb{R}}

\newcommand{\Hkin}{\mathcal H_{\rm kin}}
\newcommand{\Hphys}{\mathcal H_{\rm phys}}
\newcommand{\Hclk}{\mathcal H_{\rm clk}}

\newcommand{\Hgrav}{\mathcal H_{\mathrm{grav}}}

\newcommand{\kin}{\mathrm{kin}}
\newcommand{\phys}{\mathrm{phys}}
\newcommand{\clock}{\mathrm{clock}}

\newcommand{\Sym}{\mathrm{Sym}}

\newcommand{\Proj}{\mathsf P}

\newcommand{\HN}{\mathcal H_{N}}

\newcommand{\lie}{\pounds}
\newcommand{\hateq}{\mathrel{\mathop {\widehat=} }}

\title{\Large Smooth horizons from topology change in canonical quantum gravity}
\author[1]{Venkatesa Chandrasekaran\thanks{venc@stanford.edu}}
\date{\today}

\affil[1]{\small \it Leinweber Institute for Theoretical Physics, Stanford University, Stanford, CA 94305, USA}

\begin{document}
\maketitle

\begin{abstract}
We propose a resolution of the firewall paradox in JT gravity by incorporating topology change into canonical quantization under relational time evolution. The gravitational Hamiltonian acts on the black hole interior through a pair of pants interaction, mapping between a single interior sector and a connected two interior sector. To describe dynamics in the interior while keeping track of the exterior, we pass to an extended phase space description obtained by splitting the bulk Hilbert space across the event horizon. The split introduces boost edge modes at the horizon, which the Hawking modes become gravitationally dressed to. Covariance of the resulting crossed product algebra provides a precise gravitational realization of the firewall: a one sided boost of the interior edge mode relative to the exterior holding the matter fixed, or equivalently, a relative phase between the interior and exterior Hawking partners holding the edge modes fixed. Although each topology changing transition is exponentially suppressed, evolution over a Page time causes the connected two interior branch to dominate. One of these is the naive semiclassical interior, which we show carries a nontrivial one sided boost upon gluing the interior back to the exterior, and hence a firewall. The other interior is shown to be a zero mode of the one sided boost generator. Gluing the interior back to the exterior quotients by the gravitational constraints, which annihilates the firewall branch. On the surviving branch, we show the horizon vacuum measurement and the early radiation purity measurement become the same Dirac observable. Equivalently, we show that Page time dynamics induces a large diffeomorphism on the connected branch under which the operator algebra of the interior Hawking partner and that of the decoded early radiation are identified.
\end{abstract}

\tableofcontents

\section{Introduction}
The firewall paradox formulates the black hole information problem as a sharp conflict between three claims that each look compelling in semiclassical gravity: first, the Hawking radiation of an old black hole should purify itself unitarily; second, low energy effective field theory should remain valid outside the horizon; and third, an infalling observer should encounter a smooth vacuum rather than ``drama'' at the horizon \cite{AMPS2013,Apologia2013}. In the original AMPS version of the paradox, one considers a late Hawking mode $b$ of an old black hole, distills from the early radiation $E$ a subsystem $e_b$ that purifies $b$ using a decoding unitary $D$ \cite{HaydenPreskill2007}, verifies that purification by a measurement via the projector $\Pi_{be_b}$, and then asks whether an infalling observer can also verify that $b$ is maximally entangled with an interior partner $\tilde b$ via the projector $\Pi_{\tilde b b}$ \cite{AMPS2013,Apologia2013,HaydenPreskill2007,HarlowHayden2013}. Monogamy naively forbids both measurements from succeeding.

There exist myriad arguments on putative resolutions to the AMPS paradox. One can take the paradox at face value and conclude that old horizons necessarily have firewalls, or more generally that the semiclassical vacuum is not realized in typical black hole microstates \cite{AMPS2013,Apologia2013,MarolfPolchinski2013,BoussoDoublePurity2013}. One can instead question the operational accessibility of the protocol, as in the Harlow--Hayden argument \cite{HarlowHayden2013}. The line of thought most relevant for the present paper, however, is the family of proposals in which the interior is not an independent tensor factor at all, but is roughly speaking identified with the early radiation. In early forms this was expressed as refinement of black hole complementarity, and later sharpened by slogans such as $A=R_B$ \cite{Susskind1993Complementarity,SusskindThorlaciusUglum1993,LoweEtAl1995,BoussoComplementarityNotEnough2013}. In AdS/CFT this idea led to the Papadodimas--Raju construction of mirror operators and state dependent bulk boundary maps \cite{PapadodimasRaju2013Infalling,PapadodimasRaju2014PRL,PapadodimasRaju2014StateDependent,PapadodimasRaju2016Remarks,Harlow2014AspectsPR}, to ER$=$EPR \cite{MaldacenaSusskind2013}, and more recently to formulations in which gravitational constraints and the holography of information imply that the exterior already contains a complete encoding of what semiclassical gravity would have labeled the interior \cite{RajuLessons2022,LaddhaPrabhuRajuShrivastava2021,ChowdhuryPapadoulakiRaju2021,RajuSplitProperty2022}. In parallel, bulk reconstruction recast the same theme in the language of quantum error correction, entanglement wedge reconstruction, and recovery maps, culminating in explicit Petz map reconstructions and observer subsystem formulations of bulk time evolution \cite{AlmheiriDongHarlow2015,DongHarlowWall2016,ChenPeningtonSalton2020,BahiruVardian2023,JafferisLamprou2022,deBoerJafferisLamprou2022}.

The island formula and the gravitational path integral made these ideas more precise by showing that, after the Page time, part of the interior is indeed contained in the radiation through replica wormhole contributions to entanglement wedge reconstruction \cite{AlmheiriEngelhardtMarolfMaxfield2019,AlmheiriMahajanMaldacenaZhao2020,AlmheiriHartmanMaldacenaShaghoulianTajdini2020,Penington2020,PeningtonShenkerStanfordYang2022,StanfordYang2022,IliesiuEtAl2024,BlommaertChenNomura2024,AkersPenington2022,AkersEtAl2024}. But despite this progress, there still doesn't exist a principled resolution directly in the setting of the original AMPS experiment. The firewall paradox is really a question about sequential measurements made by infalling observers on naively independent gravitational subsystems, and hence relies on Lorentzian dynamics, neither of which can be accessed directly via Euclidean path integral methods. Moreover, the aforementioned results do not furnish a purely gravitational mechanism by which the AMPS paradox is resolved dynamically; rather, they hint at it kinematically.\footnote{Though with the exception of \cite{StanfordYang2022}, which does offer a Lorentzian mechanism by which firewalls, in the sense of high energy shocks at the horizon, can be understood, namely by including higher genus contributions to the Euclidean path integral and then analytically continuing to Lorentzian.}

More recently, building off the ideas above, Almheiri identified an elegant kinematical mechanism based upon the following ambiguity: which black hole interior, among the multiple candidate interiors introduced by the first measurement, does the second measurement actually land on \cite{AlmheiriFSP2025}? The paper uses quantum circuits to demonstrate how landing on the naive interior results in a nontrivial probability of seeing a firewall, corresponding to a ket $\tilde b$ contracting with a bra $\tilde b$ in the calculation of the probability $\langle \Psi_H|\Pi_{\tilde b b}D^{\dagger}\Pi_{be_b}D|\Psi_H\rangle$ in the Hawking state $|\Psi_H\rangle$. But the decoding unitary $D$ creates a second candidate interior labeled by an additional $\tilde b$ leg; if we allow the $e_b$ leg to contract with that additional $\tilde b$ leg, then the AMPS experiment always succeeds with probability one. But where would such a contraction rule arise from in quantum gravity? The argument given therein is essentially that entanglement wedge reconstruction implies $\tilde b \simeq e_b$.

\begin{figure}[t]
\centering
\begin{tikzpicture}[x=0.92cm,y=0.92cm,>=Latex,line cap=round,line join=round,every node/.style={font=\small}]
  \tikzset{
    bdry/.style={very thick},
    eow/.style={very thick,magenta!75!black},
    hor/.style={densely dashed,thick},
    slice/.style={black!75!black,thick},
    timedot/.style={fill=black,draw=black},
    cornerdot/.style={fill=red!75!black,draw=blue!75!black},
    modeR/.style={red!75!black,thick,decorate,decoration={snake,amplitude=1.35pt,segment length=5pt}},
    modeB/.style={blue!75!black,thick,decorate,decoration={snake,amplitude=1.35pt,segment length=5pt}},
    distillbox/.style={draw=black!70,rounded corners=2pt,fill=gray!8,inner sep=2.5pt},
    braceann/.style={decorate,thin}
  }

  \coordinate (TL) at (0.92, 2.95);
  \coordinate (BL) at (0.92,-2.95);
  \coordinate (TR) at (6.05, 2.95);
  \coordinate (BR) at (6.05,-2.95);
  \coordinate (Bif) at (2.62, 0.00);

  \draw[bdry] (TL) -- (TR);
  \draw[bdry] (TR) -- (BR);
  \draw[bdry] (BR) -- (BL);
  \path[name path=eowbrane] (TL) .. controls (0.18, 1.48) and (0.18,-1.48) .. (BL);
  \draw[eow] (TL) .. controls (0.18, 1.48) and (0.18,-1.48) .. (BL);

  \draw[hor,name path=hup] (Bif) -- (TR);
  \draw[hor] (Bif) -- (BR);
  \fill (Bif) circle (1.7pt);

  \coordinate (Tzero) at (6.05,-1.58);
  \coordinate (Tone)  at (6.05,-0.25);
  \coordinate (ebox)  at (6.0,-0.57);

  \fill[timedot] (Tzero) circle (1.5pt);
  \fill[timedot] (Tone) circle (1.5pt);

  \node[red!75!black] at (6.42,-1.58) {$E$};


  \node[left] at ($(Tzero)+(0.0,-0.35)$) {$T_0$};
  \node[left] at ($(Tone)+(0.0,0.35)$) {$T_1$};

  \node[scale=0.8] at (5.77,-0.91) {$\left\{\vphantom{\begin{matrix}x\\x\\x\\x\end{matrix}}\right.$};
  \node[left] at (5.75,-0.91) {$S_0$};

  \coordinate (Ttwo) at ($(Bif)!0.64!(TR)$);
  \filldraw[cornerdot] (Ttwo) circle (1.5pt);

  \coordinate (SR) at (6.02,1.56);
  \path[name path=sliceraw]
    (-0.05,1.08)
      .. controls (2.00,1.22) and (3.55,1.82) .. (Ttwo)
      .. controls (5.05,1.98) and (5.55,1.70) .. (SR);
  \path[name intersections={of=eowbrane and sliceraw, by=SliceLeft}];
  \draw[slice]
    (SliceLeft)
      .. controls (2.00,1.22) and (3.55,1.82) .. (Ttwo)
      .. controls (5.05,1.98) and (5.55,1.70) .. (SR);

  \coordinate (PairIn)  at (3.86,1.44);
  \coordinate (PairOut) at (4.86,1.44);

  \draw[modeB] (PairIn) -- ++(0.62,0.62);
  \node[blue!75!black,above left=1pt and 1pt] at ($(PairIn)+(0.5,0.4)$) {$\tilde b$};

  \draw[modeR] (PairOut) -- ++(0.62,0.62);
  \node[red!75!black,above right=1pt and 1pt] at ($(PairOut)+(0.52,0.4)$) {$b$};

  \node[black!75!black] at (1.90,1.75) {$\Sigma_{T_2}$};
  \node[black!75!black,above left=1pt and 1pt] at ($(Ttwo) + (0.35, 0.05)$) {$T_2$};
\end{tikzpicture}
\caption{Penrose diagram of the AMPS setup for a one sided black hole in JT gravity with a dynamical end of the world brane. The interval from $T_0$ to $T_1$ denotes the $\mathcal{O}({S_0})$ evolution needed to reach the Page time. We model an already old black hole with early radiation $E$ of size $e^{S_0}$ collected in an external reservoir. In our analysis the matter sector consists only of probe degrees of freedom, e.g.\ internal spin states, and couples to gravity only through gravitational dressing. It is solely the gravitational field that directly evolves under time evolution in our toy model. The later slice $\Sigma_{T_2}$ intersects the future horizon at the event labeled $T_2$. The outgoing late mode $b$ and its interior partner $\tilde b$ are identified in a neighborhood of this horizon crossing.} \label{fig:evapdiagrampenrose}
\end{figure}
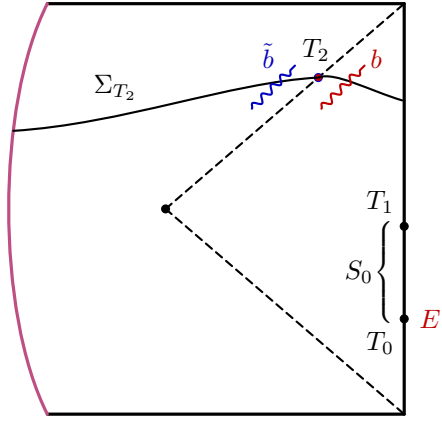

While this is certainly compelling, it is not clear when and why such a rule would be applicable; that is, how does gravity actually implement this rule dynamically? Our goal in this paper is to answer this question as much as possible from first principles using canonical quantum gravity, thereby providing an explicit bulk Lorentzian mechanism by which the AMPS paradox gets resolved. Canonical quantum gravity lends itself naturally to this problem, because the paradox is fundamentally about how a sequence of measurements is represented in a constrained gravitational theory: what postselected branch is prepared by the first purity measurement, how that branch is represented in the physical Hilbert space after imposing the Hamiltonian constraint, and which Dirac observable the infalling observer actually measures on it.\footnote{Throughout this paper when we say ``infalling observer'' we really have in mind the single observer that performs the entire AMPS experiment, from distillation $\rightarrow$ purity measurement $\rightarrow$ infall $\rightarrow$ horizon vacuum measurement.} Once phrased this way, one is led almost immediately to a relational description with an explicit clock subsystem carried by the infalling observer, conditional states at fixed clock reading, and Hamiltonian evolution with respect to the clock that can move support between topological sectors with different interior branch configurations.

From this perspective, topology change in canonical quantum gravity is the correct setting for the problem. To say that the distillation measurement has prepared a branch, and that the infalling observer's near horizon measurement acts on that branch, is to say that the physical Hilbert space must accommodate sectors with different interior branch configurations and must provide a notion of sequential Dirac observables relative to an observer clock. Topology change in canonical quantum gravity is, of course, far from fully understood. In a completely general theory of gravity, it is not at all clear how to incorporate topology change into the dynamics; we base our approach on \cite{MarolfRAQ,MarolfMaxfield2020,CasaliMarolfMaxfieldRangamani2021} and especially \cite{Maxfield:2022sio, Penington:2023dql}. A natural starting point is to write down an interaction Hamiltonian that can mix sectors of different topology or different numbers of connected components, and enters into Schr\"odinger evolution of the physical bulk state via deparametrization of the Hamiltonian constraint with respect to a bulk clock degree of freedom.

JT gravity makes this otherwise formidable problem tractable, especially in the black hole setting relevant for AMPS \cite{PeningtonShenkerStanfordYang2022, Penington:2023dql} (see Fig.~\ref{fig:evapdiagrampenrose}). We work with a one sided black hole in JT gravity whose interior ends on a dynamical end of the world brane \(B\). The brane replaces the second asymptotic region by a timelike worldline, thus allowing for an on shell one boundary phase space, as in \cite{GaoJafferisKolchmeyer2022}. Let \(\gamma\) denote the worldline of the infalling observer, and let \(\gamma_T:=\gamma(T)\) be the event selected by the observer clock reading \(T\). The clock conditioned state is defined on a (unique) smooth interval \(\Sigma_T\) running from \(\gamma_T\) to \(B\). This interval intersects the future horizon at a distinguished corner \(C_T=\Sigma_T\cap\mathscr H\). We then pass over to an extended Hilbert space description \cite{DonnellyFreidel2016} by splitting across $C_T$ into black hole interior and exterior. This introduces a pair of left and right boost edge modes at the split corners, and the resulting one interior sector is labeled by the corner boost together with the end of the world brane label and the hard mode label $\tilde b$ that appears in the AMPS experiment. This allows us to quantize each fixed interior sector explicitly, attach the finite dimensional probe matter labels only after quantization, and finally introduce topology change in the language of third quantization as an interaction between different topological sectors built from black hole interior creation and annihilation operators.

It should also be clear what is, and is not, being claimed here. We are neither claiming to have a resolution of the AMPS paradox in a general theory of quantum gravity, nor a fully general derivation of topology change in canonical quantum gravity. We are also not working in a fully realistic model of black hole evaporation. Rather, we study the problem within a simplified JT gravity setting wherein the salient ingredients can all be written down explicitly: a relational infalling observer clock subsystem, a nonperturbatively gauge invariant Lorentzian slice, a pair of pants interaction Hamiltonian that moves support between interior sectors under relational Schr\"odinger time evolution, and coarse graining over microscopic degrees of freedom. We therefore view this paper as demonstrating a proof of concept: it exhibits a concrete Lorentzian mechanism of gravitational dynamics by which a toy model of the AMPS experiment is resolved purely within canonical quantum gravity. The claim is that the mechanism exists and can be derived explicitly in a solvable setting.

The first ingredient is relational time evolution. We enlarge the kinematical Hilbert space by a clock subsystem and impose a single Hamiltonian constraint, so that physical states are obtained by refined algebraic quantization and group averaging \cite{MarolfRAQ, Chandrasekaran:2022cip}. This provides a clean algebraic notion of relational time and Schr\"odinger evolution via deparametrization of the full Hamiltonian constraint \cite{PageWootters1983}. In the case of an infalling observer in JT gravity, the proper time available to the observer is only $\mathcal O(1)$ in AdS units, so the corresponding clock states are effectively nonorthogonal; relational time must therefore be described by a positive operator valued measure rather than by sharp projective measurements.\footnote{While there is a single observer throughout the AMPS experiment, the nature of the clock changes between the purity measurement phase and the horizon vacuum phase. What changes is the operational clock POVM available to that observer. During the first phase, the clock has resolution of order \(e^{-S_0}\) because the observer follows the orbits of the asymptotic timelike Killing field. But in the second phase, after the observer jumps in, the clock becomes the local pointer variable accessible within the infalling observer's causal diamond. Since that diamond has only \(\mathcal{O}(1)\) duration in AdS units, the resolution becomes \(d_{\rm clock}=\mathcal{O}(1)\).}

The second ingredient is the extended Hilbert space construction \cite{DonnellyFreidel2016}, wherein we factorize the global one boundary Hilbert space across the horizon $\mathscr{H}$ by introducing gravitational edge modes (namely, the boost edge mode in the setting of JT gravity) at the corner $C_T$. Before the cut is made, \(\Sigma_T\) is a smooth one boundary interval. The gravitational constraints tie the two sides of the horizon together at \(C_T\), so there is no canonical factorization into independent interior and exterior Hilbert spaces. But for the AMPS calculation, wherein one wants to study the entangled state of the exterior Hawking mode \(b\) with an interior partner \(\tilde b\) before and after topology change acts on the interior part of the slice, we need to be able to treat the interior and exterior independently during intermediate stages of the calculation. To this aim, we excise a small interval of size $\varepsilon$ around \(C_T\), obtaining an interior interval \(\Sigma_T^{-}\) ending on a left horizon cut and an exterior interval \(\Sigma_T^{+}\) ending on a right horizon cut. The two cuts carry independent corner (boost) frames before gluing the interior back to the exterior via Marsden Weinstein (symplectic) reduction \cite{MarsdenWeinstein1974}.

The conjugate pairs which parametrize the corner phase space can be written as follows
\begin{align}
&J:=\mathscr{A}_L+\mathscr{A}_R, \ q:=\frac{s_L-s_R}{2}
\\
&K:=\mathscr{A}_R-\mathscr{A}_L, \ s:=-\frac{s_L+s_R}{2},
\end{align}
where $\mathscr{A}_{L,R}$ are the area operators of the respective corners, and $s_{L,R}$ the conjugate boost angles. The variable \(s\) is the smooth two sided boost deformation of $\Sigma_T$ across the horizon corner, and \(K\) is its conjugate two sided boost generator. The pair \((q,J)\) is introduced only in the extended phase space. The variable \(q\) is the gluing phase, equivalently the one sided boost discontinuity across the horizon cut, and \(J\) is its conjugate charge i.e.\ the generator of one sided boosts. Gluing imposes \(J=0\) and quotients by shifts of \(q\). Thus \((s,K)\) remains in the smooth one boundary phase space, while \((q,J)\) is annihilated by symplectic reduction.  

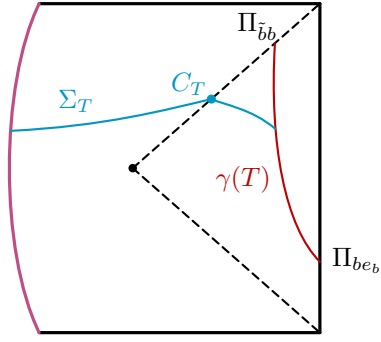
\begin{figure}[t]
\centering
\begin{tikzpicture}[x=0.75cm,y=0.75cm,>=Latex,line cap=round,line join=round,every node/.style={font=\small}]
  \tikzset{
    bdry/.style={very thick},
    eow/.style={very thick,magenta!75!black},
    hor/.style={densely dashed,thick},
    slice/.style={cyan!75!black,thick},
    world/.style={red!75!black, thick},
    cornerdot/.style={fill=cyan!75!black,draw=cyan!75!black},
    patch/.style={fill=blue!10,draw=blue!45!black,thin}
  }

  \coordinate (TL) at (0.90, 2.90);
  \coordinate (BL) at (0.90,-2.90);
  \coordinate (TR) at (5.85, 2.90);
  \coordinate (BR) at (5.85,-2.90);
  \coordinate (Bif) at (2.55, 0.00);

  \draw[bdry] (TL) -- (TR);
  \draw[bdry] (TR) -- (BR);
  \draw[bdry] (BR) -- (BL);

  \path[name path=eowpath]
    (TL) .. controls (0.20, 1.45) and (0.20,-1.45) .. (BL);
  \draw[eow]
    (TL) .. controls (0.20, 1.45) and (0.20,-1.45) .. (BL);

  \draw[hor] (Bif) -- (TR);
  \draw[hor] (Bif) -- (BR);
  \fill (Bif) circle (1.6pt);

  \coordinate (C1) at ($(Bif)!0.42!(TR)$);

  \coordinate (Tin) at (5.85,-1.65);
  \node[anchor=west,inner sep=1pt] at ($(Tin)+(0.16,0.02)$) {$\Pi_{be_b}$};

  \coordinate (Whor) at ($(Bif)!0.76!(TR)$);

  \path[name path=worldpath]
    (Tin) .. controls (5.25,-1.05) and (4.95,0.40) .. (Whor)
    coordinate[pos=0.70] (Send);
  \draw[world]
    (Tin) .. controls (5.25,-1.05) and (4.95,0.40) .. (Whor);

  \path[name path=slicepath]
    (-0.20,0.62)
      .. controls (2.05,0.74) and (3.10,1.02) .. (C1)
      .. controls (4.15,1.12) and (4.65,1.05) .. (Send);

  \path[name intersections={of=eowpath and slicepath, by=SL}];

  \draw[slice]
    (SL)
      .. controls (2.05,0.74) and (3.10,1.02) .. (C1)
      .. controls (4.15,1.12) and (4.65,1.05) .. (Send);

  \filldraw[cornerdot] (C1) circle (1.5pt);

  \node[cyan!75!black] at (1.55,1.18) {$\Sigma_T$};
  \node[text=cyan!75!black,above left=1pt and 0pt] at ($(C1)+(0.10,-0.15)$) {$C_T$};
  \node at (4.75,2.5) {$\Pi_{\tilde b b}$};
  \node[red!75!black, inner sep=0.5pt] at (4.5,-0.20) {$\gamma(T)$};
\end{tikzpicture}
\caption{Penrose diagram of the one sided black hole in JT gravity with a dynamical end of the world brane. The red curve $\gamma(T)$ is the infalling observer's worldline. The label $\Pi_{be_b}$ marks the early radiation purity measurement while $\Pi_{\tilde b b}$ is the horizon vacuum measurement. The cyan curve is the observer dressed slice $\Sigma_T$: it starts on the end of the world brane, intersects the future horizon $\mathscr H$ at the distinguished corner $C_T=\Sigma_T\cap\mathscr H$, and terminates on the observer worldline. The portion of $\Sigma_T$ to the left of $C_T$ is the interior part of the slice, while the portion to the right of $C_T$ is the exterior part ending on the observer.}
\label{fig:penroseinfall}
\end{figure}

We depict the Lorentzian kinematics in \cref{fig:penroseinfall}. The red curve is the infalling observer worldline \(\gamma\), and the red endpoint is the event \(\gamma_T\) selected by the clock. The cyan curve is the full interval \(\Sigma_T\), running from the end of the world brane to the observer. The point \(C_T\) is the horizon corner. The early radiation purity measurement is an exterior operation, while the horizon vacuum measurement is an operator dressed to the same full interval and then rewritten in the extended Hilbert space description.

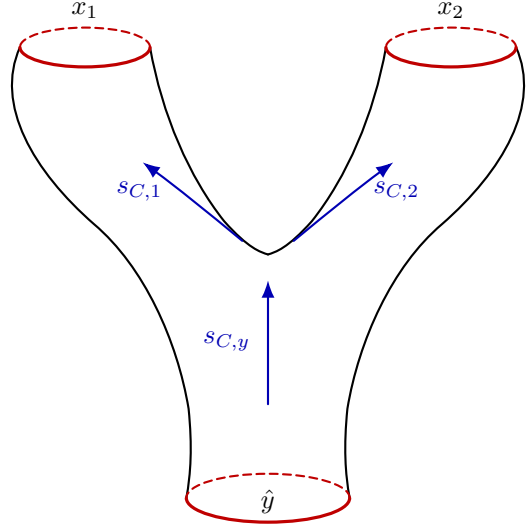
\begin{figure}[t]
\centering
\begin{tikzpicture}[x=1cm,y=1cm,>=Latex,line cap=round,line join=round,every node/.style={font=\small}]
  \tikzset{
    surf/.style={thick},
    rimvis/.style={very thick,red!75!black},
    rimhid/.style={red!75!black,densely dashed,thick},
    boost/.style={blue!70!black,->,thick},
  }

  \coordinate (TL) at (-2.42, 2.86);
  \coordinate (TR) at ( 2.42, 2.86);
  \coordinate (B)  at ( 0.00,-3.10);
  \coordinate (V)  at ( 0.00, 0.12);

  \def\rxt{0.86}
  \def\ryt{0.26}
  \def\rxb{1.08}
  \def\ryb{0.32}

  \draw[surf] ($(TL)+(-\rxt,0)$)
      .. controls (-3.62,2.10) and (-3.10,1.22) .. (-2.30,0.52)
      .. controls (-1.64,-0.04) and (-1.20,-1.02) .. (-1.05,-1.92)
      .. controls (-1.00,-2.42) and (-1.02,-2.78) .. ($(B)+(-\rxb,0)$);

  \draw[surf] ($(TR)+(\rxt,0)$)
      .. controls (3.62,2.10) and (3.10,1.22) .. (2.30,0.52)
      .. controls (1.64,-0.04) and (1.20,-1.02) .. (1.05,-1.92)
      .. controls (1.00,-2.42) and (1.02,-2.78) .. ($(B)+(\rxb,0)$);

  \draw[surf] ($(TL)+(\rxt,0)$)
      .. controls (-1.40,1.96) and (-1.04,1.14) .. (-0.58,0.56)
      .. controls (-0.26,0.20) and (-0.10,0.15) .. (V);

  \draw[surf] ($(TR)+(-\rxt,0)$)
      .. controls (1.40,1.96) and (1.04,1.14) .. (0.58,0.56)
      .. controls (0.26,0.20) and (0.10,0.15) .. (V);

  \draw[rimhid] ($(TL)+(\rxt,0)$) arc[start angle=0,end angle=180,x radius=\rxt,y radius=\ryt];
  \draw[rimvis] ($(TL)+(-\rxt,0)$) arc[start angle=180,end angle=360,x radius=\rxt,y radius=\ryt];

  \draw[rimhid] ($(TR)+(\rxt,0)$) arc[start angle=0,end angle=180,x radius=\rxt,y radius=\ryt];
  \draw[rimvis] ($(TR)+(-\rxt,0)$) arc[start angle=180,end angle=360,x radius=\rxt,y radius=\ryt];

  \draw[rimhid] ($(B)+(\rxb,0)$) arc[start angle=0,end angle=180,x radius=\rxb,y radius=\ryb];
  \draw[rimvis] ($(B)+(-\rxb,0)$) arc[start angle=180,end angle=360,x radius=\rxb,y radius=\ryb];

  \node[above] at ($(TL) + (0.0, 0.25) $) {$\hat{x}_1$};
  \node[above] at ($(TR) + (0.0, 0.25) $){$\hat{x}_2$};
  \node[below] at ($(B) + (0.0, 0.25)$) {$\hat{y}$};

  \draw[boost] (0.00,-1.86) -- (0.00,-0.22);
  \node[blue!70!black,left] at (-0.10,-1.02) {$s_{C,y}$};

  \draw[boost] (-0.34,0.30) .. controls (-0.80,0.68) and (-1.22,1.02) .. (-1.66,1.34);
  \node[blue!70!black,left] at (-1.26,0.96) {$s_{C,1}$};

  \draw[boost] (0.34,0.30) .. controls (0.80,0.68) and (1.22,1.02) .. (1.66,1.34);
  \node[blue!70!black,right] at (1.26,0.96) {$s_{C,2}$};
\end{tikzpicture}
\caption{Pair of pants topology changing vertex in the extended Hilbert space description of the black hole interior. The lower horizon slot $\hat y$ is the parent interior leg and the upper slots $\hat x_1,\hat x_2$ are the two daughter interior legs. Here a hatted label denotes the complete one leg data: $\hat x_i=(s_{C,i},x_i),\ x_i=(\tilde r_i,\tilde i_i)$,  and similarly for $\hat{y}$, so each slot carries the horizon corner boost frame together with the end of the world brane label and the interior hard mode label. The red rims denote the horizon cuts of these legs, and the black surface is a schematic representation of the local cubic vertex that splits one interior leg into two, with the inverse process giving the $2\to1$ half of the doubled kernel. The blue arrows label the boost edge mode variables $s_{C,y},s_{C,1},s_{C,2}$ at the corresponding horizon corners. Locality at the horizon imposes the boost composition law $s_{C,y}=s_{C,1}+s_{C,2}$.}
\label{fig:pairofpants}
\end{figure}

The third ingredient is topology change itself, which is introduced only after passing to the extended Hilbert space. After quantization, we introduce an interaction Hamiltonian that dynamically changes the number of interior legs. An interior leg means the interior part of the observer dressed slice $\Sigma_T$ after cutting across the horizon, with one endpoint at the left horizon cut and the other endpoint on the end of the world brane. After quantization, a complete one leg label has the form $\hat x=(s_C,x), \ x=(\tilde r,\tilde i)$. Here \(s_C\) is the horizon corner boost frame, \(\tilde i\) is the interior hard mode label for the interior partner $\tilde b$, and \(\tilde r\) is the internal end of the world brane label corresponding to the remaining interior radiation. The exterior corner frame, the late mode \(b\), and the early radiation \(E\) are common to all sectors. The leading order interaction is a cubic pair of pants Hamiltonian. It maps one interior leg to two interior legs, and its adjoint maps two interior legs back to one. In the topological expansion of JT gravity the coupling scales as \(\lambda\sim e^{-S_0}\). The vertex is local at the horizon corner in the following sense: its universal edge mode part is the quantization of the \(SO(1,1)\) group multiplication map, so in rapidity coordinates the daughter and parent corner frames obey \(s_{C,y}=s_{C,1}+s_{C,2}\). Fig.~\ref{fig:pairofpants} depicts this local topology changing interaction.

\begin{figure}[!t]
\captionsetup{width=0.93\linewidth}
\centering
\begin{tikzpicture}[x=0.90cm,y=0.90cm,>=Latex,line cap=round,line join=round,every node/.style={font=\small}]
  \tikzset{
    slice/.style={cyan!75!black,very thick},
    cut/.style={blue!75!black,very thick},
    eowdot/.style={fill=magenta!75!black,draw=magenta!75!black},
    obsdot/.style={fill=red!75!black,draw=red!75!black},
    hdot/.style={fill=blue!75!black,draw=blue!75!black},
    flow/.style={-{Latex[length=2.0mm,width=1.45mm]},line width=0.75pt,draw=black!70},
    panellabel/.style={font=\bfseries},
    lab/.style={font=\small},
    slabel/.style={font=\scriptsize},
    qlabel/.style={font=\small,text=blue!75!black,fill=white,inner sep=1.2pt}
  }

  \def\FlowLen{0.78}

  \def\ya{0.00}
  \def\yb{-2.40}
  \def\yc{-5.35}
  \def\yd{-8.0}
  \def\ye{-10.90}

  \begin{scope}[shift={(0,\ya)}]
    \node[panellabel,anchor=west] at (-3.10,0.72) {(a)};

    \draw[slice] (-2.55,0) -- (2.55,0);
    \filldraw[eowdot] (-2.55,0) circle (2.2pt);
    \filldraw[hdot]    (0,0)    circle (1.9pt);
    \filldraw[obsdot]  (2.55,0) circle (2.2pt);

    \node[lab,text=cyan!75!black] at (-1.02,0.33) {$\Sigma_T$};
    \node[lab,text=blue!75!black] at (0,0.35) {$C_T$};
  \end{scope}

  \draw[flow] (0,-0.55) -- ++(0,-\FlowLen)
    node[midway,right=4pt,fill=white,inner sep=1pt] {$\mathcal C_{\mathscr H}^{\varepsilon}$};

  \begin{scope}[shift={(0,\yb)}]
    \node[panellabel,anchor=west] at (-3.10,0.88) {(b)};

    \draw[slice] (-2.90,0) -- (-0.58,0);
    \draw[cut]   (-0.58,-0.38) -- (-0.58,0.38);

    \draw[cut]   (0.58,-0.38) -- (0.58,0.38);
    \draw[slice] (0.58,0) -- (2.90,0);

    \filldraw[eowdot] (-2.90,0) circle (2.2pt);
    \filldraw[obsdot] ( 2.90,0) circle (2.2pt);

    \node[slabel] at (-1.70,-0.36) {$\Sigma_T^{-}$};
    \node[slabel] at ( 1.70,-0.36) {$\Sigma_T^{+}$};

    \node[lab,text=blue!75!black] at (-0.58,0.82) {$C_T^{-}$};
    \node[lab,text=blue!75!black] at ( 0.58,0.82) {$C_T^{+}$};
  \end{scope}

  \draw[flow] (0,-3.20) -- ++(0,-\FlowLen);

  \begin{scope}[shift={(0,\yc)}]
    \node[panellabel,anchor=west] at (-4.15,1.05) {(c) interior $1\to2\to1$};

    \filldraw[eowdot] (-3.35,0) circle (2.0pt);
    \draw[slice] (-3.35,0) -- (-2.12,0);
    \draw[cut]   (-2.12,-0.38) -- (-2.12,0.38);

    \draw[flow] (-1.82,0) -- ++(\FlowLen,0);
    \node[slabel,above=1pt] at (-1.43,0) {$V_{\rm pants}$};

    \filldraw[eowdot] (-0.60, 0.52) circle (2.0pt);
    \draw[slice] (-0.60, 0.52) -- (0.82, 0.52);
    \draw[cut]   ( 0.82, 0.12) -- (0.82, 0.92);

    \filldraw[eowdot] (-0.60,-0.52) circle (2.0pt);
    \draw[slice] (-0.60,-0.52) -- (0.82,-0.52);
    \draw[cut]   ( 0.82,-0.92) -- (0.82,-0.12);

    \draw[flow] (1.12,0) -- ++(\FlowLen,0);
    \node[slabel,above=1pt] at (1.51,0) {$V_{\rm pants}^{\dagger}$};

    \filldraw[eowdot] (2.16,0) circle (2.0pt);
    \draw[slice] (2.16,0) -- (3.39,0);
    \draw[cut]   (3.39,-0.38) -- (3.39,0.38);
  \end{scope}

  \draw[flow] (0,-6.18) -- ++(0,-\FlowLen);

  \begin{scope}[shift={(0,\yd)}]
    \node[panellabel,anchor=east] at (-5.05,0.92) {(d)};
    \node[panellabel] at (-2.95,0.92) {direct};
    \node[panellabel] at ( 2.95,0.92) {exchange};

    \coordinate (Dgap) at (-2.95,0);

    \draw[cut] ($(Dgap)+(-0.38,-0.24)$) -- ($(Dgap)+(-0.38,0.24)$);
    \draw[cut] ($(Dgap)+( 0.38,-0.24)$) -- ($(Dgap)+( 0.38,0.24)$);

    \filldraw[eowdot] ($(Dgap)+(-2.15,-0.42)$) circle (2.1pt);
    \draw[slice] ($(Dgap)+(-2.15,-0.42)$) -- ($(Dgap)+(-0.38,0)$);

    \draw[slice] ($(Dgap)+(0.38,0)$) -- ($(Dgap)+(2.15,0)$);
    \filldraw[obsdot] ($(Dgap)+(2.15,0)$) circle (2.1pt);

    \node[qlabel] at ($(Dgap)+(0,0.52)$) {$q\neq0$};

    \coordinate (Xgap) at (2.95,0);

    \draw[cut] ($(Xgap)+(-0.38,-0.24)$) -- ($(Xgap)+(-0.38,0.24)$);
    \draw[cut] ($(Xgap)+( 0.38,-0.24)$) -- ($(Xgap)+( 0.38,0.24)$);

    \filldraw[eowdot] ($(Xgap)+(-2.15,0)$) circle (2.1pt);
    \draw[slice] ($(Xgap)+(-2.15,0)$) -- ($(Xgap)+(-0.38,0)$);

    \draw[slice] ($(Xgap)+(0.38,0)$) -- ($(Xgap)+(2.15,0)$);
    \filldraw[obsdot] ($(Xgap)+(2.15,0)$) circle (2.1pt);

    \node[qlabel] at ($(Xgap)+(0,0.52)$) {$q=0$};
  \end{scope}

  \draw[flow] (0,-8.92) -- ++(0,-\FlowLen)
    node[midway,right=4pt,fill=white,inner sep=1pt] {$\mathcal G_{\mathscr H}^{\varepsilon}$};

  \begin{scope}[shift={(0,\ye)}]
    \node[panellabel] at (0,0.78) {(e) exchange};

    \draw[slice] (-2.55,0) -- (2.55,0);
    \filldraw[eowdot] (-2.55,0) circle (2.2pt);
    \filldraw[hdot]    (0,0)    circle (1.9pt);
    \filldraw[obsdot]  (2.55,0) circle (2.2pt);

    \node[qlabel] at (0,0.36) {$q=0$};
  \end{scope}

\end{tikzpicture}
\vspace{-0.35em}
\caption{Cutting, topology change, and gluing. (a) shows the full observer dressed interval $\Sigma_T$ from the end of the world brane, shown by the magenta endpoint, to the observer event $\gamma(T)$, shown by the red endpoint, with horizon corner $C_T$. (b) applies the cutting map $\mathcal C_{\mathscr H}^{\varepsilon}$, producing an interior interval $\Sigma_T^{-}$ and an exterior interval $\Sigma_T^{+}$ separated by cutoff $\varepsilon$. (c) shows the interior topology changing process $1\to2\to1$, generated by the pair of pants Hamiltonian $V_{\rm pants}$. (d) shows the two candidate channels before gluing. The variable $q$ is the boost discontinuity across the split horizon cuts. The direct channel carries $q\neq0$, while the exchange channel has $q=0$. The gluing map $\mathcal G_{\mathscr H}^{\varepsilon}$ keeps the exchange channel and yields the smooth surviving glued state in (e).}
\label{fig:1dhorizonrep}
\end{figure}
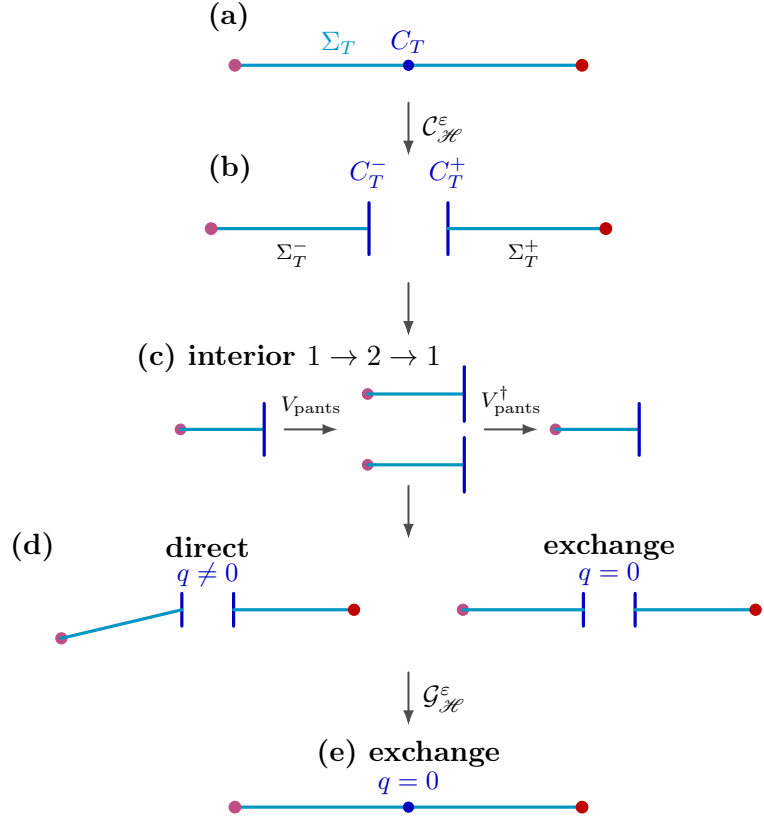

The physical picture is summarized in Fig.~\ref{fig:1dhorizonrep}. One starts from the smooth interval \(\Sigma_T\), applies the cutting map \(\mathcal C_{\mathscr H}^{\varepsilon}\), evolves the interior sector through the \(1\to2\to1\) topology changing process generated by a pair of pants interaction Hamiltonian \(\hat{V}_{\rm pants}\), and obtains two candidate contraction channels before gluing. The direct channel carries \(q\neq0\), so it has a one sided boost discontinuity across the cut. The exchange channel has \(q=0\). The gluing map \(\mathcal G_{\mathscr H}^{\varepsilon}\) keeps only the exchange channel and returns a smooth one boundary state. 

Importantly, $q\neq 0$ corresponds to a firewall. The point is that the interior mode \(\tilde b\) needs to be gravitationally dressed. On the full observer dressed slice \(\Sigma_T\), this dressing is fixed by the clock reading $T$. After we split the global one boundary phase space across the horizon, the same insertion is instead gravitationally dressed to the left boost frame \(s_L\), while the exterior mode is dressed to the right boost frame \(s_R\). But this means the $\tilde b$ matter algebra does not commute with the $s_L$ edge mode algebra. In other words, the interior algebra on the extended Hilbert space is the crossed product algebra \cite{Chandrasekaran:2022cip, Chandrasekaran:2022eqq, Chandrasekaran:2026pnc}
\begin{align}
\widehat{\mathfrak B}_{\tilde b}
=
{\rm End}(\mathcal H_{\tilde b})\rtimes_{s_L}\mathbb R .
\end{align}
But since $s_L = q-s$, and $s$ remains fixed whenever we hold the exterior field configuration fixed, it follows that we send $q = 0$ to $q\neq 0$ under the $s_L$ automorphism above. By the covariant structure of a crossed product algebra, this means we can view a $q \neq 0$ state as equivalently a one sided boost of $\tilde b$ relative to $b$, which is exactly a firewall since such a state is projected out by $\Pi_{\tilde b b}$.

The topology changing Hamiltonian prepares the connected branch used in the AMPS calculation. By connected branch we mean the component of the clock conditioned state in which the original interior leg has undergone a pair of pants transition and the interior sector contains two daughter legs. This is the gravitational version of the additional candidate interior in the quantum circuit model of \cite{AlmheiriFSP2025}. The first AMPS measurement does not merely project out the early radiation. In the gravitational theory it is evaluated after relational evolution has had time to transfer support from the one leg sector into this connected two leg sector. The reason this branch is relevant is precisely that the later horizon vacuum measurement has two possible daughter legs on which it can act.

\begin{figure}[t]
\centering
\begin{tikzpicture}[x=1cm,y=1cm,>=Latex,line cap=round,line join=round,every node/.style={font=\small}]
  \tikzset{
    ket/.style={blue!70!black,thick},
    bra/.style={red!70!black,thick},
    tick/.style={black!45,thin},
    vertex/.style={fill=black,draw=black,circle,inner sep=1.4pt},
    op/.style={draw=black!65,fill=gray!7,rounded corners=2pt,inner sep=3pt},
    overlap/.style={black!55,densely dashed,thick}
  }

  \coordinate (ketL) at (0,0);
  \coordinate (ketR) at (9.4,0);
  \coordinate (braL) at (0,1.75);
  \coordinate (braR) at (9.4,1.75);

  \draw[blue!20,line width=5pt] (2.10,0) -- (3.15,0);
  \draw[red!20,line width=5pt] (6.30,1.75) -- (7.35,1.75);

  \draw[ket,->] (ketL) -- (ketR);
  \draw[bra,->] (braR) -- (braL);

  \foreach \x in {0,1.05,2.10,3.15,4.20,5.25,6.30,7.35,8.40,9.40} {
    \draw[tick] (\x,-0.10) -- (\x,0.10);
    \draw[tick] (\x,1.65) -- (\x,1.85);
  }

  \draw[black!70,thick] (ketR) -- (braR);
  \node[op,right=3pt] at ($(ketR)!0.5!(braR)$) {$P_{\rm conn}^{(2)}$};

  \node[left=4pt] at (ketL) {$|\eta(\Psi_{\rm H})\rangle$};
  \node[left=4pt] at (braL) {$\langle\eta(\Psi_{\rm H})|$};

  \node[below=4pt] at (ketL) {$T_0$};
  \node[below=4pt] at (ketR) {$T_1$};

  \node[below=7pt] at (2.625,0) {$I_j$};
  \node[above=7pt] at (6.825,1.75) {$I_{j'}$};

  \coordinate (vket) at (2.63,0);
  \coordinate (vbra) at (6.82,1.75);

  \filldraw[vertex] (vket) circle (1.7pt);
  \filldraw[vertex] (vbra) circle (1.7pt);

  \node[below=19pt,align=center] at (vket)
  {$T\in I_j$};

  \node[above=19pt,align=center] at (vbra)
  {$T'\in I_{j'}$};

  \draw[overlap] (vket) -- node[fill=white,inner sep=1.5pt,sloped,above] {$\langle T'|T\rangle$} (vbra);
\end{tikzpicture}
\caption{Schwinger Keldysh contour for the inclusive connected branch probability. The interval from \(T_0\) to \(T_1\) is of order the Page time $\sim S_0$ and is divided into timefolds \(I_j\) of size \(O(1)\). At leading order, a pair of pants vertex at \(T\in I_j\) on the forward branch is paired with its adjoint at \(T'\in I_{j'}\) on the backward branch. Group averaging produces the clock propagator \(\langle T'|T\rangle\) when computing the correlator. The coincident timefold terms yield the local transition rate, while separated timefolds are suppressed by the clock propagator, since $\langle T'|T\rangle \sim e^{-S_0}$ before the observer jumps into the black hole. The insertion at the turn around projects onto the connected two leg sector.}
\label{fig:intro_sk_pconn}
\end{figure}
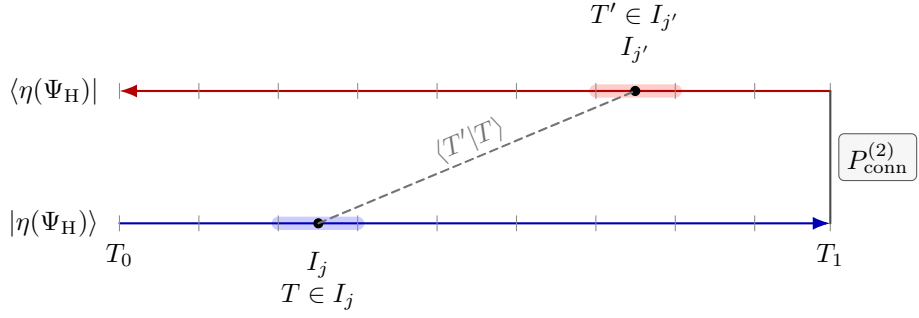

Between \(T_0\) and \(T_1\), relational evolution is generated by
\begin{align}
\hat H_{\rm grav}
=
\hat K+\lambda \hat V_{\rm pants}.
\end{align}
Here \(\hat K\) is the number preserving boost Hamiltonian on the interior leg sectors, while \(\widehat V_{\rm pants}\) maps one interior leg to two and its adjoint maps two legs back to one. The branch probability is computed as an inclusive probability on a Schwinger Keldysh contour,
\begin{align}
P_{\rm conn}(T_1)
=
\langle \eta(\Psi_{\rm H})|
U(T_1,T_0)^{\dagger}
P_{\rm conn}^{(2)}
U(T_1,T_0)
|\eta(\Psi_{\rm H})\rangle_{\rm phys},
\end{align}
where 
\begin{align}
&|\Phi\rangle_{\tilde b b}
:=
\frac{1}{\sqrt{d_b}}
\sum_{i=1}^{d_b}
|\tilde i\rangle_{\tilde b}\otimes |i\rangle_b, \ |\chi_0\rangle_{\tilde R E}
:=
\sum_{\tilde r,a}
c_{\tilde r a}
|\tilde r\rangle_{\tilde R}\otimes |a\rangle_E,
\\
&|\Psi_{\rm H}(T)\rangle
:=
\int \Dd s\,
\Psi_{\rm H}(T;s)
|T;s\rangle
\otimes
|\Phi\rangle_{\tilde b b}
\otimes
|\chi_0\rangle_{\tilde R E},
\end{align}
is the semiclassical Hawking state \cite{AlmheiriFSP2025, Penington2020, PeningtonShenkerStanfordYang2022, AlmheiriHartmanMaldacenaShaghoulianTajdini2020, AlmheiriEngelhardtMarolfMaxfield2019, AlmheiriMahajanMaldacenaZhao2020}, $\eta(\Psi_H)$ is the rigging map (group average), \(P_{\rm conn}^{(2)}\) projects onto the connected two leg sector, and $U(T_1, T_0)$ is the relational time evolution unitary which to leading order in $\lambda$ is just $\hat{K}$. Expanding the contour in \(\lambda\), the first nonzero contribution is second order in the pair of pants vertex. The two vertices appear on the two sides of the inclusive contour, and relational time group averaging suppresses long range interference between distinct timefolds. See \cref{fig:intro_sk_pconn}. This reduces the growth of the connected branch to an ordinary rate equation,
\begin{align}
&\frac{dP_{\rm conn}}{dT}
=
\Lambda(T)\bigl(1-P_{\rm conn}(T)\bigr), \ \Lambda(T)
\sim
|\lambda|^2\sigma(T) k\rho_{\rm eff},
\end{align}
where \(\rho_{\rm eff}\) is the effective density of the boost generator, \(k=\dim\mathcal H_{\tilde E}\) is the effective number of
unobserved environment degrees of freedom traced out in the inclusive connected
branch probability computation, and \(\sigma(T)\) is an \(\mathcal O(1)\) time dependent
coefficient. In the old black hole regime, the pair of pants coupling scales as \(\lambda\sim e^{-S_0}\), while the number of relevant radiation channels and the effective density of connected daughter states each scale as $e^{S_0}$. Moreover, it takes Hamiltonian evolution of order $\sim S_0$ to reach the Page time. Thus the suppression of a single topology changing vertex is compensated by the large density of available connected states over the Page time evolution. In the $S_0 \gg 1$ limit, the probability of tunneling from a one leg branch to a connected two leg branch becomes approximately unity by the Page time,
\begin{align}
P_{\rm conn}(T_1)
=
1-\mathcal O(e^{-cS_0})
\end{align}
for some positive constant \(c\) of order one. This is the gravitational branch on which the purity measurement is evaluated.

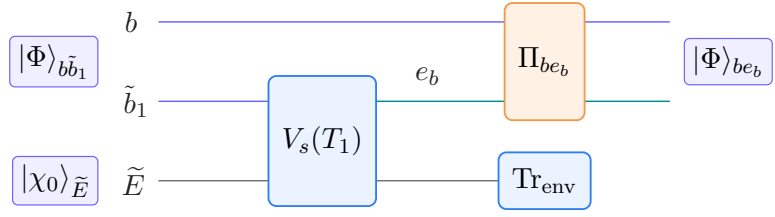
\begin{figure}[t]
\centering

\definecolor{figBlue}{HTML}{2F80ED}
\definecolor{figBlueLight}{HTML}{EAF3FF}
\definecolor{figTeal}{HTML}{008C8C}
\definecolor{figTealLight}{HTML}{E7F7F7}
\definecolor{figOrange}{HTML}{F2994A}
\definecolor{figOrangeLight}{HTML}{FFF2E5}
\definecolor{figPurple}{HTML}{6C63FF}
\definecolor{figPurpleLight}{HTML}{F0EFFF}
\definecolor{figGray}{HTML}{737373}
\definecolor{figText}{HTML}{222222}

\begin{tikzpicture}[
  x=1cm,
  y=1cm,
  wire/.style={
    line width=0.55pt,
    line cap=round
  },
  bwire/.style={wire, draw=figPurple},
  ewire/.style={wire, draw=figTeal},
  auxwire/.style={wire, draw=figGray},
  box/.style={
    draw=figText,
    fill=white,
    line width=0.45pt,
    align=center,
    inner sep=3pt,
    minimum height=0.62cm,
    rounded corners=2pt
  },
  channelbox/.style={
    draw=figBlue,
    fill=figBlueLight,
    line width=0.70pt,
    align=center,
    inner sep=5pt,
    rounded corners=3pt
  },
  measbox/.style={
    draw=figOrange,
    fill=figOrangeLight,
    line width=0.70pt,
    align=center,
    inner sep=5pt,
    rounded corners=3pt
  },
  statebox/.style={
    draw=figPurple,
    fill=figPurpleLight,
    line width=0.45pt,
    align=center,
    inner sep=3pt,
    rounded corners=2pt
  },
  smalllabel/.style={
    font=\scriptsize,
    text=figGray
  },
  lab/.style={
    text=figText
  }
]

\def\xin{0.00}
\def\xVleft{1.55}
\def\xVright{2.78}
\def\xPileft{4.65}
\def\xPiright{5.55}
\def\xout{6.75}

\def\yb{1.05}
\def\ye{0.00}
\def\yaux{-1.05}

\draw[bwire] (\xin,\yb) -- (\xPileft,\yb);
\draw[bwire] (\xin,\ye) -- (\xVleft,\ye);
\draw[auxwire] (\xin,\yaux) -- (\xVleft,\yaux);

\draw[ewire] (\xVright,\ye) -- (\xPileft,\ye);
\draw[auxwire] (\xVright,\yaux) -- (5.10,\yaux);

\draw[bwire] (\xPiright,\yb) -- (\xout,\yb);
\draw[ewire] (\xPiright,\ye) -- (\xout,\ye);

\node[channelbox, minimum width=1.23cm, minimum height=1.72cm]
  at (2.16,-0.525) {$V_s(T_1)$};

\node[measbox, minimum width=0.92cm, minimum height=1.55cm]
  at (5.10,0.525) {$\Pi_{b e_b}$};

\node[channelbox, minimum width=0.92cm, minimum height=0.75cm]
  at (5.10,\yaux) {$\rm{Tr}_{\rm{env}}$};

\node[lab, anchor=east] at (-0.15,\yb) {$b$};
\node[lab, anchor=east] at (0.0,\ye) {$\tilde b_1$};
\node[lab, anchor=east] at (-0.05,\yaux) {$\widetilde E$};

\node[statebox, anchor=east] at (-0.80,0.525)
  {$|\Phi\rangle_{b\tilde b_1}$};

\node[statebox, anchor=east] at (-0.80,\yaux)
  {$|\chi_0\rangle_{\widetilde E}$};

\node[lab, above=0.08cm] at (3.55,\ye) {$e_b$};

\node[statebox, anchor=west] at (6.94,0.525)
  {$|\Phi\rangle_{b e_b}$};

\end{tikzpicture}

\caption{
Circuit representation of the connected branch purity measurement.
The input pair is \( |\Phi\rangle_{b\tilde b_1}\), with unobserved state
\( |\chi_0\rangle_{\widetilde E}\). The map \(V_s(T_1)\) acts only on
\(\tilde b_1\otimes \widetilde E\), producing the recovered subsystem \(e_b\)
and an environment, while \(b\) is unchanged. The state immediately after the
channel is \( |\Psi_s(T_1)\rangle\). After tracing out the environment, the physical
purity measurement is \(\Pi_{b e_b}\) on \(b\otimes e_b\), and the output legs
indicate the postselected branch.
}
\label{fig:connected_branch_channel}
\end{figure}

The distillation step is then analyzed on the connected branch already prepared by the topology changing gravitational dynamics described above. Fix a connected two leg branch and a value of the smooth two sided boost \(s\). Let \(\tilde E\) denote the unobserved environment, consisting of the remaining early radiation together with the interior labels not measured by the later horizon vacuum measurement. The distillation operator and relational time evolution unitary restricted to this branch define a recovery channel
\begin{align}
V_s(T_1)
&:=
D\,U_{{\rm conn},s}(T_1,T_0):
\mathcal H_{\tilde b_1}\otimes\mathcal H_{\tilde E}
\to
\mathcal H_{e_b}\otimes\mathcal H_{\rm env}.
\end{align}
The only input from the Hayden Preskill protocol \cite{HaydenPreskill2007, YoshidaKitaev2017} that we make use of in this paper is decoupling of the complementary channel: the unobserved environment carries negligible information about the late mode \(b\). Since \(b\) initially purifies the first daughter hard mode $\tilde b$, this implies that there is a branch dependent unitary $U_s:\mathcal H_{\tilde b_1}\to \mathcal H_{e_b}$ such that
\begin{align}
(1_b\otimes U_s^{\dagger})
\Pi_{b e_b}
(1_b\otimes U_s)
&=
\frac{1}{d_b}
\sum_{i,j=1}^{d_b}
|i\rangle_b\langle j|
\otimes
|\tilde i\rangle_{\tilde b_1}\langle \tilde j|.
\end{align}
Thus the successful purity projection contracts the \(b e_b\) indices and maps the index originally labeled by \(\tilde b_1\) onto that of the recovered subsystem \(e_b\), up to the decoding error. This is the sense in which the purity measurement supplies the index structure that the later horizon measurement must reproduce. See \cref{fig:connected_branch_channel}.

This leads us to our final result. Let $U(T_2,T_1)$ denote the relational time evolution operator between the two clock readings $T_1$ and $T_2$ measured by the infalling observer, with the same Hamiltonian constraint used to define the physical clock conditioned state. The horizon vacuum insertion at \(T_2\) may be pulled back by this relational evolution and compared with the purity insertion at \(T_1\). On the connected postselected branch one finds, in physical matrix elements on the relevant states,
\begin{align}
U(T_2,T_1)^{\dagger}
\Pi_{\tilde b b}(T_2)
U(T_2,T_1)
\sim_{\rm phys}
\Pi_{be_b}(T_1)
+
\mathcal O(e^{-cS_0}).
\end{align}
Thus the horizon vacuum measurement and the early radiation purity measurement define the same Dirac observable on the connected branch, once both are written in the extended Hilbert space and returned to the one boundary Hilbert space by gluing. The direct firewall contraction is not a smooth one boundary observable. The exchange contraction is. In this way, the paradox is resolved neither by modifying the rules of quantum mechanics nor by declaring the AMPS experiment meaningless, but rather by formulating the experiment correctly in a relational gravitational Hilbert space with dynamical topology change.

There is an equivalent algebraic way to state the result. As we've discussed, near the horizon, gravitational dressing
turns the interior hard mode algebra into the crossed product $\widehat{\mathfrak B}_{\tilde b}$ so that a relative phase between \(\tilde b b\) may also be represented as a discontinuous one sided boost between the left and right corner frame. After a Page time, the
topology changing Hamiltonian produces an analogous but global crossed product structure on
the connected branch. The relevant automorphism is no longer the small horizon boost
automorphism \(\alpha^-\), but a large diffeomorphism \(\Theta_s\) generated by relational
gravitational time evolution. It exchanges the interior Hawking partner representative with the
decoded early radiation representative. \cref{sec:emergent_nonlocality} makes this precise by showing that
\begin{align}
\mathfrak A_{\rm Page}(s)
=
\left(
\mathcal A_{\tilde b_1}(s)
\oplus
\mathcal A_{e_b}(s)
\right)
\rtimes_{\Theta_s}\mathbb Z_2,
\end{align}
where $\Theta_s$ satisfies $\Theta_s(B_{ij}(s)\oplus 0)=0\oplus E^{(s)}_{ij}$ and $\Theta_s(0\oplus E^{(s)}_{ij})=B_{ij}(s)\oplus 0$. Thus the relation \(\tilde b\simeq e_b\) is not merely a kinematical property of the gravitational Hilbert space. It is the result of a large diffeomorphism generated dynamically by relational time evolution under the topology changing Hamiltonian, with accuracy controlled by the Hayden Preskill protocol \cite{HaydenPreskill2007}.

\begin{figure}[t]
\centering
\begin{tikzpicture}[x=1cm,y=1cm,>=Latex,line cap=round,line join=round,every node/.style={font=\small}]
  \tikzset{
    surf/.style={thick},
    rimvis/.style={very thick,red!75!black},
    rimhid/.style={red!75!black,densely dashed,thick},
    dress/.style={orange!80!black,very thick},
    dresspath/.style={orange!80!black,densely dashed,thick},
    conn/.style={black!70,thick},
    modemark/.style={orange!85!black,decorate,decoration={snake,amplitude=1.1pt,segment length=3.6pt},thick},
    kernellabel/.style={font=\scriptsize,fill=white,inner sep=1.2pt,text opacity=1,fill opacity=0.92}
  }

  \newcommand{\DrawBottomPants}{%
    \draw[surf] (-0.82,-3.05)
      .. controls (-0.88,-2.58) and (-0.94,-1.98) .. (-1.02,-1.42)
      .. controls (-1.10,-0.88) and (-1.34,-0.42) .. (-1.74,-0.20);
    \draw[surf] ( 0.82,-3.05)
      .. controls ( 0.88,-2.58) and ( 0.94,-1.98) .. ( 1.02,-1.42)
      .. controls ( 1.10,-0.88) and ( 1.34,-0.42) .. ( 1.74,-0.20);
    \draw[surf] (0.00,-1.85)
      .. controls (0.00,-1.62) and (-0.10,-1.34) .. (-0.30,-1.05)
      .. controls (-0.52,-0.72) and (-0.72,-0.48) .. (-0.90,-0.20);
    \draw[surf] (0.00,-1.85)
      .. controls (0.00,-1.62) and ( 0.10,-1.34) .. ( 0.30,-1.05)
      .. controls ( 0.52,-0.72) and ( 0.72,-0.48) .. ( 0.90,-0.20);
    \draw[rimhid] (0.82,-3.05) arc[start angle=0,end angle=180,x radius=0.82,y radius=0.18];
    \draw[rimvis] (-0.82,-3.05) arc[start angle=180,end angle=360,x radius=0.82,y radius=0.18];
    \draw[rimhid] (-0.90,-0.20) arc[start angle=0,end angle=180,x radius=0.42,y radius=0.12];
    \draw[rimvis] (-1.74,-0.20) arc[start angle=180,end angle=360,x radius=0.42,y radius=0.12];
    \draw[rimhid] ( 1.74,-0.20) arc[start angle=0,end angle=180,x radius=0.42,y radius=0.12];
    \draw[rimvis] ( 0.90,-0.20) arc[start angle=180,end angle=360,x radius=0.42,y radius=0.12];
  }

  \newcommand{\DrawTopPants}{%
    \draw[surf] (-0.82,3.05)
      .. controls (-0.88,2.58) and (-0.94,1.98) .. (-1.02,1.42)
      .. controls (-1.10,0.88) and (-1.34,0.42) .. (-1.74,0.20);
    \draw[surf] ( 0.82,3.05)
      .. controls ( 0.88,2.58) and ( 0.94,1.98) .. ( 1.02,1.42)
      .. controls ( 1.10,0.88) and ( 1.34,0.42) .. ( 1.74,0.20);
    \draw[surf] (0.00,1.85)
      .. controls (0.00,1.62) and (-0.10,1.34) .. (-0.30,1.05)
      .. controls (-0.52,0.72) and (-0.72,0.48) .. (-0.90,0.20);
    \draw[surf] (0.00,1.85)
      .. controls (0.00,1.62) and ( 0.10,1.34) .. ( 0.30,1.05)
      .. controls ( 0.52,0.72) and ( 0.72,0.48) .. ( 0.90,0.20);
    \draw[rimhid] (-0.90,0.20) arc[start angle=0,end angle=180,x radius=0.42,y radius=0.12];
    \draw[rimvis] (-1.74,0.20) arc[start angle=180,end angle=360,x radius=0.42,y radius=0.12];
    \draw[rimhid] ( 1.74,0.20) arc[start angle=0,end angle=180,x radius=0.42,y radius=0.12];
    \draw[rimvis] ( 0.90,0.20) arc[start angle=180,end angle=360,x radius=0.42,y radius=0.12];
    \draw[rimhid] (0.82,3.05) arc[start angle=0,end angle=180,x radius=0.82,y radius=0.18];
    \draw[rimvis] (-0.82,3.05) arc[start angle=180,end angle=360,x radius=0.82,y radius=0.18];
  }

  \begin{scope}[shift={(-3.5,0)}]
    \DrawBottomPants
    \DrawTopPants

    \draw[modemark] (-0.98,-0.96) -- (-0.69,-0.67);
    \node[orange!85!black,left] at (-1.1,-0.86) {$\tilde b(\hat x_1)$};

    \draw[dresspath] (-0.71,-0.69) .. controls (-1.00,-0.43) and (-1.22,-0.25) .. (-1.32,-0.20);
    \draw[dress] (-1.32,-0.20) -- (-1.32,0.20);
    \draw[conn]  ( 1.32,-0.20) -- ( 1.32,0.20);

    \node[left]  at (-1.98,-0.5) {$\hat x_1$};
    \node[right] at ( 1.98,-0.5) {$\hat x_2$};
    \node[left]  at (-2.02, 0.5) {$\hat x'_1$};
    \node[right] at ( 2.02, 0.5) {$\hat x'_2$};
    \node[left] at (-1.10,-3.18) {$\hat y$};
    \node[left] at (-1.14, 3.18) {$\hat y'$};
    \node[font=\bfseries] at (0,3.68) {(a) direct channel};
    \node[kernellabel,text=orange!85!black] at (-2.4,0.04) {$B_{T}(1',1)$};
    \node[kernellabel,text=black!70]         at ( 2.4,0.04) {$G(2',2)$};
  \end{scope}

  \begin{scope}[shift={(3.5,0)}]
    \DrawBottomPants
    \DrawTopPants

    \draw[modemark] (-0.98,-0.96) -- (-0.69,-0.67);
    \node[orange!85!black,left] at (-1.1,-0.86) {$\tilde b(\hat x_1)$};

    \draw[dresspath] (-0.71,-0.69) .. controls (-1.00,-0.43) and (-1.22,-0.25) .. (-1.32,-0.20);
    \draw[dress] (-1.32,-0.20) .. controls (-0.26,0.02) and (0.30,0.10) .. (1.32,0.20);
    \draw[conn]  ( 1.32,-0.20) .. controls ( 0.26,0.02) and (-0.30,0.10) .. (-1.32,0.20);

    \node[left]  at (-1.98,-0.5) {$\hat x_1$};
    \node[right] at ( 1.98,-0.5) {$\hat x_2$};
    \node[left]  at (-2.02, 0.5) {$\hat x'_1$};
    \node[right] at ( 2.02, 0.5) {$\hat x'_2$};
    \node[left] at (-1.10,-3.18) {$\hat y$};
    \node[left] at (-1.14, 3.18) {$\hat y'$};
    \node[font=\bfseries] at (0,3.68) {(b) exchange channel};
    \node[kernellabel,text=orange!85!black] at ( 2.4,0.04) {$B_{T}(2',1)$};
    \node[kernellabel,text=black!70]         at (-2.4,0.04) {$G(1',2)$};
  \end{scope}
\end{tikzpicture}
\caption{Bra and ket representation of the connected branch inner product. The lower pair of pants is the ket preparation $1\to 2$, while the upper inverted pair of pants is the bra partner in the inclusive correlator, so the probability is represented schematically by a doubled $1\to 2\to 1$ geometry. The orange squiggle denotes the dressed interior mode $\tilde b(\hat x_1)$, the orange contraction line is the dressed kernel $B_T$, and the black line is the propagator $G$. The two panels are the geometric realizations of the direct and exchange contractions of \cref{subsec:AMPS_warmup_contractions}: the direct channel preserves the daughter assignment, while the exchange channel crosses it.} \label{fig:directvsexchange}
\end{figure}
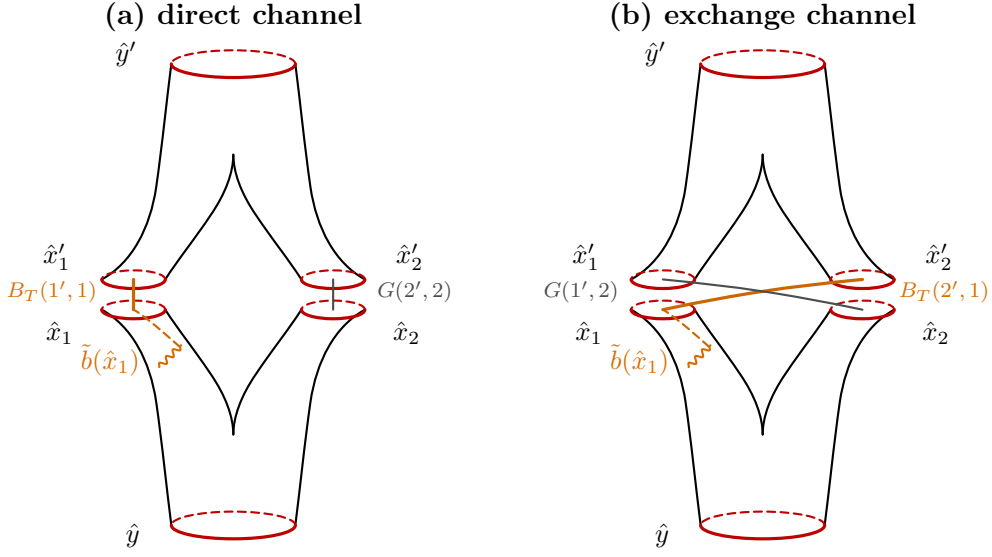

The overarching viewpoint of this paper also clarifies more sharply how our results provide a canonical quantum gravity understanding of recent ``observer rule'' proposals. Harlow \emph{et al} argue that the combined system of observer + bulk in quantum gravity can be described by a distinguished Hilbert space of dimension $\sim e^{S_{\rm obs}}$, with exponentially small errors \cite{HarlowUsatyukZhao2026}. Antonini \emph{et al}, by contrast, derive an extended Hilbert space directly from the gravitational path integral, where the effective observer description emerges after summing over geometries and coarse graining over nonperturbative data that the observer does not control \cite{AntoniniEtAl2025}. In black hole evaporation the early radiation provides a natural environment relative to which such a description is meaningful, and indeed our formalism contains coarse graining over the remaining early radiation as well as interior radiation degrees of freedom.

The basic mechanism from which the ``observer rule'' emerges is that group averaging over the Hamiltonian constraint produces a clock propagator $\langle T|T'\rangle$, so every topology changing contribution additionally inherits a clock overlap weight of order $e^{-S_{\rm clock}}$ from the finite clock Hilbert space of dimension $d_{\rm clock}$. In the semiclassical limit of an almost ideal clock, that is when $d_{\rm clock}\rightarrow \infty$, these overlaps suppress all nontrivial contractions and one recovers a regime in which only the trivial contraction structure survives. For a genuinely quantum mechanical clock, however, for which $d_{\rm clock} \sim \mathcal{O}(1)$, the overlaps are nonzero and the gravitationally dressed amplitudes retain nontrivial contractions. In this sense our JT gravity construction yields a canonical quantum gravity derivation of when one should expect effectively classical observer rules and when intrinsically quantum mechanical observer rules become important. It is also closest in spirit to the construction of \cite{AntoniniEtAl2025}.

The rest of the paper develops this story in stages. In \cref{subsec:AMPS_warmup_contractions} we begin with a simple toy model of AMPS, motivated by \cite{AlmheiriFSP2025}, that isolates the direct and exchange contraction patterns in a purely quantum mechanical setting without gravity. In \crefrange{sec:RAQ_bridge}{sec:relational_time_vs_foliation} we review refined algebraic quantization and relational time evolution, emphasizing the observer clock subsystem, the Hamiltonian constraint, and the clock propagator. In \cref{app:jtphasespace} we construct the global one sided black hole phase space in JT gravity and the subsequent extended phase space after cutting across the horizon, including the conjugate corner pairs \((s,K)\) and \((q,J)\). In \cref{sec:RAQobserver} we quantize the extended phase space, add the probe matter Hilbert spaces, define the cutting and gluing maps, construct the dressed horizon vacuum projector, and introduce the pair of pants Hamiltonian on the interior leg sectors. Finally, in \cref{sec:AMPS} we compute the AMPS correlator and show that the gluing map projects onto the zero mode of the one sided boost in the exchange channel while annihilating the firewall in the direct channel.

\section{A toy model of AMPS}
\label{subsec:AMPS_warmup_contractions}

Before diving into canonical gravity, it is useful to isolate the purely quantum mechanical tensor contraction mechanism at play in the AMPS experiment. This section identifies the precise ambiguity that we would like canonical quantum gravity to resolve. In the quantum circuit language elucidated by Almheiri in \cite{AlmheiriFSP2025}, the basic question is: after the purity measurement has succeeded, which interior does the horizon vacuum measurement actually land on? This is of course closely related to the final state proposal and other postselection models of black hole evaporation \cite{HorowitzMaldacena2004FinalState,GottesmanPreskill2004FinalState,LloydPreskill2014FinalState,BoussoStanford2014FinalState}.

Concretely, we first model distillation by a decoding unitary
\begin{align}
&D:{\cal H}_E\rightarrow {\cal H}_{E'}\otimes {\cal H}_{e_b}
\end{align}
which acts instantaneously on the early radiation and extracts the candidate purifier \(e_b\). Let \(d_b:=\dim {\cal H}_b=\dim {\cal H}_{e_b}\). For any two \(d_b\)-dimensional systems \(X,Y\), write
\begin{align}
&|\Phi\rangle_{XY}
:=
\frac{1}{\sqrt{d_b}}\sum_{i=1}^{d_b}|i\rangle_X|i\rangle_Y,
\
\Pi^{(1)}_{XY}:=|\Phi\rangle\langle\Phi|_{XY}.
\label{eq:warmup_max_ent_projector_def}
\end{align}
We denote the purity measurement outcome by \(n\) and the horizon vacuum measurement outcome by \(m\). Thus \(n=1\) means that the distilled subsystem $e_b$ really purifies the outgoing late mode $b$, and \(m=1\) means that the same outgoing late mode is maximally entangled with its interior partner $\tilde b$. In this section, superscripts such as \(\Pi^{(1)}\) label the respective successful measurement outcome. Similarly, the state \(|\Psi_{n=1}\rangle\) below is the normalized state after conditioning on the first measurement outcome.

The quantity we compute is the conditional success probability for the AMPS measurement, namely the probability that the horizon vacuum measurement is successful, given that the early radiation purity outcome has already been deemed successful. In an ordinary quantum mechanics calculation this would be
\begin{align}
&p_{\parallel}(m=1\mid n=1)
=
\frac{
\left\|
\bigl(\Pi^{(1)}_{\tilde b b}\otimes \mathbf 1_{e_bE'}\bigr)
\bigl(\Pi^{(1)}_{b e_b}\otimes \mathbf 1_{\tilde bE'}\bigr)
D|\Psi\rangle
\right\|^2
}{
\left\|
\bigl(\Pi^{(1)}_{b e_b}\otimes \mathbf 1_{\tilde bE'}\bigr)
D|\Psi\rangle
\right\|^2
}.
\label{eq:warmup_conditional_probability_target}
\end{align}
The fundamental question is whether the index contractions in the inner product have to correspond to those of the computation above, or whether the state prepared by the first postselection allows an exchange contraction in which the recovered subsystem \(e_b\) is treated as the candidate interior mode instead of the original $\tilde b$ mode, at the level of indices. These two contraction patterns give parametrically different answers.

The actual early radiation purity measurement on \(b\otimes E\) is $\Xi^{(1)}_{bE}= D^\dagger \bigl(\Pi^{(1)}_{b e_b}\otimes \mathbf1_{E'}\bigr)D$. Given a pre distillation density matrix \(\rho_{bE}\), define \(\widehat\rho=D\rho_{bE}D^\dagger\) and \(\widehat\rho_{b e_b}=\Tr_{E'}\widehat\rho\). Then the success probability of the purity measurement is exactly
\begin{align}
&p_{\rm pur}
=
\Tr\!\left[\bigl(\Pi^{(1)}_{b e_b}\otimes \mathbf 1_{E'}\bigr)\widehat\rho\right]
=
\Tr\!\left[\Pi^{(1)}_{b e_b}\,\widehat\rho_{b e_b}\right]
=
\langle\Phi|\widehat\rho_{b e_b}|\Phi\rangle.
\label{eq:warmup_purity_exact}
\end{align}
Assume now the standard Hayden Preskill decoding protocol: after the Page time, and for a sufficiently deep decoding unitary, the extracted subsystem \(e_b\) purifies \(b\) up to a small decoding error \(\varepsilon_{\rm dist}\) \cite{HaydenPreskill2007}. Equivalently, assume \(\langle\Phi|\widehat\rho_{b e_b}|\Phi\rangle \ge 1-\varepsilon_{\rm dist}\), with \(0<\varepsilon_{\rm dist}\ll 1\). Then \eqref{eq:warmup_purity_exact} gives immediately \(p_{\rm pur}\ge 1-\varepsilon_{\rm dist}\).

Condition now on the successful purity outcome \(n=1\). The corresponding normalized postselected state is
\begin{align}
&|\Psi_{n=1}\rangle
=
\frac{
\bigl(\Pi^{(1)}_{b e_b}\otimes \mathbf 1_{\tilde bE'}\bigr)
D|\Psi\rangle
}{
\sqrt{p_{\rm pur}}
}.
\label{eq:warmup_postselected_state_exact}
\end{align}
In the ideal limit \(\varepsilon_{\rm dist}\to 0\), this reduces to $|\Psi_{n=1}\rangle=|\Phi\rangle_{b e_b}\otimes |\chi\rangle_{\tilde b E'}$ where $|\chi\rangle_{\tilde b E'}$ is the state of the remaining environment
\begin{align}
&|\chi\rangle_{\tilde b E'}
=
\sum_{k,a}\chi_{ka}\,|k\rangle_{\tilde b}|a\rangle_{E'},
\
\sum_{k,a}|\chi_{ka}|^2=1.
\label{eq:warmup_chi_expansion}
\end{align}
The key point is that after the successful purity measurement, the state factorizes as a maximally entangled pair on \(b\otimes e_b\) times an otherwise arbitrary state on \(\tilde b\otimes E'\).

The horizon vacuum measurement is given by the projector $|\Phi\rangle{}_{\tilde b b}\langle\Phi|$ where
\begin{align}
|\Phi\rangle_{\tilde b b}
=
\frac{1}{\sqrt{d_b}}
\sum_{i=1}^{d_b}{}_{\tilde b}|i\rangle_{\tilde b}|i\rangle_b.
\label{eq:warmup_late_maximally_entangled_bra}
\end{align}
Strictly speaking, the projector has a unique ordinary action in a fixed tensor product Hilbert space. The subscripts \(\parallel\) and \(\times\) below therefore do not denote two different ordinary inner products. They denote two different Wick contraction patterns within the inner product. The direct contraction is what we get from ordinary quantum mechanics. The exchange pattern is the index contraction whose gravitational origin will be derived later.

The direct contraction pairs the \(b\) bra with the \(b\) ket and the \(\tilde b\) bra with the original semiclassical interior \(\tilde b\) ket:
\begin{align}
&\contraction[1.4ex]{{}_{\tilde b}\langle i|\,}{{}_b\langle i|\,}{}{|j\rangle_b\,}
\contraction[2.8ex]{}{{}_{\tilde b}\langle i|\,}{{}_b\langle i|\,|j\rangle_b\,|j\rangle_{e_b}\,}{|k\rangle_{\tilde b}}
{}_{\tilde b}\langle i|\,{}_b\langle i|\,|j\rangle_b\,|j\rangle_{e_b}\,|k\rangle_{\tilde b}\,|a\rangle_{E'}
=
\delta_{ij}\delta_{ik}\,|j\rangle_{e_b}|a\rangle_{E'}.
\label{eq:warmup_direct_wick_contraction}
\end{align}
This gives
\begin{align}
&{}_{\tilde b b}\langle\Phi|\Psi_{n=1}\rangle_{\parallel}
=
\frac{1}{d_b}
\sum_{i,a}\chi_{ia}\,|i\rangle_{e_b}|a\rangle_{E'}.
\label{eq:warmup_direct_amplitude}
\end{align}
Therefore
\begin{align}
&p_{\parallel}(m=1\mid n=1)
=
\bigl\|{}_{\tilde b b}\langle\Phi|\Psi_{n=1}\rangle_{\parallel}\bigr\|^2
=
 d_b^{-2}.
\label{eq:warmup_direct_probability}
\end{align}
This is the channel in which the second measurement lands on the original semiclassical interior. Conditioned on the purity measurement having succeeded, it implies a firewall.

The exchange contraction instead pairs the \(\tilde b\) bra index of $|\Phi\rangle{}_{\tilde b b}\langle\Phi|$ to the recovered subsystem \(e_b\). We emphasize that the contraction between $\tilde b$ and \(e_b\) in what follows is merely a posited index contraction, not the result of the ordinary rules of plain quantum mechanics:
\begin{align}
&\contraction[1.4ex]{{}_{\tilde b}\langle i|\,}{{}_b\langle i|\,}{}{|j\rangle_b\,}
\contraction[2.8ex]{}{{}_{\tilde b}\langle i|\,}{{}_b\langle i|\,|j\rangle_b\,}{|j\rangle_{e_b}}
{}_{\tilde b}\langle i|\,{}_b\langle i|\,|j\rangle_b\,|j\rangle_{e_b}\,|k\rangle_{\tilde b}\,|a\rangle_{E'}
=
\delta_{ij}\,|k\rangle_{\tilde b}|a\rangle_{E'}.
\label{eq:warmup_exchange_wick_contraction}
\end{align}
Then
\begin{align}
&{}_{\tilde b b}\langle\Phi|\Psi_{n=1}\rangle_{\times}
=
\sum_{k,a}\chi_{ka}\,|k\rangle_{\tilde b}|a\rangle_{E'}
=
|\chi\rangle_{\tilde b E'}.
\label{eq:warmup_exchange_amplitude}
\end{align}
Hence
\begin{align}
&p_{\times}(m=1\mid n=1)
=
\bigl\|{}_{\tilde b b}\langle\Phi|\Psi_{n=1}\rangle_{\times}\bigr\|^2
=
1.
\label{eq:warmup_exchange_probability}
\end{align}

The exchange contraction is therefore the mechanism that would make the AMPS experiment succeed. But in the calculation above it has simply been named as a tensor-network rule. Nothing in the ordinary Hilbert space calculation explains why the second measurement should be routed through \(e_b\) rather than through the original \(\tilde b\), or why the first measurement should have produced such a candidate interior in the first place. These are precisely the questions that gravity must answer. The rest of the paper provides such a derivation in canonical quantum gravity: the first measurement is represented by relational time evolution in a theory whose Hilbert space contains sectors with different numbers of black hole interiors obtained from splitting the Hilbert space into interior and exterior factors across the horizon, and the second measurement is a gravitationally dressed observable whose action is fixed only after gluing the extended Hilbert space back across the horizon. In this way the exchange contraction is not an ad hoc modification of quantum mechanics, but rather a direct consequence of relational topology changing dynamics in canonical quantum gravity.

\section{Refined algebraic quantization}
\label{sec:RAQ_bridge}

Canonical quantum gravity is a constrained system \cite{MarolfRAQ, HeldMaxfielddSJT, Rovelli1991Observable}. One begins with a kinematical Hilbert space $\Hkin$ and a set of constraints $\{\hat C_I\}$ whose kernel defines the physical state space in the Dirac sense, \(\hat C_I|\Psi\rangle_{\phys}=0\) for all \(I\). The essential technical point is that this condition typically has distributional solutions: the relevant part of the spectrum often includes $0$ in the continuous spectrum, so nontrivial solutions are not normalizable in the $\Hkin$ norm. Dirac quantization constructs \(\Hphys\) with a well defined inner product and a corresponding algebra of Dirac observables \cite{MarolfRAQ, HeldMaxfielddSJT, GiuliniMarolf1999Uniqueness}.

In this paper we will implement the constraint using refined algebraic quantization and group averaging \cite{MarolfRAQ, HeldMaxfielddSJT, GiuliniMarolf1999Uniqueness, Rovelli1991Observable}, because it makes the quotient structure of the physical theory manifest and connects to the path integral viewpoint.

\subsection{Constraints and Dirac observables}
\label{subsec:constraints_gauge_dirac_observables}

For clarity, we first focus on a single self adjoint constraint $\hat C$ generating a one parameter unitary group \(U(\tau)\equiv \ee^{-\ii \tau \hat C}\), with \(\tau\in\R\), interpreted as a gauge redundancy, for example time reparametrizations in the simplest Wheeler--DeWitt setting. Physical states are invariant under this group, \(U(\tau)|\Psi\rangle_{\phys}=|\Psi\rangle_{\phys}\Leftrightarrow \hat C|\Psi\rangle_{\phys}=0\) \cite{Rovelli1991Observable, Dittrich2006Canonical, Dittrich2007Partial}. An operator $\hat O$ is a Dirac observable if it is gauge invariant, that is, \(U(\tau)\hat O U(\tau)^{-1}=\hat O\Leftrightarrow \comm{\hat C}{\hat O}=0\) on an appropriate domain.
Commuting with the constraint is naturally understood in terms of quotients along gauge orbits: two kinematical operators can differ on $\Hkin$ while becoming identical on $\Hphys$ once null directions are removed.

\subsection{Rigged Hilbert space and group averaging}
\label{subsec:RAQ_group_averaging_general}

Refined algebraic quantization begins by choosing a dense test space $\Phi\subset \Hkin$ and forming a rigged Hilbert space \(\Phi \subset \Hkin \subset \Phi^\ast\), where $\Phi^\ast$ denotes continuous anti linear functionals on $\Phi$ \cite{MarolfRAQ}. The rigging map, or group average, is defined distributionally by
\begin{align}
&  \eta:\Phi\to\Phi^\ast,
  \;
  \eta\bigl(|\psi\rangle\bigr)
  :=
  \int_{-\infty}^{+\infty}\Dd\tau \, \ee^{-\ii\tau \hat C}\,|\psi\rangle,
  \label{eq:RAQ_rigging_general}
\end{align}
and the induced physical inner product is
\begin{align}
&  \braket{\eta(\phi)}{\eta(\psi)}_{\phys}
  :=
  \int_{-\infty}^{+\infty}\Dd\tau \,
  \matrixel{\phi}{\ee^{-\ii\tau \hat C}}{\psi}_{\kin}.
  \label{eq:RAQ_inner_product_general}
\end{align}
Formally one may package the group average into an improper projector \cite{MarolfRAQ, HeldMaxfielddSJT, GiuliniMarolf1999Uniqueness}
\begin{align}
&  \Proj \equiv\int_{-\infty}^{+\infty}\Dd\tau \,\ee^{-\ii\tau \hat C},
  \;
  \braket{\eta(\phi)}{\eta(\psi)}_{\phys}=\matrixel{\phi}{\Proj}{\psi}_{\kin},
  \label{eq:improper_projector_general}
\end{align}
with the understanding that $\Proj$ is a distribution, since its square contains the infinite gauge volume. Operationally, $\Proj$ is the correct object to insert inside matrix elements.\footnote{In this paper, refined algebraic quantization is always implemented at the level of the physical matrix element $\langle \phi|\Proj\,\hat O\,|\psi\rangle$. Thus there is only one group average for each physical amplitude. We do not first construct separately rigged bra and ket wavefunctions and then perform a second averaging after passing to a clock representation. When this physical matrix element is later written in the observer clock basis, the group average appears through the clock propagator \(\langle T'|T\rangle\) to be defined below. For an inclusive correlator, this gives one clock propagator linking the bra and ket clock arguments. If instead one writes the kernel of a relational operator inserted at clock reading \(T\) between external clock labels \(T'\) and \(T''\), the same construction appears as the pair \(\langle T'|T\rangle \langle T|T''\rangle\). Those two factors are not two independent rigging maps; they are the bra side and ket side clock propagators of a single relational insertion.}

For several constraints $\hat C_I$ generating a gauge group $G$, \eqref{eq:RAQ_rigging_general} and \eqref{eq:RAQ_inner_product_general} generalize by averaging over $G$ with an appropriate Haar measure. In the JT gravity setting below, the horizon cut construction will supply a canonical diffeomorphism invariant description of the slice data, so the remaining group averaging will effectively reduce to a single one parameter constraint of the clock + bulk form discussed next.

\subsection{Null states}
\label{subsec:null_quotient_operator_equivalence}

The sesquilinear form \eqref{eq:RAQ_inner_product_general} is positive semidefinite on $\Phi$ but can be degenerate. This produces null vectors:
\begin{align}
&  \mathcal N
  :=\Bigl\{|\psi\rangle\in\Phi\colon\braket{\eta(\psi)}{\eta(\psi)}_{\phys}=0\Bigr\}.
  \label{eq:null_ideal_def_general}
\end{align}
The physical Hilbert space is the completion of $\eta(\Phi)$ modulo this null ideal,
\begin{align}
&  \Hphys =\overline{\eta(\Phi)/\eta(\mathcal N)}.
  \label{eq:Hphys_as_quotient_general}
\end{align}
We write \(\hat O_1 \sim_{\phys} \hat O_2\) to mean that $\hat O_1$ and $\hat O_2$ have the same matrix elements between physical states:
\begin{align}
&  \matrixel{\eta(\phi)}{\hat O_1-\hat O_2}{\eta(\psi)}_{\phys}=0,
  \ \forall\,\phi,\psi\in\Phi.
  \label{eq:phys_operator_equiv_def_general}
\end{align}
Equivalently, $\hat O_1-\hat O_2$ lies in the null ideal of operators, namely operators that act trivially in all physical matrix elements.

Group averaging implies that operators related by constraint flow are equivalent on $\Hphys$,
\begin{align}
&  \hat O \sim_{\phys} \hat O_s := \ee^{-\ii s\hat C}\,\hat O\,\ee^{\ii s\hat C},
  \;
  s\in\R,
  \label{eq:RAQ_conjugation_invariance_general}
\end{align}
specifically in the sense that their matrix elements between physical states agree,
\begin{align}\label{eq:RAQconjLemmaME}
&\matrixel{\eta(\phi)}{\hat O}{\eta(\psi)}_{\rm phys}
=
\matrixel{\eta(\phi)}{\hat O_s}{\eta(\psi)}_{\rm phys},
\ \forall\,\phi,\psi\in\Phi.
\end{align}
Equivalently, commutators with the constraint vanish in $\Hphys$ matrix elements:
\begin{align}\label{eq:RAQcommNull}
&\matrixel{\eta(\phi)}{\comm{\hat C}{\hat O}}{\eta(\psi)}_{\rm phys}=0.
\end{align}
The proof of this is simple. Using the definition of the physical inner product,
\begin{align}
&\matrixel{\eta(\phi)}{\hat O_s}{\eta(\psi)}_{\rm phys}
=
\int_{-\infty}^{+\infty}\Dd\tau\,\matrixel{\phi}{\ee^{-\ii(\tau+s)\hat C}\,\hat O\,\ee^{\ii s\hat C}}{\psi}_{\rm kin}.
\end{align}
Now change integration variable $\tau'=\tau+s$; since the integral is over all $\R$,
\begin{align}
&\int_{-\infty}^{+\infty}\Dd\tau\,(\cdots)=\int_{-\infty}^{+\infty}\Dd\tau'\,(\cdots),
\end{align}
so
\begin{align}
&\matrixel{\eta(\phi)}{\hat O_s}{\eta(\psi)}_{\rm phys}
=
\int_{-\infty}^{+\infty}\Dd\tau'\,\matrixel{\phi}{\ee^{-\ii\tau'\hat C}\,\hat O}{\psi'}_{\rm kin},
\end{align}
with \(\ket{\psi'}:=\ee^{\ii s\hat C}\ket{\psi}\).
But $\eta(\psi')=\eta(\psi)$ because group averaging projects onto the $\hat C=0$ sector, which establishes \eqref{eq:RAQconjLemmaME}. Differentiating with respect to \(s\) at \(s=0\) yields \eqref{eq:RAQcommNull}. In the language of canonical quantum gravity, this is a formal version of the statement that different clock readings of an operator insertion are gauge equivalent when time evolution is pure gauge \cite{MarolfRAQ, Dittrich2007Partial}.

\section{Relational time evolution}
\label{sec:relational_time_vs_foliation}

In order to discuss dynamics and time ordered correlators in a constrained theory, we have to introduce an explicit clock subsystem to which bulk operators are dressed \cite{PageWootters1983, Rovelli1991Time, Dittrich2007Partial, Marolf1995AlmostIdeal}. For the infalling observer the natural ideal label is the local ingoing affine coordinate \(v\) of the null congruence through the observer's worldline. This coordinate is regular at the future horizon, and in the highly boosted limit used below it is affine in the observer proper time along the whole segment relevant for the AMPS experiment. We first describe the ideal deparametrized construction and then replace it by the finite resolution clock actually carried by the observer.

Consider the extended kinematical Hilbert space \(\Hkin=\Hclk\otimes\Hgrav\), with an ideal local clock basis satisfying
\begin{align}
&\hat v\,|v\rangle
=
v\,|v\rangle,
\\
&\langle v'|v\rangle
=
\delta(v'-v),
\\
&\int_{\mathbb R}dv\,|v\rangle\langle v|
=
\mathbf 1,
\label{eq:idealclockconds}
\end{align}
and with conjugate generator \(\hat P_T=-\ii\partial_v\), so that \(\ee^{-\ii a\hat P_T}|v\rangle=|v+a\rangle\) and \(\ee^{\ii a\hat P_T}|v\rangle=|v-a\rangle\). Parametrized evolution is obtained by imposing a single constraint,
\begin{align}
&  \hat C
  =
  \hat P_T + \hat H_{\rm grav}(v)
  \approx 0,
  \label{eq:C_tau_with_redshift}
\end{align}
where the dependence on \(v\) includes the calibration of the clock with respect to the local affine parameter. Physically, this means that time evolution is defined relationally with respect to the infalling observer's clock, with \(\hat H_{\rm grav}(v)\) determining the coupling between the clock and the rest of the system. Then the Dirac constraint \(\hat C|\Psi\rangle_{\phys}=0\) is equivalent, in the \(v\) representation, to a Schr\"odinger equation
\begin{align}
&  \ii\partial_v \Psi(v)=\hat H_{\rm grav}(v)\Psi(v),
  \label{eq:relational_Schrodinger_general}
\end{align}
and group averaging reproduces the corresponding relational evolution \cite{Rovelli1991Time, Rovelli2002Partial, Dittrich2006Canonical}.

At this stage it is important to distinguish the ideal affine representation from the operational clock measurement used later for the infalling observer. We can define an ideal deparametrized state by \(|\psi(v)\rangle := \langle v|\Psi\rangle\). This is a useful representation of a physical state, but it is not the state conditioned on an actual finite resolution clock measurement. The latter is described by a POVM density \(\mathbb{E}_T\ge 0\), where \(T\) is the observer's recorded clock value after calibration to the affine parameter.
For a physical density matrix \(\rho_{\rm phys}\), the corresponding unnormalized bulk state conditioned on the observer recording \(T\) is \(\rho(T):=\operatorname{Tr}_{\Hclk}[(\mathbb{E}_T\otimes \bm{1})\rho_{\rm phys}]\).
Inserting the ideal clock resolution of the identity, one may also write
\begin{align}
&\rho(T)
=
\int dv' dv''\,
\langle v''|\mathbb{E}_T|v'\rangle\,
|\psi(v')\rangle\langle\psi(v'')|.
\end{align}
This is the precise sense in which a finite observer clock smears relational time \cite{PageWootters1983, Marolf1995AlmostIdeal}. With this distinction in place, \eqref{eq:relational_Schrodinger_general} has the general solution \begin{align}|\psi(v)\rangle=\mathcal T\exp[-\ii\int_{v_0}^{v} dv'\,\hat H_{\rm grav}(v')]|\psi(v_0)\rangle.\end{align}
In the next section we will derive the analogue of this for the actual clock subsystem carried by the infalling observer.

\subsection{Infalling clock subsystem\label{sec:infallclocksub}}

For an infalling clock, we now argue that the states $|T\rangle$ are necessarily effectively nonorthogonal (see also \cite{Marolf1995AlmostIdeal}). In JT gravity the local metric sector is fixed to be locally AdS$_2$ by the dilaton equation of motion, so the only intrinsic length and time scale in the geometry is the AdS$_2$ radius (see \cref{app:jtphasespace}). The parameter $S_0$ multiplies a purely topological term and does not enter the local equations of motion; in particular, $S_0$ does not determine the location of the horizon. Instead, the horizon is specified relationally by a dilaton level set \(\Phi\hateq \Phi_h\), with $\Phi_h>0$ fixed by the ADM energy, equivalently the temperature above extremality, together with the asymptotic dilaton boundary condition. In other words $\Phi_h$ remains fixed as $S_0\to\infty$, so $\Phi_h=\mathcal O(1)$ in $S_0$ counting.

The AdS$_2$ black hole metric takes the form
\begin{align}
&\Dd s^2 = -(r^2-r_h^2)\,\Dd t^2 + \frac{\Dd r^2}{r^2-r_h^2},
\;
\Phi(r)=\Phi_h\,\frac{r}{r_h},
\label{eq:JT_bh_metric_dilaton}
\end{align}
for which the surface $\Phi=0$ sits at $r=0$. This is not a curvature singularity, but it is the natural end of space where the dilaton vanishes and semiclassical JT gravity ceases to be reliable. Because the metric is locally AdS$_2$, an infalling observer has only a finite
amount of proper time between falling in from the asymptotic region and
reaching the ``singularity''. For the AMPS calculation it is
useful to choose the infalling observer to be a
highly boosted timelike probe. Let
\(\tau\) denote the observer proper time and write $f(r):=r^2-r_h^2$. For an ingoing radial timelike geodesic with conserved Killing energy
\(E_\gamma\), we have
\begin{align}
E_\gamma=f(r)\frac{dt_\gamma}{d\tau},
\
\frac{dr_\gamma}{d\tau}
=
-\sqrt{E_\gamma^2-f(r_\gamma)}.
\end{align}
The proper time from the finite cutoff asymptotic boundary \(r=r_c\) to the surface \(r=0\)
is therefore
\begin{align}
\Delta \tau(E_\gamma)
=
\int_0^{r_c}
\frac{dr}{\sqrt{E_\gamma^2+r_h^2-r^2}}
=
\sin^{-1}
\left(
\frac{r_c}{\sqrt{E_\gamma^2+r_h^2}}
\right),
\end{align}
where the expression is real for $E_\gamma^2\ge r_c^2-r_h^2$.
The limiting value \(E_\gamma^2=r_c^2-r_h^2\) is the infalling observer released from
rest in the asymptotic region and gives \(\Delta\tau=\pi/2\). For the highly
boosted infalling observer, i.e.\  $E_\gamma^2\gg r_c^2-r_h^2$,
one instead has
\begin{align}
\Delta\tau(E_\gamma)
=
\frac{r_c}{E_\gamma}
\left[
1+O\!\left(\frac{r_c^2}{E_\gamma^2}\right)
\right].
\end{align}
In either case the duration is $\mathcal{O}(1)$ in AdS units, that is, it does not scale with
\(S_0\). Thus the infalling observer has access only to a finite causal
diamond.

Now, introduce the tortoise coordinate, with a cutoff \(r=r_c\) so that
\(r^*(r_c)=0\),
\begin{align}
r^*(r)
:=
\int_{r_c}^{r}\frac{d\tilde r}{f(\tilde r)}
=
\frac{1}{2r_h}
\log\left|
\frac{r-r_h}{r+r_h}
\frac{r_c+r_h}{r_c-r_h}
\right|,
\end{align}
and define the regular ingoing coordinate $v:=t+r^*(r)$.
Along the same ingoing geodesic, we have
\begin{align}
\frac{dv_\gamma}{d\tau}
=
\frac{1}{E_\gamma+\sqrt{E_\gamma^2-f(r_\gamma)}}.
\end{align}
In the highly boosted regime this becomes
\begin{align}
\frac{dv_\gamma}{d\tau}
=
\frac{1}{2E_\gamma}
\left[
1+O\!\left(
\frac{r_c^2-r_h^2}{E_\gamma^2}
\right)
\right].
\end{align}
Thus \(v\) is an affine parameter in the observer proper time to leading order in the highly boosted limit. We absorb the constant Jacobian into the calibration of the clock and use \(T\) to denote the corresponding ideal clock reading. The
actual observer clock is finite resolution. We model it by an effective clock
Hilbert space \({\cal H}_{\rm clk}\) carrying a bounded momentum band in
\(\hat P_T\). This means the overlap of physical clock states is not a delta function because the finite causal
diamond only supports a finite band in the spectrum of the clock.

The finite resolution of the clock restricts us to a finite microcanonical band $W_{\rm clock}$ in the conjugate momentum $\hat P_T$. A convenient choice of seed state supported on this band is \( |\varphi\rangle:=\Delta P_T^{-1/2}\int_{W_{\rm clock}}\Dd p\,|p\rangle\),
where $\{|p\rangle\}$ are the eigenstates of $\hat{P}_T$, and we define the associated time states by translation,
\begin{align}
&|T\rangle :=\ee^{-\ii T \hat P_T}\,|\varphi\rangle.
\label{eq:bandlimited_time_states}
\end{align}
The overlap then coincides with the Fourier transform of the band,
\begin{align}
&\langle T'|T\rangle
=\frac{1}{\Delta P_T}\int_{W_{\rm clock}}\Dd p\,\ee^{-\ii p(T-T')},
\label{eq:kernel_band}
\end{align}
so the clock has intrinsic time resolution $\Delta T\sim \Delta P_T^{-1}=\mathcal O(1)$. Because the family $\{|T\rangle\}$ is overcomplete on $\Hclk$, clock readings are described by a POVM. The covariant POVM measure associated to \eqref{eq:bandlimited_time_states} is \(\mathbb{E}_T:=(\Delta P_T/2\pi)|T\rangle\langle T|\), which satisfies the exact normalization \(\int_{-\infty}^{\infty}\Dd T\,\mathbb{E}_T=\mathbf 1_{\clock}\) on the observer clock.
This is the precise sense in which our clock provides a fuzzy relational ordering variable but does not define a sharp foliation parameter \cite{Marolf1995AlmostIdeal}.

We now derive one of the key ingredients of this paper. The observer's clock is not a purely kinematical label introduced after deparametrization; once the Hamiltonian constraint is imposed and the ideal affine parameter label is replaced by the finite clock algebra actually accessible to the infalling observer, canonical quantum gravity acquires a new dynamical vertex factor from group averaging, on top of the standard topology weighting vertex factor in JT gravity: the clock propagator \(\langle T'|T\rangle\).

Start from a seed state in the ideal local clock description, \( |\psi_0\rangle = |\phi\rangle_v\otimes |\chi\rangle\).
The corresponding physical state is
\begin{align}
&|\eta(\psi_0)\rangle
=
\int_{-\infty}^{\infty} d\tau\,
\ee^{-\ii\tau(\hat P_T+\hat H_{\rm grav}(v))}
|\phi\rangle_{v}\otimes |\chi\rangle.
\label{eq:clock_group_average_minus_sign}
\end{align}
Write the ideal local clock seed in the \(v\) basis as \( |\phi\rangle_v = \int_{-\infty}^{\infty} dv_0\, f(v_0)\,|v_0\rangle\). With the convention above,
\begin{align}
&\ee^{-\ii\tau \hat P_T}|v_0\rangle=|v_0+\tau\rangle.
\end{align}
The full constraint exponential acts by the corresponding relational gravitational propagator,
\begin{align}
&\ee^{-\ii\tau(\hat P_T+\hat H_{\rm grav})}
\bigl(|v_0\rangle\otimes |\chi\rangle\bigr)
=
|v_0+\tau\rangle\otimes U_{\rm grav}(v_0+\tau,v_0)|\chi\rangle,
\label{eq:constraint_flow_forward_clock}
\end{align}
where
\begin{align}
&U_{\rm grav}(v',v_0)
:=
\mathcal T\exp\!\left[-\ii\int_{v_0}^{v'} d\bar v\,\hat H_{\rm grav}(\bar v)\right].
\label{eq:relational_grav_propagator_def}
\end{align}
Thus the group averaged state may be written as
\begin{align}
&|\eta(\psi_0)\rangle
=
\int d\tau \int dv_0\,
f(v_0)\,
|v_0+\tau\rangle
\otimes
U_{\rm grav}(v_0+\tau,v_0)|\chi\rangle.
\label{eq:group_averaged_clock_orbit_forward}
\end{align}

Now evaluate this physical state with the infalling observer's clock. The natural rank one POVM amplitude at clock reading \(T\) is
\begin{align}
&|\psi(T)\rangle
:=
\sqrt{\frac{\Delta P_T}{2\pi}}\,
\langle T|\eta(\psi_0)\rangle.
\end{align}
Substituting the group averaged state gives
\begin{align}
&|\psi(T)\rangle
=
\sqrt{\frac{\Delta P_T}{2\pi}}
\int d\tau \int dv_0\,
f(v_0)\,
\langle T|v_0+\tau\rangle\,
U_{\rm grav}(v_0+\tau,v_0)|\chi\rangle.
\label{eq:fuzzy_clock_state_tau_integral}
\end{align}
The matrix element \(\langle T|v_0+\tau\rangle\) is exactly the clock propagator. Changing variables from \((v_0,\tau)\) to \((v_0,v')\) with \(v'=v_0+\tau\), we obtain
\begin{align}
&|\psi(T)\rangle
=
\sqrt{\frac{\Delta P_T}{2\pi}}
\int dv_0\,dv'\,
f(v_0)\,
\langle T|v'\rangle\,
U_{\rm grav}(v',v_0)|\chi\rangle.
\label{eq:fuzzy_clock_state_v_integral}
\end{align}
Choosing the reference clock seed to be the locally affine state \(|v_0=0\rangle\), this reduces to
\begin{align}
&|\psi(T)\rangle
=
\sqrt{\frac{\Delta P_T}{2\pi}}
\int dv'\,
\langle T|v'\rangle\,
U_{\rm grav}(v',0)|\chi\rangle.
\label{eq:fuzzy_clock_state_final}
\end{align}
This is the precise sense in which the nontrivial clock propagator arises from group averaging. Group averaging produces the integral over the gauge orbit parameter \(\tau\), while the observer's finite clock resolution converts the ideal locally affine translation orbit into the overlap \(\langle T|v'\rangle\).

In the rest of the paper \( |\psi(T)\rangle\) denotes the state conditioned on the observer's clock measurement.

\subsection{Lorentzian topology change does not obstruct relational time}

The standard obstruction to Lorentzian topology change is an obstruction to realizing a topology changing history as a single smooth globally hyperbolic Lorentzian spacetime admitting a smooth foliation by Cauchy slices \cite{Geroch1967Topology, Tipler1977Singularities, Horowitz1991TopologyChange, Borde1994TopologyChange, DowkerSurya1998TopologyChange}. By contrast, the canonical framework we use consists of:
\begin{enumerate}[leftmargin=2em]
\item A physical Hilbert space $\Hphys$ defined by the constraint and its quotient
\item A relational clock $T$ that labels conditional states and relational observables
\item A Hamiltonian interaction that mixes topological sectors in a baby universe or Fock space decomposition
\end{enumerate}
None of these ingredients requires that there exist a single smooth Lorentzian manifold with a global time function whose level sets interpolate through the topology changing event. Indeed, in a strict ADM setting with fixed $\Sigma$, topology change is forbidden by construction \cite{Geroch1967Topology, Tipler1977Singularities, Horowitz1991TopologyChange, Borde1994TopologyChange, DowkerSurya1998TopologyChange}. We instead work in an enlarged kinematics where distinct spatial topologies, or different numbers of interior legs, appear as orthogonal sectors of $\Hkin$, and topology change is implemented by an interaction that maps between these sectors.

Concretely, if we write the gravitational Hilbert space as a direct sum over topological sectors, schematically ordered by the number of interior legs and/or boundary components, \(\Hgrav \simeq \bigoplus_{N\ge 0}\HN\), then topology change can be represented by an interaction term in the Hamiltonian which, at leading order in the topological expansion, is cubic in creation and annihilation operators in a baby universe Fock space representation:
\begin{align}
&\hat H_{\rm grav} = \hat H_{0} + \lambda\,\hat V_{\text{pants}} + \cdots,
\;
\hat V_{\text{pants}}:\HN \to \mathcal{H}_{N\pm 1},
\label{eq:topology_change_Hamiltonian}
\end{align}
so that relational evolution mixes the sectors $\HN$ as a function of the clock reading $T$ \cite{MarolfMaxfield2020, CasaliMarolfMaxfieldRangamani2021, Maxfield:2022sio, Post:2022UniverseFieldJT}. In the one sided black hole JT gravity setting studied below, the relevant number preserving piece is the two sided boost generator $\hat K$ about a cut of the horizon, written in the corner variables of \cref{app:jtphasespace}. The topology changing vertex is not an event in a smooth foliation; rather, it is the action of $\hat V_{\text{pants}}$ on $\Hgrav$ at a clock reading $T$. 

One may nevertheless attempt a semiclassical reconstruction of spacetime histories from transition amplitudes. In such reconstructions, a topology changing history is typically represented by geometries with corners or defects \cite{Usatyuk:2022afj}, or equivalently by gluing rules that fail to be smooth across the join or split. This is fully compatible with the canonical description, because the canonical theory never assumes the existence of a smooth foliation time through the join or split. The relational clock $T$ continues to label conditional amplitudes even when no global Lorentzian time function exists. And as we have already seen, in the relational picture the clock is a quantum system which contributes clock propagator factors due to the Hamiltonian constraint.

\section{Gravitational phase space}
\label{app:jtphasespace}

We now turn to the gravitational phase space on which the
two AMPS measurements are actually defined. The previous sections described how
physical states are obtained by group averaging, how the infalling observer's
finite resolution clock defines conditional states, and how topology change can
be represented as a Hamiltonian interaction between sectors. On a smooth global one boundary slice, however, the gravitational phase space does not factorize across the horizon. The gravitational constraints tie the two sides of the horizon
together, so there is no canonical factorization into an independent interior
phase space and an independent exterior phase space. But we need such a factorization in order to make sense of topology changing dynamics in the black hole interior. Furthermore, we need to identify the gravitational degrees of freedom to which bulk observables are dressed. In a gravitational theory the question ``which subsystem is $\tilde b$?'' has no invariant meaning until the
operator has been specified relationally. As we will see, the extended phase space (edge mode) formalism \cite{DonnellyFreidel2016, Speranza2018LocalPhaseSpace, Chandrasekaran:2026pnc} solves both of these problems simultaneously.

The construction proceeds in two steps. First we describe the one sided black hole
phase space of JT gravity on observer dressed slices $\Sigma_T$, running from the clock reading $T$ to
the dynamical end of the world brane. Since JT gravity has no local propagating
bulk degrees of freedom, the reduced gravitational phase space is finite
dimensional: after imposing the brane boundary condition and fixing the
observer worldline, the remaining gravitational degree of freedom is the
relative time shift. We do not include the matter labels $b,\tilde b,E, \tilde R$ as part of the classical phase space. They will instead be introduced as finite dimensional probe quantum systems with internal degrees of freedom (e.g.\ spin) only
after the gravitational phase space has been quantized.

Second, we introduce an extended phase space by cutting $\Sigma_T$ across the horizon. The cut produces an interior partial slice and an exterior
partial slice with independent left and right boost edge modes at the respective corners. Equivalently, the extended phase space is parametrized by a smooth two sided boost conjugate pair $(s,K)$ alongside a one sided boost discontinuity conjugate pair $(q, J)$. The smooth one boundary theory is recovered by
Marsden--Weinstein reduction \cite{MarsdenWeinstein1974}: impose $J=0$ and quotient by shifts of $q$. In other words, $q$ is a relative boost
discontinuity that measures the extent to which the interior and exterior fail to be glued back together smoothly. Hence a nonzero value of $q$ is the classical
precursor of the firewall.

The main point of this section is that gravitational dressing forces this picture. Once an interior insertion is dressed to the left corner frame and
an exterior insertion is dressed to the right corner frame, the hard mode
algebra and the boost edge mode algebra are related by covariance rather than
by an ordinary tensor product. In other words, the extended subregion algebra
is a crossed product. Thus, a relative boost phase between the interior and exterior may be represented either on the
$\tilde b$ hard mode or on the one sided boost discontinuity edge mode $q$.

\subsection{One sided black hole phase space}
\label{sec:contclocklabel}

We now specialize the general relational framework to the one sided black hole JT gravity setup depicted in \cref{fig:penroseinfall}. The physical state at clock time \(T\) is defined on the full observer dressed slice \(\Sigma_T\), which runs from the infalling observer worldline to the dynamical end of the world brane. In this subsection we describe only the smooth global one sided phase space of that slice. The extended phase space resulting from cutting across the horizon, which will be needed for the AMPS calculation, is introduced only in the next subsection.

In this subsection we keep the probe matter sector in the classical vacuum. In particular, we do not introduce additional bulk matter phase space coordinates. The AMPS labels \(\tilde b\), \(b\), the early radiation, and the end of the world brane label sector enter only after the extended phase space has been quantized. This keeps the classical discussion minimal and makes clear that the probe matter labels are purely intrinsic (quantum mechanical) degrees of freedom attached to the field configurations on the chosen slice, not extra classical bulk phase space coordinates.

We work with JT gravity on a one sided black hole spacetime \(M\) with a dynamical end of the world brane \(\mathcal B\), an infalling observer worldline \(\gamma\), and a future horizon \(\mathscr H\). The observer carries a probe clock. The clock reading \(T\) selects the event \(\gamma_T:=\gamma(T)\).
The observer is not a gravitational boundary. It is a relational marker used to choose a slice and to define a clock variable.

The Lorentzian action is
\begin{align}
&I
=
-S_0\chi(M)
-
\frac{1}{16\pi G}
\int_M d^2x\sqrt{-g}\,\Phi(R+2)
-
\frac{1}{8\pi G}
\int_{\mathcal B} ds\,(\Phi K-\mu).
\label{eq:sec5_EOW_action}
\end{align}
Here \(K=\nabla_a n^a\), \(n^a\) is the outward pointing unit normal to \(\mathcal B\), and \(ds\) is the proper length element on the brane. The brane variation is
\begin{align}
&\delta I\big|_{\mathcal B}
=
-
\frac{1}{8\pi G}
\int_{\mathcal B} ds
\left[
K\,\delta\Phi
+
\frac12
(n^a\nabla_a\Phi-\mu)h^{bc}\delta h_{bc}
\right],
\label{eq:sec5_EOW_variation}
\end{align}
where \(h_{ab}\) is the induced metric on \(\mathcal B\). Since the brane is dynamical, neither \(\delta\Phi\) nor \(\delta h_{ab}\) is fixed at \(\mathcal B\). The variational principle gives
\begin{align}
&K=0,
\\
&n^a\nabla_a\Phi=\mu.
\label{eq:sec5_EOW_boundary_conditions}
\end{align}
Thus \(\mathcal B\) is a timelike geodesic and its location is fixed by the dilaton boundary condition.

The slice \(\Sigma_T\) is defined by firing from \(\gamma_T\) and requiring that it land orthogonally on the end of the world brane. If \(t^a\) is the future directed unit normal to \(\Sigma_T\), this condition is
\begin{align}
&t^a n_a\big|_{\Sigma_T\cap\mathcal B}=0.
\label{eq:sec5_brane_orthogonal_slice}
\end{align}
For each solution in the family under consideration, the event \(\gamma_T\) together with \eqref{eq:sec5_brane_orthogonal_slice} fixes \(\Sigma_T\) as a gauge equivalence class. Local deformations of the embedded curve that vanish at \(\gamma_T\) and at \(\Sigma_T\cap\mathcal B\) are generated by spacetime diffeomorphisms in degeneracy directions of the presymplectic form, so they do not define additional phase space data. The orthogonality condition also removes any would be brane boost angle. Indeed, the total brane contribution to the presymplectic potential is proportional to \eqref{eq:sec5_EOW_boundary_conditions}, up to endpoint conventions. Hence a tilt of the endpoint along \(\mathcal B\) corresponds to a degeneracy of the symplectic form. The brane therefore carries no boost edge mode in the observer dressed phase space. After the horizon cut is introduced below, the same local wiggles remain pure gauge because the quotient removes diffeomorphisms that vanish at all  corners. The only new boundary data are the independent horizon corner boosts.

The bulk equations are
\begin{align}
&R+2=0,
\\
&\nabla_a\nabla_b\Phi=g_{ab}\Phi.
\label{eq:sec5_bulk_EOM}
\end{align}
Define
\begin{align}
&\Phi_h^2:=\Phi^2-(\nabla\Phi)^2.
\label{eq:sec5_Phi_h_def}
\end{align}
This scalar is constant on shell. Indeed,
\begin{align}
&\nabla_a\Phi_h^2
=
2\Phi\nabla_a\Phi
-
2\nabla^b\Phi\nabla_a\nabla_b\Phi
=
0.
\label{eq:sec5_Phi_h_constant}
\end{align}
All on shell field configurations contain the timelike Killing vector \(\xi^a:=\epsilon^{ab}\nabla_b\Phi\).
Using \eqref{eq:sec5_bulk_EOM},
\begin{align}
&\nabla_{(a}\xi_{b)}
=
\epsilon_{c(a}\nabla_{b)}\nabla^c\Phi
=
\Phi\,\epsilon_{c(a}\delta_{b)}^c
=
0,
\\
&\lie_\xi\Phi
=
\epsilon^{ab}\nabla_a\Phi\nabla_b\Phi
=
0.
\label{eq:sec5_xi_Killing}
\end{align}
Its norm is
\begin{align}
&\xi^2
=
-(\nabla\Phi)^2
=
\Phi_h^2-\Phi^2.
\label{eq:sec5_Killing_norm}
\end{align}
Therefore the Killing horizon is the dilaton level surface
\begin{align}
&\mathscr H
=
\{\xi^2=0\}
=
\{\Phi=\Phi_h\}.
\label{eq:sec5_horizon_level_set}
\end{align}
The slice $\Sigma_T$ intersects this horizon at a unique point, \(C_T:=\Sigma_T\cap\mathscr H\).

Before introducing the cut, it is useful to recall the smooth one sided phase space. In the exterior region we may use \(\Phi\) itself as the radial coordinate and define a Killing time \(u\) by \(\xi^a\nabla_a u=1\).
The metric then takes the intrinsic static form
\begin{align}
&ds^2
=
-(\Phi^2-\Phi_h^2)\,du^2
+
\frac{d\Phi^2}{\Phi^2-\Phi_h^2}.
\label{eq:sec5_intrinsic_static_solution}
\end{align}
Let \(\tau_{\mathcal B}\) be proper time along \(\mathcal B\), and let \(U^a\) be the brane tangent. Since \(K=0\), \(U^a\) obeys the geodesic equation. The Killing energy of the brane is \(E_{\mathcal B}:=-\xi_aU^a\).
It is conserved because \(\xi^a\) is Killing and \(U^a\) is geodesic. The brane frame decomposition of the dilaton gradient gives
\begin{align}
&(\nabla\Phi)^2
=
(n^a\nabla_a\Phi)^2
-
(U^a\nabla_a\Phi)^2
=
\mu^2-\dot\Phi_{\mathcal B}^{\,2},
\label{eq:sec5_brane_gradient_decomposition}
\end{align}
where \(\dot\Phi_{\mathcal B}:=U^a\nabla_a\Phi\). Combining this with \eqref{eq:sec5_Phi_h_def} gives
\begin{align}
&\dot\Phi_{\mathcal B}^{\,2}
=
\Phi_h^2+\mu^2-\Phi_{\mathcal B}^{\,2}.
\label{eq:sec5_brane_radial_motion}
\end{align}
In the static coordinates \eqref{eq:sec5_intrinsic_static_solution},
\begin{align}
&E_{\mathcal B}
=
(\Phi_{\mathcal B}^{\,2}-\Phi_h^2)\frac{du_{\mathcal B}}{ds},
\label{eq:sec5_E_brane_static}
\end{align}
and the normalization \(U^aU_a=-1\) gives
\begin{align}
&E_{\mathcal B}^{\,2}
=
\dot\Phi_{\mathcal B}^{\,2}
+
\Phi_{\mathcal B}^{\,2}
-
\Phi_h^2.
\label{eq:sec5_E_brane_squared}
\end{align}
Using \eqref{eq:sec5_brane_radial_motion}, we find \(E_{\mathcal B}^{\,2}=\mu^2\).
The orientation of the brane fixes \(E_{\mathcal B}=\mu\). Hence
\begin{align}
\frac{du_{\mathcal B}}{ds}
=
\frac{\mu}{\Phi_{\mathcal B}^{\,2}-\Phi_h^2}, \
\left(\frac{d\Phi_{\mathcal B}}{ds}\right)^2
=
\Phi_h^2+\mu^2-\Phi_{\mathcal B}^{\,2}.
\label{eq:sec5_brane_solution_first_order}
\end{align}
Equivalently,
\begin{align}
&u_{\mathcal B}(\Phi)
=
u_0
\pm
\int^\Phi
\frac{\mu\,d\Phi'}
{\left((\Phi')^2-\Phi_h^2\right)
\sqrt{\Phi_h^2+\mu^2-(\Phi')^2}}.
\label{eq:sec5_brane_u0_intrinsic}
\end{align}
The constant \(u_0\) is the intrinsic one sided time shift. It is the remaining integration constant specifying the location of the brane geodesic relative to the Killing time coordinate. Once \(\Phi_h\) and \(\mu\) are fixed, all brane trajectories differ only by this shift.

The Hamiltonian \(H_{\rm JT}\) conjugate to \(u_0\) is obtained from the covariant phase space charge associated with the intrinsic Killing field \(\xi^a\). The local surface charge variation is
\begin{align}
&h_\xi(\delta)
=
\delta Q_\xi
-
Q_{\delta\xi}
-
\xi\cdot\Theta(\delta).
\label{eq:sec5_field_dependent_kxi}
\end{align}
The term \(Q_{\delta\xi}\) is included because \(\xi^a=\epsilon^{ab}\nabla_b\Phi\) is field dependent. Let \(h_{ab}:=\delta g_{ab}\), \(h:=g^{ab}h_{ab}\), and \(h^{ab}:=g^{ac}g^{bd}h_{cd}\). For JT gravity one may take \(\Theta(\delta)=\epsilon_a\theta^a(\delta)\), with
\begin{align}
&\theta^a(\delta)
=
\frac{1}{16\pi G}
\left[
\Phi
\left(
\nabla^a h-\nabla_b h^{ab}
\right)
+
h^{ab}\nabla_b\Phi
-
h\nabla^a\Phi
\right],
\label{eq:sec5_metric_theta_covariant}
\end{align}
and the Noether charge zero form is
\begin{align}
&Q_{\zeta}
=
-\frac{1}{16\pi G}
\epsilon_{ab}
\left(
\Phi\nabla^a\zeta^b
+
2\zeta^a\nabla^b\Phi
\right).
\label{eq:sec5_JT_Noether_charge_zero_form}
\end{align}
Using
\begin{align}
&\nabla_a\xi_b
=
\nabla_a\left(\epsilon_b{}^{c}\nabla_c\Phi\right)=
-\Phi\epsilon_{ab},
\label{eq:sec5_nabla_xi}
\end{align}
together with
\begin{align}
&\delta\xi^a
=
-\frac12 h\xi^a+\epsilon^{ab}\nabla_b\delta\Phi,
\\
&\delta\epsilon_{ab}
=
\frac12 h\epsilon_{ab},
\\
&\delta\left((\nabla\Phi)^2\right)
=
-h^{ab}\nabla_a\Phi\nabla_b\Phi
+
2\nabla^a\Phi\nabla_a\delta\Phi,
\label{eq:sec5_covariant_identities_for_k}
\end{align}
a direct substitution into \eqref{eq:sec5_field_dependent_kxi} gives
\begin{align}
&16\pi G\,h_\xi(\delta)
=
2\Phi\,\delta\Phi
+
h^{ab}\nabla_a\Phi\nabla_b\Phi
-
2\nabla^a\Phi\nabla_a\delta\Phi
=
\delta\left(\Phi^2-(\nabla\Phi)^2\right).
\end{align}
Therefore
\begin{align}
&h_\xi(\delta)
=
\frac{1}{16\pi G}\delta\Phi_h^2.
\label{eq:sec5_kxi_delta_Casimir}
\end{align}

This local charge variation becomes the Hamiltonian on the interval only after we specify the relative endpoint flow. Let \(\zeta^a\) be a vector field whose boundary values are
\begin{align}
&\zeta^a\big|_{\mathcal B}
=
\alpha_{\mathcal B}\xi^a,
\\
&\zeta^a\big|_{\gamma_T}
=
\alpha_{\gamma}\xi^a.
\label{eq:sec5_relative_endpoint_generator}
\end{align}
The extension of \(\zeta^a\) into the interior is pure gauge and changes the generator only by bulk constraints. On shell,
\begin{align}
&\delta H_\zeta
=
\int_{\partial\Sigma_T} h_\zeta(\delta)
=
\alpha_{\mathcal B}h_\xi(\delta)\big|_{\mathcal B\cap\Sigma_T}
-
\alpha_{\gamma}h_\xi(\delta)\big|_{\gamma_T}.
\label{eq:sec5_endpoint_charge_variation}
\end{align}
Therefore
\begin{align}
&\delta H_\zeta
=
(\alpha_{\mathcal B}-\alpha_{\gamma})
\frac{1}{16\pi G}
\delta\Phi_h^2.
\label{eq:sec5_relative_charge_variation}
\end{align}
The common endpoint transformation \(\alpha_{\mathcal B}=\alpha_\gamma\) is pure gauge on the finite interval, since its charge vanishes on all solutions. The remaining degree of freedom is the relative flow parameter \(\alpha_{\mathcal B}-\alpha_\gamma\), normalized by \(\alpha_{\mathcal B}-\alpha_\gamma=1\). Choosing the zero of energy at \(\Phi_h=0\), one obtains
\begin{align}
&H_{\rm JT}
=
\frac{\Phi_h^2}{16\pi G}.
\label{eq:sec5_HJT_Phi_h}
\end{align}
This charge is conjugate to the relative Killing time shift \(u_0\) between the observer endpoint and the end of the world brane endpoint. Thus
\begin{align}
&\Theta_{\rm red}
=
H_{\rm JT}\,\delta u_0,
\label{eq:sec5_reduced_potential_one_sided}
\end{align}
and
\begin{align}
&\Omega_{\rm red}
=
\delta H_{\rm JT}\wedge\delta u_0.
\label{eq:sec5_reduced_symplectic_one_sided}
\end{align}
Hence the smooth one sided JT plus end of the world brane phase space is \(\Gamma_{\rm red}\simeq T^*\mathbb R_{u_0}\), with \(\{u_0,H_{\rm JT}\}=1\).

Adding the infalling clock yields \(\Gamma=\Gamma_{\rm red}\times T^*\mathbb R_T\), with
\begin{align}
&\Omega
=
\delta H_{\rm JT}\wedge\delta u_0
+
\delta P_T\wedge\delta T.
\label{eq:sec5_clock_extended_phase_space}
\end{align}

\subsection{Extended phase space and crossed product algebra}
\label{sec:classical_horizon_extension}

The AMPS calculation needs two kinds of data that are absent from the smooth chart by itself. First, the interior mode \(\tilde b\) and the exterior mode \(b\) must be described as operators dressed to opposite sides of the horizon corner on the same observer dressed slice. Second, after topology change is turned on, the interior portion of that slice must be promoted to a leg on which creation and annihilation operators act. Both steps require a subregion description. On the smooth slice, however, the gravitational constraint ties the interior and exterior dressings together at the single corner \(C_T\), so there is no canonical factorization
\begin{align}
&\Gamma(T)
\not\simeq
\Gamma_{\rm int}(T)\times \Gamma_{\rm ext}(T).
\label{eq:sec5_no_naive_factorization}
\end{align}
The extended phase space is therefore an enlargement of the same one boundary theory in which the two sides of the horizon cut carry independent corner frames before gluing. This leads to a four dimensional corner phase space. One configuration variable is the smooth two sided boost \(s\) of the observer dressed slice. The other is \(q\), the discontinuous jump in boost angle between the left and right horizon cuts. A nonzero \(q\) is precisely the classical precursor to the firewall: it prevents the two sides from being glued smoothly back together across the horizon. Although \(q\) is written symmetrically in the two corner frames, the AMPS topology changing calculation keeps the exterior corner frame, together with \(b\) and \(E\), fixed across the interior leg sectors. Thus a nonzero \(q\) in that calculation is a gluing phase, i.e.\ a one sided boost of an interior daughter leg with the exterior held fixed. The charge conjugate to this discontinuity is the one sided boost generator \(J\). In the smooth one boundary phase space the pair $(q,J)$ is absent. It is introduced only so that the interior and exterior algebras can be represented separately before the final gluing step.

The smooth observer dressed slice at clock time \(T\) carries the reduced one sided phase space
\begin{align}
&\Gamma(T)
\simeq
T^*\mathbb R_{u_0}
\times
T^*\mathbb R_T,
\\
&\Omega
=
\delta H_{\rm JT}\wedge\delta u_0
+
\delta P_T\wedge\delta T.
\label{eq:sec5_global_phase_space_again}
\end{align}
The horizon point \(C_T=\Sigma_T\cap\mathscr H\) is a field dependent corner of this slice. Following Donnelly and Freidel, we replace this field dependent corner by a fixed corner on a reference slice and promote the map from the reference slice into spacetime to phase space data \cite{DonnellyFreidel2016,Speranza2018LocalPhaseSpace}. In the black hole language of the subregion algebra construction, this gives the corner data needed to dress the two sides of the horizon cut separately before symplectic reduction glues them back into the smooth one boundary phase space.

Choose a reference slice \(\bar\Sigma_T\) with coordinate \(\sigma\) such that \(\bar C_T=\{\sigma=0\}\). For each \(\varepsilon>0\), excise the reference interval \(I_\varepsilon:=(-\varepsilon,\varepsilon)\), define the split reference subregions \(\bar\Sigma_{T,-}^{\varepsilon}:=\{\sigma\le -\varepsilon\}\) and \(\bar\Sigma_{T,+}^{\varepsilon}:=\{\sigma\ge \varepsilon\}\), and denote their split reference corners by \(\bar C_T^-:=\{\sigma=-\varepsilon\}\) and \(\bar C_T^+:=\{\sigma=\varepsilon\}\).
The regulator is a short reference interval centered at the horizon corner. It thickens the single cut into two nearby  corners, which allows the interior and exterior partial slices to carry independent corner frames before the symplectic quotient reconstructs the smooth slice.

Let \(\bar{\mathscr H}\) be the reference future horizon generator through \(\bar C_T\), with affine parameter \(\bar u\) chosen so that \(\bar C_T=\{\bar u=0\}\). The respective horizon copies are \(\bar{\mathscr H}_L^\varepsilon:=\{\bar u\le -\varepsilon\}\) and \(\bar{\mathscr H}_R^\varepsilon:=\{\bar u\ge \varepsilon\}\),
with endpoints \(\bar C_T^-\) and \(\bar C_T^+\). These are the one dimensional horizon segments whose images will appear in the corner symplectic potential.

The extended configuration space is described by two embeddings, fixed by the same relational slice prescription away from the split corners, \(X_-^{\varepsilon}:\bar\Sigma_{T,-}^{\varepsilon}\to M\) and \(X_+^{\varepsilon}:\bar\Sigma_{T,+}^{\varepsilon}\to M\). Their images are \(\Sigma_T^-:=X_-^{\varepsilon}(\bar\Sigma_{T,-}^{\varepsilon})\) and \(\Sigma_T^+:=X_+^{\varepsilon}(\bar\Sigma_{T,+}^{\varepsilon})\), while the corner images are \(C_T^-:=X_-^{\varepsilon}(\bar C_T^-)\) and \(C_T^+:=X_+^{\varepsilon}(\bar C_T^+)\). Restricting the embeddings to the respective horizon copies gives \(X_L^\varepsilon:=X_-^\varepsilon|_{\bar{\mathscr H}_L^\varepsilon}\) and \(X_R^\varepsilon:=X_+^\varepsilon|_{\bar{\mathscr H}_R^\varepsilon}\), with images \(\mathscr H_L^\varepsilon:=X_L^\varepsilon(\bar{\mathscr H}_L^\varepsilon)\) and \(\mathscr H_R^\varepsilon:=X_R^\varepsilon(\bar{\mathscr H}_R^\varepsilon)\). Equivalently, \(\Sigma_T^\varepsilon=\Sigma_T^-\cup\Sigma_T^+\),
where \(\Sigma_T^-\) is the interior partial slice ending on \(\mathcal B\) and \(\Sigma_T^+\) is the exterior partial slice containing the observer clock. The limit \(\varepsilon\to0\) is taken after gluing.

The  extended phase space is
\begin{align}
&\widehat\Gamma_\varepsilon(T)
:=
\Gamma_-^\varepsilon(T)
\times
\Gamma_+^\varepsilon(T),
\label{eq:sec5_regulated_extended_phase_space}
\end{align}
with
\begin{align}
&\Gamma_\pm^\varepsilon(T)
:=
\left\{
\left(
X_\pm^{\varepsilon\,*}g,
X_\pm^{\varepsilon\,*}\Phi,
X_\pm^\varepsilon
\right)
\right\}
\big/
\mathrm{Diff}_0(\bar\Sigma_{T,\pm}^{\varepsilon}).
\label{eq:sec5_subregion_phase_spaces}
\end{align}
Here \(\mathrm{Diff}_0(\bar\Sigma_{T,\pm}^{\varepsilon})\) denotes diffeomorphisms whose induced spacetime action vanishes at every  corner of the corresponding partial slice. These are the local wiggles of the interior and exterior embeddings; they remain pure gauge after the cut. The new data introduced by the extension are the independent corner boosts at \(C_T^-\) and \(C_T^+\). Diffeomorphisms with support at those corners act as surface symmetries and remain Hamiltonian on \(\widehat\Gamma_\varepsilon(T)\). In JT gravity there are no local propagating bulk degrees of freedom, so on the constraint surface the symplectic form is entirely a sum of boundary and corner terms. Before the constraints are imposed it is still useful to write the formal covariant phase space decomposition
\begin{align}
&\widehat\Omega_\varepsilon
=
\Omega_{{\rm bulk},-}^\varepsilon
+
\Omega_{{\rm bulk},+}^\varepsilon
+
\Omega_{\mathscr H}^{(1),\varepsilon},
\label{eq:sec5_regulated_extended_symplectic_form}
\end{align}
Here \(\Omega_{{\rm bulk},\pm}^\varepsilon\) denotes the covariant symplectic current integrated over the two partial slices before solving the constraints. On the JT gravity constraint surface this contribution is exact and reduces to the endpoint term already described by \(\delta H_{\rm JT}\wedge\delta u_0\), together with the vanishing brane and observer endpoint contributions fixed above. Thus no local bulk mode is being added by \(\Omega_{{\rm bulk},\pm}^\varepsilon\). The genuinely new term produced by cutting across the horizon is the corner term \(\Omega_{\mathscr H}^{(1),\varepsilon}\), derived below from the pullback of the covariant symplectic potential to the horizon.

After the corner term is derived, the gluing data are organized by
\begin{align}
&J
:=
\mathscr A_L+\mathscr A_R,
\\
&K
:=
\mathscr A_R-\mathscr A_L,
\\
&q
:=
\frac{s_L-s_R}{2},
\\
&s
:=
-\frac{s_L+s_R}{2}.
\label{eq:sec5_extended_gluing_variables_formal}
\end{align}
The pair \((s,K)\) is the smooth two sided corner pair. The variable \(s\) is the boost of the observer dressed glued slice, and \(K\) is the two sided boost generator conjugate to \(s\). The gluing orbit shifts the one sided boost discontinuity \(q\) while leaving \(s\) fixed. This is the gauge orbit removed by gluing. In the later topology change calculation the same one sided boost discontinuity coordinate is represented in a frame where the exterior corner frame is fixed and the shift is carried by the interior daughter frame. Since \(\iota_{\partial_q}\widehat\Omega_{\mathscr H}^{(1),\varepsilon}=-\delta J\), the moment map is \(\mu_\varepsilon:=J\), with the sign convention used below. The classical gluing map is the Marsden Weinstein reduction at zero moment map. Its constraint surface is
\begin{align}
&\mathcal C_\varepsilon
:=
\mu_\varepsilon^{-1}(0)
=
\{J=0\}
\subset
\widehat\Gamma_\varepsilon(T).
\label{eq:sec5_classical_gluing_surface}
\end{align}
and the reduced phase space is the quotient
\begin{align}
&\Gamma(T)
:=
\mathcal C_\varepsilon
\big/
\mathbb R_{q}.
\label{eq:sec5_classical_glued_phase_space}
\end{align}
This is the formal meaning of gluing in the classical calculation: impose the constraint \(J=0\), the zero moment map condition for smooth gluing across the horizon, then identify points along the gauge orbit generated by \(J\), equivalently along shifts of the one sided boost discontinuity \(q\). The smooth two sided corner pair \((s,K)\) survives this reduction, while the conjugate pair \((q,J)\) is removed.
If \(\iota_\varepsilon:\mathcal C_\varepsilon\hookrightarrow\widehat\Gamma_\varepsilon(T)\) is the inclusion and \(\pi_\varepsilon:\mathcal C_\varepsilon\to\Gamma(T)\) is the quotient map, then
\begin{align}
&\iota_\varepsilon^*\widehat\Omega_\varepsilon
=
\pi_\varepsilon^*\Omega.
\label{eq:sec5_symplectic_reduction_identity}
\end{align}
The chosen relational dressing determines a section of the quotient,
\begin{align}
&\mathsf c_\varepsilon:
\Gamma(T)
\to
\mathcal C_\varepsilon,
\\
&\pi_\varepsilon\circ\mathsf c_\varepsilon
=
\mathbf 1_{\Gamma(T)}.
\label{eq:sec5_classical_cut_section}
\end{align}
An extended observable \(\widehat{\mathcal O}_\varepsilon\) is converted back to an observable on the smooth slice by
\begin{align}
&\mathcal O
=
\pi_\varepsilon
\circ
\widehat{\mathcal O}_\varepsilon
\circ
\mathsf c_\varepsilon.
\label{eq:sec5_classical_cut_glue_observable}
\end{align}
So we start from a global observer dressed state, pass to the extended phase space before topology changing and local interior calculations, and apply the gluing map before comparing with a one boundary Dirac observable.

We now derive the horizon symplectic potential directly from the covariant expression \eqref{eq:sec5_metric_theta_covariant}. Choose a normalized null frame \((\ell^a,n^a)\) satisfying
\begin{align}
&\ell^2=0,\;
n^2=0,\;
\ell\cdot n=-1,
\\
&g_{ab}=-2\ell_{(a}n_{b)},\;
\epsilon_{ab}=2\ell_{[a}n_{b]}.
\label{eq:sec5_null_frame}
\end{align}
In two dimensions the normalized null frame has structure group \(SO(1,1)\). Since \(\ell^2=0\), \(\nabla_a\ell^b\) is orthogonal to \(\ell_b\), and in two dimensions every vector orthogonal to \(\ell_b\) is proportional to \(\ell^b\). The coefficient is the \(SO(1,1)\) spin connection, equivalently the normal bundle connection of the null frame in the conventions commonly used for gravitational edge modes and null boundaries \cite{DonnellyFreidel2016}:
\begin{align}
&\nabla_a\ell^b=-\omega_a\ell^b,
\\
&\omega_a=-n_b\nabla_a\ell^b.
\label{eq:sec5_omega_null_frame}
\end{align}
Under a local boost \(\ell^a\mapsto \ee^{-\lambda}\ell^a\), \(n^a\mapsto \ee^{\lambda}n^a\), the connection shifts as \(\omega\mapsto\omega+d\lambda\). This is a redundancy in the null dyad used to evaluate the symplectic potential. In other words, a variation of the null dyad contains, besides the metric variation, an arbitrary vertical \(SO(1,1)\) rotation. For the calculation below we fix this gauge freedom by choosing the field dependence of the null dyad to contain no such additional frame rotation,
\begin{align}
&\delta\ell^a=-\frac12 h^a{}_b\ell^b,
\\
&\delta n^a=-\frac12 h^a{}_b n^b.
\label{eq:sec5_null_frame_variation_choice}
\end{align}
In this gauge, we have
\begin{align}
&\delta\omega_a
=
-\delta n_b\nabla_a\ell^b
-
n_b\nabla_a\delta\ell^b
-
n_b\delta\Gamma^b{}_{ac}\ell^c
=
\frac12 n_b\ell^c\nabla_a h^b{}_c
-
n_b\ell^c\delta\Gamma^b{}_{ac},
\end{align}
where the terms proportional to \(\omega_a\) cancel. Using
\begin{align}
&\delta\Gamma^b{}_{ac}
=
\frac12 g^{bd}
\left(
\nabla_a h_{cd}
+
\nabla_c h_{ad}
-
\nabla_d h_{ac}
\right),
\end{align}
we obtain
\begin{align}
\delta\omega_a
=
\frac12 \epsilon^{cd}\nabla_d h_{ac}.
\label{eq:sec5_delta_omega_covariant}
\end{align}
Using \(\epsilon_{ab}\epsilon^{cd}=-\delta_a^c\delta_b^d+\delta_a^d\delta_b^c\),
this implies
\begin{align}
&2\delta\omega_b
=
\epsilon_{ab}
\left(
\nabla^a h-\nabla_c h^{ac}
\right).
\label{eq:sec5_metric_to_spin_identity}
\end{align}

We only need the pullback of the symplectic potential to the horizon. On \(\mathscr H\), the relational horizon location \(\Phi\hateq\Phi_h\) implies \(\nabla_a\Phi\hateq \alpha\,\ell_a\) for some scalar \(\alpha\), and therefore \(\ell^a\nabla_a\Phi\hateq0\). The variation preserves the null signature of the horizon, so \(\delta\left((\nabla\Phi)^2\right)\hateq0\).
But we have that \(\delta((\nabla\Phi)^2)=-h^{ab}\nabla_a\Phi\nabla_b\Phi+2\nabla^a\Phi\nabla_a\delta\Phi\),
while \(\delta\Phi\hateq\delta\Phi_h\) is constant along the horizon generator, so \(h^{ab}\nabla_a\Phi\nabla_b\Phi\hateq0\).
Hence the pullback of the dilaton gradient terms in \eqref{eq:sec5_metric_theta_covariant} vanishes:
\begin{align}
&\iota_{\mathscr H}^*
\Big[
\epsilon_a
\left(
h^{ab}\nabla_b\Phi
-
h\nabla^a\Phi
\right)
\Big]
=0.
\label{eq:sec5_dilaton_gradient_terms_vanish}
\end{align}
Combining \eqref{eq:sec5_metric_theta_covariant}, \eqref{eq:sec5_metric_to_spin_identity}, and \eqref{eq:sec5_dilaton_gradient_terms_vanish}, we find
\begin{align}
&\iota_{\mathscr H}^*\Theta(\delta)
=
\frac{1}{8\pi G}\Phi_h\,\iota_{\mathscr H}^*\delta\omega.
\label{eq:sec5_spin_connection_theta}
\end{align}
This is the horizon symplectic potential that enters the corner calculation.

Let \(X_L\) and \(X_R\) denote the embedding maps of the two horizon copies into \(M\). They are not new dynamical fields. They only specify which copy of the  cut is being pulled back. The structure group is \(SO(1,1)\simeq \mathbb R\), so the spin connection has no commutator term in its curvature, i.e.\ it is an abelian connection. On a one dimensional horizon every pulled back one form is locally exact. Thus, after choosing a boost frame at one endpoint of each  copy, there exist boost potentials \(s_L\) and \(s_R\) such that \(X_L^*\omega\hateq ds_L\) and \(X_R^*\omega\hateq ds_R\). A change of endpoint frame shifts \(s_L\) or \(s_R\) by a constant, which is exactly the corner boost transformation generated by the corresponding corner charge. Using \eqref{eq:sec5_spin_connection_theta},
\begin{align}
&\Theta_{\mathscr H}^{\varepsilon}(\delta)
=
\frac{1}{8\pi G}
\int_{\mathscr H_L^\varepsilon\cup\mathscr H_R^\varepsilon}
\Phi\,\delta\omega=
\frac{\Phi_h}{8\pi G}
\left[
\int_{\mathscr H_L^\varepsilon} d(\delta s_L)
+
\int_{\mathscr H_R^\varepsilon} d(\delta s_R)
\right].
\end{align}
The common regulator endpoint cancels between the two copies because they come with opposite orientations. Only the two corner endpoints remain. With the orientations inherited from \(\Sigma_T^-\) and \(\Sigma_T^+\), define the oriented corner charges \(\mathscr A_L:=-\Phi_{C_T^-}/(8\pi G)\) and \(\mathscr A_R:=\Phi_{C_T^+}/(8\pi G)\).
Then
\begin{align}
&\Theta_{\mathscr H}^{\varepsilon}(\delta)
=
\mathscr A_L\,\delta s_L
-
\mathscr A_R\,\delta s_R,
\label{eq:sec5_corner_potential_LR}
\end{align}
and therefore
\begin{align}
&\widehat\Omega_{\mathscr H}^{(1),\varepsilon}
=
\delta\mathscr A_L\wedge\delta s_L
-
\delta\mathscr A_R\wedge\delta s_R.
\label{eq:sec5_single_extended_symplectic}
\end{align}

Using the gluing variables of \eqref{eq:sec5_extended_gluing_variables_formal}, \eqref{eq:sec5_corner_potential_LR} becomes
\begin{align}
&\Theta_{\mathscr H}^{\varepsilon}(\delta)
=
J\delta q
+
K\delta s,
\label{eq:sec5_corner_potential_rewritten}
\end{align}
and
\begin{align}
&\widehat\Omega_{\mathscr H}^{(1),\varepsilon}
=
\delta J\wedge\delta q
+
\delta K\wedge\delta s.
\label{eq:sec5_corner_symplectic_rewritten}
\end{align}
Thus the moment map constraint \(J=0\) removes the pair \((q,J)\) and leaves \(\Omega_{\rm red}^{(1)}=\delta K\wedge\delta s\). On a smooth glued slice, \(K=\Phi_h/(4\pi G)\).
The reduced corner chart and the global chart therefore describe the same one boundary phase space. Equating the reduced one forms gives \(H_{\rm JT}\,\delta u_0=K\,\delta s\) on a branch with fixed \(\Phi_h\).

We now write the relational constraint on the extended slice. Let \(\xi_T^a\) be the vector field that advances the observer dressed slice when the observer clock is advanced. Its restrictions to the two sides are \(\xi_T^{-a}\) and \(\xi_T^{+a}\). The local bulk constraints are
\begin{align}
C_{\pm}[\xi_T^{\pm}]
=
\int_{\Sigma_T^{\pm}}
\left(
N_{\pm}\mathcal H_{\perp,{\pm}}
+
N_{\pm}^x\mathcal H_{x,\pm}
\right).
\label{eq:sec5_bulk_constraints_extended}
\end{align}
These are the genuine bulk constraints. The corner charges are not part of the local bulk constraint densities. They are boundary terms required to make the subregion generators differentiable. The corner part of the differentiable extended generator is
\begin{align}
H_C[\xi_T]
=
\dot s_L\mathscr A_L
-
\dot s_R\mathscr A_R.
\label{eq:sec5_corner_generator_eta}
\end{align}
Using \eqref{eq:sec5_extended_gluing_variables_formal}, this becomes
\begin{align}
&H_C[\xi_T]
=
\dot q\,J
+
\dot s\,K.
\label{eq:sec5_corner_generator_JK}
\end{align}
Using the normalization derived in \cref{sec:infallclocksub}, the clock reading \(T\) directly measures the physical horizon boost variable, so \(\dot s=1\).
In this parametrization the flow generated by \(\dot q\) remains pure gauge, while the physical two sided flow corresponding to the observer's clock is generated by \(K\).

Because \(\Sigma_T\) ends on the observer worldline rather than the asymptotic boundary, the regulated generator contains no ADM term. The observer worldline is a relational marker rather than a gravitational boundary, and the end of the world brane contribution vanishes after imposing \(K=0\) and \(n^a\nabla_a\Phi-\mu=0\). Therefore the full regulated constraint before gluing is
\begin{align}
&\mathcal C_{\Sigma_T}^{\varepsilon}
=
P_T
+
C_-[\xi_T^-]
+
C_+[\xi_T^+]
+
\dot q\,J
+
K
\approx0.
\label{eq:sec5_extended_clock_constraint_before_gluing}
\end{align}
On the bulk constraint surface this reduces to
\begin{align}
&\mathcal C_{\Sigma_T}^{\varepsilon}
\doteq
P_T
+
\dot q\,J
+
K
\approx0.
\label{eq:sec5_extended_clock_constraint_bulk_reduced}
\end{align}
Before gluing, the term proportional to \(J\) generates shifts of the one sided boost discontinuity \(q\). The term proportional to \(K\) generates the smooth two sided boost flow of the observer dressed slice. After gluing one imposes \(J=0\) and removes the conjugate gluing phase \(q\). The smooth slice constraint is therefore
\begin{align}
&\mathcal C_{\Sigma_T}
=
P_T+K
\approx0.
\label{eq:sec5_boost_clock_constraint_after_gluing}
\end{align}

We now introduce gravitational dressing. A local matter field written at a coordinate point of \(\Sigma_T^\pm\) is not a Dirac observable, because the coordinate point moves under a diffeomorphism. To obtain an observable of the gravitational subregion one must specify where the insertion is placed relative to physical data of the solution. In the present construction the observer clock event and the end of the world brane endpoint condition fix the observer dressed slice. Once that slice and the regulator have been fixed, the only residual subregion variables are the horizon corner frames exposed by the cut. Therefore the residual dressing data of an interior insertion is \(s_L\), and the residual dressing data of an exterior insertion is \(s_R\). In our model, the observer, the clock, and the AMPS matter excitations are probes. They label relational insertions and test observables relative to these corner boost frames.

Let \(\mathfrak a_{\rm int}^0\) be the algebra of local classical observables on \(\Sigma_T^-\) before adjoining the corner frame, and let \(\mathfrak a_{\rm ext}^0\) be the corresponding exterior algebra on \(\Sigma_T^+\). A local scalar observable is not yet an observable of a gravitational subregion, because its location must be specified relationally. In the present prescription the observer dressed slice and the regulator fix the embeddings \(X_-^\varepsilon\) and \(X_+^\varepsilon\) across field configurations up to the independent corner boosts. Thus the only residual subregion data entering the dressed observables are \(s_L\) and \(s_R\). We therefore write \(\mathcal O_{\rm int}(s_L)\) and \(\mathcal O_{\rm ext}(s_R)\).
These are the observables obtained by evaluating the chosen relational insertions in the corner frames determined by the restrictions \(X_L^\varepsilon\) and \(X_R^\varepsilon\) of the extended embeddings. No additional embedding map is introduced here. The clock localization is part of the exterior dressing, and the scalar clock reading \(T\) is an internal clock label, so it has vanishing Poisson bracket with the corner boost charges.

The boost generators act on the dressed observables through their angle dependence:
\begin{align}
&\{\mathscr A_L,\mathcal O_{\rm int}(s_L)\}
=
\partial_{s_L}\mathcal O_{\rm int}(s_L),
\\
&\{\mathscr A_R,\mathcal O_{\rm ext}(s_R)\}
=
\partial_{s_R}\mathcal O_{\rm ext}(s_R).
\label{eq:sec5_dressed_boost_brackets}
\end{align}
Let \(\alpha_s^-\) and \(\alpha_s^+\) denote the corresponding automorphism groups. The classical extended algebras are
\begin{align}
&\mathfrak a_{\rm int}
=
\mathfrak a_{\rm int}^0\rtimes_{\alpha^-}\mathbb R,
\\
&\mathfrak a_{\rm ext}
=
\left(
\mathfrak a_{\rm ext}^0\otimes C^\infty(T^*\mathbb R_T)
\right)\rtimes_{\alpha^+}\mathbb R.
\label{eq:sec5_classical_crossed_products}
\end{align}
In a formal element \(FU(s)\), the product is
\begin{align}
&(FU(s))(GU(s'))
=
F\alpha_s(G)U(s+s').
\label{eq:sec5_crossed_product_rule}
\end{align}
The involution is
\begin{align}
&\bigl(FU(s)\bigr)^*
=
\alpha_{-s}(F^*)U(-s).
\label{eq:sec5_crossed_product_involution}
\end{align}
Equivalently, the algebra is generated by the dressed observables together with the boost unitaries \(U(s)\), subject to the covariance relation
\begin{align}
&U(s)FU(s)^{-1}
=
\alpha_s(F).
\label{eq:sec5_crossed_product_covariance}
\end{align}
This is the precise classical meaning of adjoining the corner boost. The boost does not commute with the dressed observable algebra, because changing the corner frame moves the relational dressing of every insertion dressed to that corner.

A covariant algebraic representation of the crossed product is obtained by using the corner frame as the base of a family of dressed configurations $C^\infty(\mathbb R_{s_L},\mathfrak a_{\rm int}^0)$. A function \(\Psi\in\mathcal C^\infty(\mathbb R_{s_L},\mathfrak a_{\rm int}^0)\) assigns to each corner frame \(s_L\) an interior classical observable written in that frame. In this representation,
\begin{align}
&\bigl(\pi_{\rm int}(F)\Psi\bigr)(s_L)
=
\alpha_{-s_L}^-(F)\Psi(s_L),
\\
&\bigl(U_L(s)\Psi\bigr)(s_L)
=
\Psi(s_L-s),
\label{eq:sec5_crossed_product_direct_integral_action}
\end{align}
and therefore
\begin{align}
&U_L(s)\pi_{\rm int}(F)U_L(s)^{-1}
=
\pi_{\rm int}\bigl(\alpha_s^-(F)\bigr).
\label{eq:sec5_crossed_product_direct_integral_covariance}
\end{align}
The exterior algebra has the same structure with \(s_R\) and \(\alpha_s^+\). The hard factor and the edge factor are therefore linked by the covariance relation rather than by a canonical tensor factorization.

The one sided boost discontinuity \(q\), equivalently the gluing phase, is conjugate to the one sided boost generator \(J\). An extended observable admits the Fourier decomposition
\begin{align}
&\widehat{\mathcal O}_\varepsilon
=
\int_{\mathbb R} d\nu\,
\ee^{\ii\nu q}\,
\mathcal O_{\varepsilon,\nu},
\\
&\{J,\mathcal O_{\varepsilon,\nu}\}
=
\nu\,\mathcal O_{\varepsilon,\nu}.
\label{eq:sec5_classical_gluing_phase_modes}
\end{align}
The zero mode is the orbit average
\begin{align}
&\widehat{\mathcal O}_{\varepsilon,0}
:=
\int_{\mathbb R}\frac{d\beta}{2\pi}\,
\ee^{\beta\{J,\cdot\}}
\widehat{\mathcal O}_\varepsilon,
\label{eq:sec5_classical_zero_mode_average}
\end{align}
and the resulting observable on the smooth slice is
\begin{align}
&\mathcal O
=
\pi_\varepsilon
\circ
\widehat{\mathcal O}_{\varepsilon,0}
\circ
\mathsf c_\varepsilon.
\label{eq:sec5_classical_glued_zero_mode_observable}
\end{align}
Thus the classical reduction first imposes \(J=0\) and then keeps the \(J\) invariant part of the extended algebra. The pair \((q,J)\) is removed, while the smooth two sided pair \((s,K)\) remains. After quantization the same Fourier decomposition becomes the edge mode statement that a hard boost phase and an edge gluing phase are the two crossed product realizations of the same dressed operator.

\section{Canonical quantization and topology change}
\label{sec:RAQobserver}

The previous section identified the classical phase space relevant for the AMPS experiment in JT gravity. The smooth one boundary phase space has one smooth two sided corner pair \((s,K)\). The extended phase space doubles the corner frame and introduces the pair \((q,J)\), which is removed classically by the gluing condition \(J=0\). Quantization changes the status of the discussion in two ways. First, the corner frame becomes quantum mechanical, so operators may carry nontrivial Fourier modes in the gluing phase \(q\) before gluing. Second, the AMPS labels \(\tilde b\), \(b\), \(E\), and \(\tilde R\) appear as purely internal quantum states on the chosen observer dressed slice.

Moreover, the algebra of observables acting on the extended phase space was shown to be a crossed product. A boost of a dressed hard operator can be represented equivalently as an action on the edge mode variable that defines its dressing. Upon quantization, a firewall corresponds precisely to a nontrivial gluing phase in this crossed product representation. Written on the hard sector, it breaks the \(\tilde bb\) vacuum entanglement. Written on the edge sector, it breaks the smooth maximally entangled state of the left and right horizon edge modes. These are equivalent representations of the same operator.

The last part of the construction is the topology changing Hamiltonian. Once the extended phase space construction has turned the interior portion of \(\Sigma_T\) into an interior leg ending at the horizon corner, the natural third quantized kinematics is the Fock space sum over any number of such legs. Creation and annihilation operators add or remove entire black hole interiors. Their bosonic Wick contractions give the direct and exchange pairings that appeared in \cref{subsec:AMPS_warmup_contractions}. The interaction Hamiltonian used below is the leading order local cubic vertex at the horizon corner: one interior leg can split into two daughter legs, or two daughter legs can join into one. The vertex preserves the local boost frame through a boost composition law, which is why the edge mode labels survive into the AMPS calculation.

\subsection{Canonical quantization of the extended phase space}
\label{sec:canquanthilb}

We now quantize the extended phase space constructed above. The gravitational corner variables obey
\begin{align}
&[\hat s_L,\hat{\mathscr A}_L]=\ii,
\\
&[\hat s_R,\hat{\mathscr A}_R]=-\ii,
\\
&[\hat T,\hat P_T]=\ii,
\label{eq:sec6_corner_clock_commutators}
\end{align}
with all other elementary commutators vanishing. Equivalently,
\begin{align}
&[\hat q,\hat J]=\ii,
\\
&[\hat s,\hat K]=\ii.
\label{eq:sec6_q_s_commutators}
\end{align}
The one leg edge mode Hilbert space is
\begin{align}
&\widehat{\mathcal H}_{\mathscr H}^{(1),\varepsilon}
=
L^2(\mathbb R,ds_L)\otimes L^2(\mathbb R,ds_R).
\label{eq:sec6_one_leg_edge_hilbert}
\end{align}
In the \((s_L,s_R)\) representation,
\begin{align}
&\hat{\mathscr A}_L=-\ii\frac{\partial}{\partial s_L},
\\
&\hat{\mathscr A}_R=\ii\frac{\partial}{\partial s_R}.
\label{eq:sec6_A_operators_s_rep}
\end{align}

At fixed clock reading \(T\), a physical one boundary state may be written in the physical corner basis as
\begin{align}
&|\Psi(T)\rangle
=
\int ds\,
\Psi(T;s)\,
|T;s\rangle.
\label{eq:sec6_global_state_s_basis}
\end{align}
At the level of canonical quantization, both constraints are imposed by group averaging. The Hamiltonian constraint is the usual refined algebraic quantization average over relational time, as in \cref{sec:RAQ_bridge}. The horizon gluing constraint is the corresponding average over the corner gauge orbit generated by \(\hat J\). We keep the two different group averages notationally distinct because they act on different factors: the first constructs relational states, while the second implements the quantum version of the classical reduction by \(J=0\) at the split horizon corners:
\begin{align}
&\Pi^{(1),\varepsilon}_{J=0}
:=
\int \frac{d\beta}{2\pi}\,\ee^{\ii\beta\hat J}
=
\delta(\hat J).
\label{eq:sec6_gluing_projector}
\end{align}
Let \(|A_L\rangle_L\) and \(|A_R\rangle_R\) be generalized eigenstates of \(\hat{\mathscr A}_L\) and \(\hat{\mathscr A}_R\). Then
\begin{align}
&\Pi^{(1),\varepsilon}_{J=0}
\bigl(|A_L\rangle_L\otimes |A_R\rangle_R\bigr)
=
\delta(A_L+A_R)\,
|A_L\rangle_L\otimes |A_R\rangle_R.
\label{eq:sec6_charge_gluing}
\end{align}
At fixed two sided boost \(s\), the regulated smooth edge mode state is
\begin{align}
&|\Omega^{\varepsilon}_{s}\rangle_{LR}
:=
\int_{\mathbb R} dA\,
f_\varepsilon(A)\,\ee^{2\ii As}\,
|A\rangle_L\otimes |-A\rangle_R,
\label{eq:sec6_regulated_edge_modes_charge}
\end{align}
where \(f_\varepsilon\) is normalized on the charge band and chosen flat on its support. This state is the regulated left right maximally entangled edge mode state associated with the smooth glued corner.

The cutting map and gluing map are
\begin{align}
&\mathcal C^{\varepsilon}_{\mathscr H}
=
\int ds\,
|\Omega^{\varepsilon}_{s}\rangle_{LR}\,
\langle s|, \label{eq:cut_map}
\\
&\mathcal G^{\varepsilon}_{\mathscr H}
=
\bigl(\mathcal C^{\varepsilon}_{\mathscr H}\bigr)^\dagger
=
\int ds\,
|s\rangle\,
{}_{LR}\!\langle\Omega^{\varepsilon}_{s}|.
\label{eq:sec6_glue_map}
\end{align}
With the regulator normalized as above,
\begin{align}
&\mathcal G^{\varepsilon}_{\mathscr H}\,
\mathcal C^{\varepsilon}_{\mathscr H}
=
\mathbf 1_{\mathcal H_{\rm phys}(T)},
\\
&\mathcal C^{\varepsilon}_{\mathscr H}\,
\mathcal G^{\varepsilon}_{\mathscr H}
=
\Pi^{(1),\varepsilon}_{\mathscr H}.
\label{eq:sec6_cut_glue_relations}
\end{align}
The projector onto the smooth image inside the  extended Hilbert space is
\begin{align}
&\Pi^{(1),\varepsilon}_{\mathscr H}
=
\int ds\,
|\Omega^{\varepsilon}_{s}\rangle_{LR}\,
{}_{LR}\!\langle\Omega^{\varepsilon}_{s}|.
\label{eq:sec6_entangled_edge_projector}
\end{align}

A general extended operator may be decomposed into Fourier modes of the gluing phase:
\begin{align}
&\hat O
=
\int d\nu\,\ee^{\ii\nu\hat q}\,\hat O_\nu,
\\
&[\hat J,\hat O_\nu]=\nu\hat O_\nu.
\label{eq:sec6_operator_zero_mode_decomposition}
\end{align}
Gluing keeps only the zero mode:
\begin{align}
&\mathcal G^{\varepsilon}_{\mathscr H}\,\hat O\,\mathcal C^{\varepsilon}_{\mathscr H}
=
\mathcal G^{\varepsilon}_{\mathscr H}\,\hat O_0\,\mathcal C^{\varepsilon}_{\mathscr H}.
\label{eq:sec6_gluing_keeps_zero_mode}
\end{align}
Thus a nonzero Fourier mode in \(q\) is a genuine gluing phase excitation of the extended Hilbert space. Gluing removes that mode unless it is combined into the smooth zero mode of a dressed observable.

The nearly lightlike clock choice yields the Hamiltonian constraint
\begin{align}
&\hat{\mathcal C}_{\Sigma_T}
=
\hat P_T+\hat K.
\label{eq:sec6_quantum_clock_constraint}
\end{align}
In the \(T\) representation,
\begin{align}
&\ii\frac{\partial}{\partial T}\Psi_T
=
\hat K\Psi_T.
\label{eq:sec6_quantum_relational_schrodinger}
\end{align}

We now add the probe Hilbert spaces. Classically the probes of Section 5 did not provide additional gravitational phase space coordinates. Quantum mechanically they carry finite internal state spaces:
\begin{align}
&\mathcal H_{\tilde b}={\rm span}\{|\tilde i\rangle\}_{i=1}^{d_b},
\\
&\mathcal H_b={\rm span}\{|i\rangle_b\}_{i=1}^{d_b},
\\
&\mathcal H_E=\mathcal H_{e_b}\otimes\mathcal H_{E'},
\\
&\mathcal H_{\tilde R}={\rm span}\{|\tilde r\rangle\}_{\tilde r=1}^{d_{\tilde R}}.
\label{eq:sec6_probe_hilbert_spaces}
\end{align}
Here \(E\) is the early radiation in the exterior bath. The label \(\tilde R\) denotes the remaining interior radiation attached to the end of the world brane in the same bookkeeping sense that \(E\) is attached to the exterior bath. The brane boost mode remains absent because the brane endpoint condition has already been fixed classically.

For one interior leg, define
\begin{align}
&\mathcal K_{\rm int}
:=
\mathcal H_{\tilde R}\otimes\mathcal H_{\tilde b},
\\
&\mathcal K_{\rm ext}
:=
\mathcal H_b\otimes\mathcal H_E.
\label{eq:sec6_probe_factors}
\end{align}
The one leg extended Hilbert space is
\begin{align}
&\widehat{\mathcal H}_{\rm kin}^{(1),\varepsilon}
=
\mathcal H_{\rm clk}
\otimes
\widehat{\mathcal H}_{\mathscr H}^{(1),\varepsilon}
\otimes
\mathcal K_{\rm int}
\otimes
\mathcal K_{\rm ext}.
\label{eq:sec6_oneleg_extended_probe_space}
\end{align}
For the topology changing calculation, only the interior factor is replicated. Thus
\begin{align}
&\widehat{\mathcal H}_{\mathscr H}^{(N),\varepsilon}
:=
L^2(\mathbb R,ds_R)
\otimes
\Sym^N\left[
L^2(\mathbb R,ds_L)
\right],
\label{eq:sec6_Nleg_horizon_space}
\end{align}
and
\begin{align}
&\widehat{\mathcal H}_{\rm kin}^{(N),\varepsilon}
=
\mathcal H_{\rm clk}
\otimes
L^2(\mathbb R,ds_R)
\otimes
\mathcal K_{\rm ext}
\otimes
\Sym^N\left[
L^2(\mathbb R,ds_L)
\otimes
\mathcal K_{\rm int}
\right].
\label{eq:sec6_Nleg_extended_probe_space}
\end{align}
The full extended Hilbert space is
\begin{align}
&\widehat{\mathcal H}_{\rm kin}^{\varepsilon}
=
\bigoplus_{N\ge0}
\widehat{\mathcal H}_{\rm kin}^{(N),\varepsilon}.
\label{eq:sec6_total_kin_space}
\end{align}
The exterior corner frame, the clock, the late mode \(b\), and the early radiation \(E\) are common to all sectors. The symmetrized factor contains the interior slots, each with its own interior boost frame and its own brane attached \(\tilde R\) label.

\subsection{Observer dressed extended Hilbert space}

Fix a clock reading $T$ and let $p(T)$ denote the event on the observer worldline selected by the clock. A relational prescription associates to it an achronal slice $\Sigma_T$ ending on the end of the world brane. The details of the prescription will not matter below, provided the same prescription is used for the state, for operator insertions, and for relational Hamiltonian evolution.

With the probe sectors adjoined, the fixed time physical one leg Hilbert space relevant for the AMPS experiment is \(\mathcal H_{\rm phys}^{(1)}(T):=\mathcal H_{\rm phys}(T)\otimes \mathcal K_{\rm int}\otimes \mathcal K_{\rm ext}\).
The corresponding one leg extended Hilbert space and \(N\) leg factors are the spaces defined in \crefrange{eq:sec6_oneleg_extended_probe_space}{eq:sec6_Nleg_extended_probe_space}.
Equivalently, if we separate the common exterior factor from the replicated interior factor, then
\begin{align}
&\widehat{\mathcal H}_{\rm kin}^{(1),\varepsilon}
=
\mathcal H_{\rm clk}
\otimes
L^2(\mathbb R,ds_R)
\otimes
\mathcal K_{\rm ext}
\otimes
\left[
L^2(\mathbb R,ds_L)
\otimes
\mathcal K_{\rm int}
\right],
\label{eq:sec6_oneleg_extended_probe_space_rewritten}
\\
&\widehat{\mathcal H}_{\rm kin}^{(N),\varepsilon}
=
\mathcal H_{\rm clk}
\otimes
L^2(\mathbb R,ds_R)
\otimes
\mathcal K_{\rm ext}
\otimes
\Sym^N\left[
L^2(\mathbb R,ds_L)
\otimes
\mathcal K_{\rm int}
\right].
\label{eq:sec6_Nleg_extended_probe_space_rewritten}
\end{align}
The asymptotic exterior remains a single common factor, while the bracketed interior factor is the part replicated when topology change shifts the number of legs.

For fixed $T$, the  edge mode factor inside the one leg extended Hilbert space is spanned by the clock labeled basis \( |T;s_L,s_R\rangle^{(1),\varepsilon}:=|T\rangle_{\rm clk}\otimes |s_L,s_R\rangle^{(1),\varepsilon}\).
It resolves the identity on the  edge mode factor:
\begin{align}
&\mathbf 1_{\widehat{\mathcal H}^{(1),\varepsilon}_{\mathscr H}}
=
\int ds_L\,ds_R\,
|s_L,s_R\rangle^{(1),\varepsilon}
{}^{(1),\varepsilon}\!\langle s_L,s_R|.
\label{eq:sec6_extended_identity}
\end{align}

The cutting map \(\mathcal C^{\varepsilon}_{\mathscr H}\) was already defined in \eqref{eq:cut_map} on the gravitational corner sector. With the probe sectors adjoined, we use the same symbol for the extension
\begin{align}
&\mathcal C^{\varepsilon}_{\mathscr H}\otimes
\mathbf 1_{\mathcal K_{\rm int}\otimes\mathcal K_{\rm ext}}:
\mathcal H_{\rm phys}^{(1)}(T)
\to
\widehat{\mathcal H}_{\rm kin}^{(1),\varepsilon}.
\label{eq:sec6_cut_map_with_probes}
\end{align}
For a factorized vector in \(\mathcal H_{\rm phys}^{(1)}(T)\), this extension acts as
\begin{align}
&
\bigl(
\mathcal C^{\varepsilon}_{\mathscr H}\otimes
\mathbf 1_{\mathcal K_{\rm int}\otimes\mathcal K_{\rm ext}}
\bigr)
\Bigl(
|\Psi(T)\rangle\otimes
|\chi\rangle\otimes
|\eta\rangle
\Bigr)
=
|\Psi(T)\rangle^{(1),\varepsilon}
\otimes
|\chi\rangle\otimes
|\eta\rangle,
\\
&|\Psi(T)\rangle^{(1),\varepsilon}
:=
\mathcal C^{\varepsilon}_{\mathscr H}|\Psi(T)\rangle
=
\int ds_L\,ds_R\,
\Psi^{(1),\varepsilon}(T;s_L,s_R)\,
|T;s_L,s_R\rangle^{(1),\varepsilon}.
\label{eq:sec6_cut_state}
\end{align}
The same convention will be used for the gluing map below.

The initial Hawking state used in the AMPS experiment is the one leg physical state whose hard horizon pair is in the ordinary maximally entangled vacuum. We write
\begin{align}
&|\Phi\rangle_{\tilde b b}
:=
\frac{1}{\sqrt{d_b}}
\sum_{i=1}^{d_b}
|\tilde i\rangle_{\tilde b}\otimes |i\rangle_b,
\label{eq:sec6_hawking_hard_pair}
\\
&|\chi_0\rangle_{\tilde R E}
:=
\sum_{\tilde r,a}
c_{\tilde r a}\,
|\tilde r\rangle_{\tilde R}\otimes |a\rangle_E,
\;
\sum_{\tilde r,a}|c_{\tilde r a}|^2=1,
\label{eq:sec6_hawking_spectator_state}
\end{align}
and
\begin{align}
&|\Psi_{\rm H}(T)\rangle
:=
\int ds\,
\Psi_{\rm H}(T;s)\,
|T;s\rangle
\otimes
|\Phi\rangle_{\tilde b b}
\otimes
|\chi_0\rangle_{\tilde R E}.
\label{eq:sec6_hawking_state_physical}
\end{align}
Its cut image is
\begin{align}
&\mathcal C^{\varepsilon}_{\mathscr H}
|\Psi_{\rm H}(T)\rangle
=
\int ds\,
\Psi_{\rm H}(T;s)\,
|\Omega^{\varepsilon}_{s}\rangle_{LR}
\otimes
|\Phi\rangle_{\tilde b b}
\otimes
|\chi_0\rangle_{\tilde R E}.
\label{eq:sec6_hawking_state_cut_image}
\end{align}
This state is smooth before the purity measurement is performed: the edge modes lie in the smooth image of $J = 0$, and the hard modes are in the ordinary \(\tilde b,b\) vacuum state.

Topology change is computed in the \(N\) leg factors of the extended Hilbert space. We denote the slot embeddings by
\begin{align}
&\iota_r^{(N),\varepsilon}:
\widehat{\mathcal H}^{(1),\varepsilon}_{\mathscr H}
\hookrightarrow
\widehat{\mathcal H}^{(N),\varepsilon}_{\mathscr H},
\\
&r=1,\ldots,N.
\label{eq:sec6_slot_embedding_def}
\end{align}
These embeddings insert the one leg interior factor into the \(r\)th interior factor of the \(N\) leg sector. The asymptotic exterior remains a single common factor and is suppressed in the notation at this stage.

\subsection{Dressed horizon vacuum projector}
\label{sec:gravdressproj}

We now use the probe matter sectors of \cref{sec:canquanthilb} to define the dressed interior operator in the extended algebra. The full slice operator is first specified relationally on \(\Sigma_T\). The cutting map rewrites that operator in the left corner frame of the extended phase space, and the gluing map returns its zero mode to the smooth one boundary Hilbert space.

Let \(\mathfrak B_{\tilde b}^0:={\rm End}(\mathcal H_{\tilde b})\) be the undressed hard algebra of the interior mode, and choose a boost eigenbasis \(|\tilde i\rangle\) for the restriction \(\hat K_{\tilde b}\) of the horizon boost generator to this hard mode, \(\hat K_{\tilde b}|\tilde i\rangle = k_i|\tilde i\rangle\). Define the operator \(B^{0}_{ij}:=|\tilde i\rangle\langle\tilde j|\), with boost charges \(k_{ij}:=k_i-k_j\).
The extended algebra in the black hole interior is the crossed product
\begin{align}
&\widehat{\mathfrak B}_{\tilde b}
=
\mathfrak B_{\tilde b}^0\rtimes_{\alpha^-}\mathbb R,
\label{eq:sec6_quantum_interior_crossed_product}
\end{align}
generated by \(\mathfrak B_{\tilde b}^0\) and the interior corner boost unitaries \(U_L(s)=\ee^{\ii s\hat{\mathscr A}_L}\), subject to
\begin{align}
&U_L(s)\,\hat O\,U_L(s)^{-1}
=
\alpha_s^-(\hat O).
\label{eq:sec6_quantum_crossed_product_covariance}
\end{align}
The exterior algebra is defined similarly with \(U_R(s)=\ee^{-\ii s\hat{\mathscr A}_R}\). In the canonical crossed product representation a state is a square integrable family of hard states over the edge mode configuration. The hard operator acts fiberwise with the dressing twisted by the edge label, while the edge unitary shifts that label. Thus the notation \(\widehat{\mathcal H}_{\rm hard}\otimes \widehat{\mathcal H}_{\rm edge}\) is only a choice of trivialization of the direct integral. The algebraic statement that survives that choice is the crossed product relation \eqref{eq:sec6_quantum_crossed_product_covariance}.

This is where the firewall appears. A smooth \(\tilde b,b\) vacuum pair is described by the hard mode vacuum projector together with the smooth edge state \(|\Omega^\varepsilon_s\rangle_{LR}\). A one sided boost discontinuity across the horizon is a nonzero Fourier mode in the variable \(q\). In the extended Hilbert space this phase may be written in the edge mode sector as \(\ee^{-\ii k_{ij}\hat q}\).
Because this operator carries nonzero \(J\) momentum whenever \(k_{ij}\neq0\), it moves the state out of the smooth image selected by \(\Pi_{\mathscr H}^{(1),\varepsilon}\). Equivalently, it breaks the left right maximally entangled edge mode state by a gluing phase. The crossed product relation represents the same phase on the interior hard mode operator by a boost. A firewall is therefore the common crossed product label of the hard pair phase and the edge gluing phase.

The extended operator corresponding to the same full slice insertion is determined by how it acts on smooth edge states. Since \(|\Omega^{\varepsilon}_{s}\rangle_{LR}\) already carries the correct corner frame, the dressed one leg operator \(\widetilde B^{(1),\varepsilon}_{ij}(T)\) is defined by
\begin{align}
&\widetilde B^{(1),\varepsilon}_{ij}(T)\,
\mathcal C^{\varepsilon}_{\mathscr H}
=
\mathcal C^{\varepsilon}_{\mathscr H}\,
\Bigl(\ee^{-\ii k_{ij}\hat s}B^{0}_{ij}\Bigr).
\label{eq:sec6_dressed_operator_cut_definition}
\end{align}
Equivalently, on a cut image state,
\begin{align}
&\widetilde B^{(1),\varepsilon}_{ij}(T)
\Bigl(
|\Omega^{\varepsilon}_{s}\rangle_{LR}\otimes |\tilde j\rangle
\Bigr)
=
\ee^{-\ii k_{ij}s}\,
|\Omega^{\varepsilon}_{s}\rangle_{LR}\otimes |\tilde i\rangle.
\label{eq:sec6_dressed_operator_action}
\end{align}
Equation \eqref{eq:sec6_dressed_operator_action} shows that the dressed operator preserves the two sided edge mode state and contributes the phase \(\ee^{-\ii k_{ij}s}\).

One may temporarily write the same phase on the left edge variable instead. In that representation,
\begin{align}
&\ee^{-\ii k_{ij}\hat s}
=
\ee^{\ii k_{ij}\hat s_L}\,\ee^{-\ii k_{ij}\hat q}.
\label{eq:sec6_same_phase_two_representations}
\end{align}
This identity is the crossed product relation in the present finite dimensional hard mode sector. The first factor acts on the left corner frame, while the second factor carries \(J\) charge:
\begin{align}
\Big[
\hat J,
\ee^{-\ii k_{ij}\hat q}
\Big]
=
-k_{ij}\,
\ee^{-\ii k_{ij}\hat q}.
\label{eq:sec6_gluing_phase_charge}
\end{align}
If \(\hat J|\nu\rangle_{q}=\nu|\nu\rangle_{q}\), then
\begin{align}
&\hat J
\ee^{-\ii k_{ij}\hat q}
|\nu\rangle_{q}
=
(\nu-k_{ij})
\ee^{-\ii k_{ij}\hat q}
|\nu\rangle_{q}.
\label{eq:sec6_gluing_phase_shift}
\end{align}
Thus a gluing phase shifts the one sided boost charge sector. The same operation may be represented on the hard mode by conjugation with the boost generator \(\hat K_{\tilde b}\):
\begin{align}
&\ee^{-\ii\varphi \hat K_{\tilde b}}
B^0_{ij}
\ee^{\ii\varphi \hat K_{\tilde b}}
=
\ee^{-\ii k_{ij}\varphi}
B^0_{ij}.
\label{eq:sec6_hard_boost_phase}
\end{align}
The crossed product identifies the hard phase \(\ee^{-\ii k_{ij}\varphi}\) with translation in the edge coordinate \(q\). The family of phase twisted hard vacuum states is
\begin{align}
&|\Phi_\varphi\rangle_{\tilde b b}
:=
\frac{1}{\sqrt{d_b}}
\sum_{i=1}^{d_b}
\ee^{-\ii k_i\varphi}
|\tilde i\rangle
\otimes
|i\rangle_b,
\label{eq:sec6_hard_firewall_state}
\end{align}
and the corresponding family of phase twisted edge states is
\begin{align}
&|\Omega_{s;\varphi}^{\varepsilon}\rangle_{LR}
:=
\ee^{\ii\varphi\hat J}
|\Omega_{s}^{\varepsilon}\rangle_{LR}.
\label{eq:sec6_edge_firewall_state}
\end{align}
Since \(\hat J\) generates translations of \(q\),
\begin{align}
&\ee^{\ii\varphi\hat J}
\hat q
\ee^{-\ii\varphi\hat J}
=
\hat q
+
\varphi.
\label{eq:sec6_q_translation}
\end{align}
The parameter \(\varphi\) therefore measures the same phase in the hard vacuum pair and in the edge mode pair. At \(\varphi=0\), the hard state is the usual maximally entangled \(\tilde b,b\) vacuum pair and the edge state lies in the smooth gluing sector. For \(\varphi\neq0\), the hard pair acquires a boost phase and the edge state is translated in the \(q\) direction.

The gluing projector extracts the zero mode under this gluing phase flow. Define the \(\nu\) mode of an operator by
\begin{align}
&\hat O_\nu
:=
\int\frac{d\beta}{2\pi}\,
\ee^{-\ii\nu\beta}
\ee^{\ii\beta\hat J}
\hat O
\ee^{-\ii\beta\hat J}.
\label{eq:sec6_operator_mode_projector_explicit}
\end{align}
Then
\begin{align}
&\Big[
\hat J,
\hat O_\nu
\Big]
=
\nu\hat O_\nu,
\\
&\hat O
=
\int d\nu\,\hat O_\nu,
\label{eq:sec6_operator_mode_properties}
\end{align}
and
\begin{align}
&\mathcal G_{\mathscr H}^{\varepsilon}
\hat O
\mathcal C_{\mathscr H}^{\varepsilon}
=
\mathcal G_{\mathscr H}^{\varepsilon}
\hat O_0
\mathcal C_{\mathscr H}^{\varepsilon}.
\label{eq:sec6_gluing_zero_mode_explicit}
\end{align}
The ordinary dressed vacuum operator contains the compensating left edge factor \(\ee^{\ii k_{ij}\hat s_L}\), so its net action is the smooth zero mode \(\ee^{-\ii k_{ij}\hat s}B^0_{ij}\). A gluing phase factor by itself is the firewall and is removed by gluing.

The exterior hard operators are \(E_{ij}:=|i\rangle_b\langle j|\in {\rm End}({\cal H}_b)\), and the horizon vacuum operator on the hard sector is
\begin{align}
&\Pi^0_{\tilde b b}
=
\frac{1}{d_b}\sum_{i,j=1}^{d_b} B^0_{ij}\otimes E_{ij}.
\label{eq:sec6_bare_maxent_projector}
\end{align}
Its operator on the extended Hilbert space is obtained by replacing the interior operator by the dressed one:
\begin{align}
&\Pi_{\tilde b b}^{(1),\varepsilon}(T)
=
\frac{1}{d_b}\sum_{i,j=1}^{d_b}
\widetilde B_{ij}^{(1),\varepsilon}(T)\otimes E_{ij}.
\label{eq:sec6_pi_equals_g_explicit}
\end{align}
Using \eqref{eq:sec6_dressed_operator_cut_definition},
\begin{align}
&\Pi_{\tilde b b}^{(1),\varepsilon}(T)\,
\mathcal C^{\varepsilon}_{\mathscr H}
=
\mathcal C^{\varepsilon}_{\mathscr H}\,
\hat O_0(T),
\label{eq:sec6_pi_zero_mode_on_cut_image}
\end{align}
where
\begin{align}
&\hat O_0(T)
:=
\frac{1}{d_b}\sum_{i,j=1}^{d_b}
\ee^{-\ii k_{ij}\hat s}\,
B^{0}_{ij}\otimes E_{ij}.
\label{eq:sec6_zero_mode_late_operator}
\end{align}
By construction, \(\hat O_0(T)\) depends only on the two sided boost \(s\) and contains no \(q\) dependence at all. It is exactly the zero mode of the horizon vacuum projector on the smooth image of the cutting map:
\begin{align}
&\mathcal G^{\varepsilon}_{\mathscr H}\,
\Pi_{\tilde b b}^{(1),\varepsilon}(T)\,
\mathcal C^{\varepsilon}_{\mathscr H}
=
\hat O_0(T).
\label{eq:sec6_glued_late_operator_equals_zero_mode}
\end{align}
This equation is the clean operator statement needed below: on the relevant smooth states, the horizon vacuum projector and gluing have the same action because both reduce to the same zero mode.

Applying \eqref{eq:sec6_glued_late_operator_equals_zero_mode} to the Hawking state \eqref{eq:sec6_hawking_state_cut_image} gives
\begin{align}
&\mathcal G^{\varepsilon}_{\mathscr H}
\Pi_{\tilde b b}^{(1),\varepsilon}(T)
\mathcal C^{\varepsilon}_{\mathscr H}
|\Psi_{\rm H}(T)\rangle
=
|\Psi_{\rm H}(T)\rangle.
\label{eq:sec6_hawking_state_vacuum_projection}
\end{align}
Thus if the purity measurement is not performed, the horizon vacuum measurement succeeds on the initial Hawking state.

\subsection{Interior leg Wick contractions}

When the full interior Hilbert space is restored, denote a complete one leg label in the extended description by \(\hat x:=(s_C,\tilde r,\tilde i)\).
The raw overlap between states with several interior slots contains disconnected pairings of those slots between bra and ket. The connected amplitudes are obtained by subtracting those disconnected pieces, exactly as in ordinary cluster decomposition.

Let \(\mathcal R_N(T',\hat x_1',\ldots,\hat x_N'\,;\,T,\hat x_1,\ldots,\hat x_N)\) denote the raw Lorentzian amplitude before subtraction. For any finite index set \(I\), write \(\Pi(I)\) for the set of all set partitions of \(I\). The connected propagators are defined by
\begin{align}
&\mathcal R_I
=
\sum_{\pi\in\Pi(I)}\prod_{B\in\pi}\mathcal A_B,
\label{eq:sec4_cluster_formula}
\end{align}
and equivalently by M\"obius inversion,
\begin{align}
&\mathcal A_I
=
\mathcal R_I
-
\sum_{\substack{\pi\in\Pi(I)\\ \pi\neq\{I\}}}
\prod_{B\in\pi}\mathcal A_B.
\label{eq:sec4_mobius_inversion}
\end{align}
For one explicit interior slot there is nothing to subtract, so \(G(\hat x',\hat x):=\mathcal A_{\{1\}}(T',\hat x'\,;T,\hat x)\).
For two interior slots,
\begin{align}
&\mathcal A_{\{1,2\}}(T';1',2'\,;T;1,2)
=
\mathcal R_{\{1,2\}}(T';1',2'\,;T;1,2)
-
G(1',1)G(2',2)
-
G(1',2)G(2',1).
\label{eq:sec4_two_leg_subtraction}
\end{align}
To leading order in the topological expansion, the extended inner product has the bosonic Fock space form
\begin{align}
&\langle T';\hat x_1',\ldots,\hat x_{N'}'|
T;\hat x_1,\ldots,\hat x_N\rangle
=
\langle T'|T\rangle\,\delta_{N,N'}
\sum_{\sigma\in S_N}
\prod_{r=1}^N G(\hat x_r',\hat x_{\sigma(r)})
+
\mathcal O(\ee^{-2S_0}).
\label{eq:sec4_fock_inner_product}
\end{align}
The two pairings in the two slot case are the direct and exchange channels used later. They are exactly the multi leg version of the two contractions already isolated in the toy model of \cref{subsec:AMPS_warmup_contractions}; gravity will resolve between them by gluing.

\subsection{Topology change from Hamiltonian dynamics}\label{sec:cubicpants}

With the extended Hilbert space in hand, we can now write down topology changing dynamics in third quantized canonical gravity. The extended Hilbert space converts the interior portion of the observer dressed slice into a dynamical open leg ending at the horizon corner. We emphasize that we are not deriving the topology changing vertex from a complete UV theory of Lorentzian gravity. We are merely writing down the leading order effective Hamiltonian on the sector wise sum over black hole interiors, with the same topological weight that appears in the JT gravity topological expansion.

A one leg label is \(\hat x=(s_C,x)\), where \(s_C\) is the corner boost frame and \(x=(\tilde r,\tilde i)\) denotes the EoW internal label together with the interior hard mode label. Let \(\mathcal H_1\) be the corresponding one leg Hilbert space with inner product kernel \(G(\hat x',\hat x)\). The higher leg sectors are fixed by the connected inner products above. To leading order in the topological expansion they are the symmetric tensor powers
\begin{align}
&\mathcal H_N:=\Sym^N\mathcal H_1,
\\
&\mathcal H_{\oplus}:=
\bigoplus_{N=0}^{\infty}\mathcal H_N.
\label{eq:sec6_third_quantized_fock_space}
\end{align}
Let \(\mathcal D\subset \mathcal H_{\oplus}\) be the dense subspace of vectors with only finitely many nonzero \(N\) leg components. For \(f\in\mathcal H_1\), define on \(\mathcal D\) the black hole interior creation and annihilation operators 
\begin{align}
&\hat A^\dagger(f)\psi_N
=
\sqrt{N+1}\,\Sym(f\otimes\psi_N),
\\
&(\hat A(f)\psi_N)(\hat x_1,\ldots,\hat x_{N-1})
=
\sqrt N
\int \Dd\hat y'\,\Dd\hat y\,
\overline{f(\hat y')}\,
G(\hat y',\hat y)\,
\psi_N(\hat y,\hat x_1,\ldots,\hat x_{N-1}).
\label{eq:sec6_smeared_creation_annihilation}
\end{align}
They obey
\begin{align}
&[\hat A(f),\hat A^\dagger(g)]
=
\langle f,g\rangle_1,
\\
&[\hat A(f),\hat A(g)]
=
[\hat A^\dagger(f),\hat A^\dagger(g)]
=
0,
\label{eq:sec6_smeared_leg_algebra}
\end{align}
where \(\langle f,g\rangle_1\) is computed with the kernel \(G\). The distributional operators are defined by \(\hat A^\dagger(f)=\int \Dd\hat x\,f(\hat x)\hat A^\dagger(\hat x)\) and \(\hat A(f)=\int \Dd\hat x\,\overline{f(\hat x)}\hat A(\hat x)\). Thus the third quantized algebra is given by
\begin{align}
&\hat A^\dagger(\hat x)|0\rangle
=
|\hat x\rangle,
\\
&[\hat A(\hat x'),\hat A^\dagger(\hat x)]
=
G(\hat x',\hat x),
\\
&[\hat A(\hat x'),\hat A(\hat x)]
=
[\hat A^\dagger(\hat x'),\hat A^\dagger(\hat x)]
=
0.
\label{eq:sec6_leg_creation_algebra}
\end{align}
These commutation relations reproduce the multi leg inner product \eqref{eq:sec4_fock_inner_product}. For example,
\begin{align}
\langle 0|
\hat A(\hat x_2')\hat A(\hat x_1')
\hat A^\dagger(\hat x_1)\hat A^\dagger(\hat x_2)
|0\rangle
=
G(\hat x_1',\hat x_1)G(\hat x_2',\hat x_2)
+
G(\hat x_1',\hat x_2)G(\hat x_2',\hat x_1).
\label{eq:sec6_two_leg_wick_from_creators}
\end{align}
The creation and annihilation operator algebra used here is therefore kinematical. It is the canonical algebra associated with the direct sum over \(N\) black hole interiors, in the same way that ordinary second quantization is the canonical algebra associated with the symmetric powers of a one particle Hilbert space. The operator algebra is defined before specifying any dynamics, namely the number preserving boost Hamiltonian and the number changing pair of pants Hamiltonian. The dynamics comes next, as an operator on \(\mathcal H_{\oplus}\).

The number preserving part of the Hamiltonian is the second quantization of the one leg boost generator:
\begin{align}
&\widehat K
=
\int \Dd\hat x'\,\Dd\hat x\,
K^{(1)}(\hat x',\hat x)
\hat A^\dagger(\hat x')\hat A(\hat x).
\label{eq:sec6_second_quantized_K}
\end{align}
The leading interaction changing the number of interiors by one has a universal
edge mode part and an effective ``microscopic'' kernel. A single interior leg ending at
the horizon carries the corner phase space
\begin{align}
\Gamma_C
&=
T^*SO(1,1)
\simeq
T^*\mathbb R,
\\
\Theta_C
&=
K\,\delta s,
\\
\Omega_C
&=
\delta K\wedge \delta s .
\label{eq:sec6_one_leg_corner_phase_space}
\end{align}
For a \(1\to2\) topology changing vertex, the two daughter legs are outgoing while the parent leg is incoming.
Thus the relevant corner phase space is
\begin{align}
\Gamma_C^{(1)}
\times
\Gamma_C^{(2)}
\times
\overline{\Gamma_C^{(y)}} ,
\end{align}
with product symplectic potential
\begin{align}
\Theta_{\rm vertex}
=
K_1\,\delta s_1
+
K_2\,\delta s_2
-
K_y\,\delta s_y .
\label{eq:sec6_vertex_corner_potential}
\end{align}
Since \(SO(1,1)\simeq \mathbb R\),
the multiplication map is
\begin{align}
m:SO(1,1)\times SO(1,1)&\longrightarrow SO(1,1),
\
g_y=g_1g_2,
\
s_y=s_1+s_2.
\label{eq:sec6_boost_group_multiplication}
\end{align}
The cubic pair of pants vertex is therefore
\begin{align}
\hat V_{\rm pants}
=
\frac12
\int \Dd \hat x_1\,\Dd \hat x_2\,\Dd \hat y\,
\Big[
C(\hat x_1,\hat x_2\mid \hat y)\,
\widehat A^\dagger(\hat x_1)
\widehat A^\dagger(\hat x_2)
\widehat A(\hat y)
+\mathrm{h.c.}
\Big].
\label{eq:sec5_full_Vpants}
\end{align}
The effective kernel is symmetric under interchange of the two daughter slots,
\begin{align}
C(\hat x_1,\hat x_2\mid \hat y)
=
C(\hat x_2,\hat x_1\mid \hat y),
\end{align}
and the boost edge mode part is
\begin{align}
C(\hat x_1,\hat x_2\mid \hat y)
=
\delta(s_{C,y}-s_{C,1}-s_{C,2})\,
C(x_1,x_2\mid y).
\label{eq:sec5_full_hatC}
\end{align}
Equation \eqref{eq:sec5_full_hatC} should be read as the group composition
law for the \(SO(1,1)\) corner frame variables, which can be thought of intuitively as a local sewing of the boost frames at the vertex. 

The gravitational Hamiltonian is therefore
\begin{align}
&\hat H_{\rm grav}
=
\widehat K+\lambda \hat V_{\rm pants}.
\label{eq:sec5_full_Hgrav}
\end{align}
We take \(\lambda\) to be real; equivalently, any phase of the coupling has been absorbed into the definition of the cubic kernel. In the usual JT topological expansion, \(\lambda\sim \ee^{-S_0}\). Thus the leading connected contribution beyond the disconnected piece is a single pair of pants transition. Any further distinction at this order is a distinction between contraction patterns on the same connected topology. Since the interaction term is perturbative, the solution is the ordinary Dyson series expansion on \(\mathcal H_{\oplus}\). Let \(U_N(T_2,T_1)\) denote the evolution generated by the number preserving Hamiltonian $\hat{K}$ on the \(N\) leg sector. The piece of \(\hat V_{\rm pants}\) that maps one leg to two legs acts by
\begin{align}
&(\hat V_{1\to2}\psi_1)(\hat x_1,\hat x_2)
=
\int \Dd\hat y\,
C(\hat x_1,\hat x_2\mid \hat y)\,
\psi_1(\hat y),
\label{eq:sec6_pants_map_one_to_two}
\end{align}
with the adjoint map taking two legs back to one leg. Therefore a state initially in the one leg sector evolves into, at first order in \(\lambda\),
\begin{align}
&\psi_2(T)
=
-\ii\lambda
\int_{T_0}^{T}\Dd T_1\,
U_2(T,T_1)\hat V_{1\to2}U_1(T_1,T_0)\psi_1(T_0)
+
\mathcal O(\lambda^2).
\label{eq:sec6_first_order_pants_solution}
\end{align}
Higher orders are iterated pair of pants transitions. Thus the creation and annihilation operators are simply a natural choice of basis for the direct sum Hilbert space over numbers of black hole interiors, while the perturbative Hamiltonian determines their transition amplitudes.

Lastly, the one leg dressed operator of \cref{sec:gravdressproj} lifts to the interior leg Fock space by second quantization:
\begin{align}
&\Pi_{\tilde b b}^{\varepsilon}(T)
:=
\int d\hat x'\,d\hat x\,
B_T^{(1),\varepsilon}(\hat x',\hat x)\,
\hat A^\dagger(\hat x')\hat A(\hat x),
\label{eq:sec6_Fock_lift_late_operator}
\end{align}
where
\begin{align}
&B_T^{(1),\varepsilon}(\hat x',\hat x)
:=
\bra{\hat x'}
\Pi_{\tilde b b}^{(1),\varepsilon}(T)
\ket{\hat x}.
\label{eq:sec6_one_leg_late_kernel}
\end{align}
The direct and exchange contractions of this one leg kernel into the connected two leg branch will be analyzed in the AMPS calculation itself, in the next section. The two contraction patterns are depicted schematically in \cref{fig:directvsexchange}.

\section{The AMPS experiment in canonical quantum gravity\label{sec:AMPS}}

We can now formulate the AMPS experiment as a sequence of two projective measurements in one relational Hilbert space. Between \(T_0\) and \(T_1\), the pair of pants Hamiltonian transfers support from the one leg sector into the connected two leg sector. On the connected branch present at \(T_1\), the branch state together with the distiller defines a branch conditioned map from the first daughter hard mode factor \(\mathcal H_{\tilde b}\) to the recovered subsystem \(e_b\) and an unobserved environment. The decoupling property of the Hayden Preskill protocol is the statement that the unobserved environment carries no appreciable information about \(b\). Equivalently, after a branch dependent basis rotation on \(e_b\), the purity operator may be represented as a dressed projector on the same hard mode index \(\tilde b\) that enters the horizon vacuum projector. The early radiation purity measurement then postselects a connected branch on which \(b\) and \(e_b\) are maximally entangled. 

Before this purity measurement the Hawking state is smooth and the horizon vacuum projector succeeds. After the purity measurement the doubled connected kernel admits two contractions. The direct contraction is the semiclassical firewall channel because it carries nonzero gluing phase \(q\). The exchange contraction is the zero mode channel. The horizon vacuum measurement is pulled back to the same relational slice and evaluated on this postselected kernel. Gluing keeps the exchange zero mode and removes the direct firewall channel. The AMPS experiment is therefore reduced to a simple question: when the one leg dressed insertion is embedded into the connected two leg kernel, which contraction survives the horizon gluing map? The exchange contraction survives and reproduces the purity operator pulled back through the distiller. The direct contraction has no smooth zero mode and is hence annihilated by gluing.

Let \(T_0<T_1<T_2\). The early radiation purity measurement is the instantaneous purity operator after distillation, \(\Xi_{bE}^{(n)}(T_1):=D^\dagger(\Pi_{b e_b}^{(n)}\otimes \mathbf 1_{E'})D\),
where \(D\) is the distillation map at \(T_1\) and \(E'\) denotes the unobserved part of the early radiation bath. The horizon vacuum measurement is
\begin{align}
&\Pi_{\tilde b b}^{(m)}(T_2)
:=
\lim_{\varepsilon\to0}
\mathcal G^{\varepsilon}_{\mathscr H}\,
\Pi_{\tilde b b}^{(m),\varepsilon}(T_2)\,
\mathcal C^{\varepsilon}_{\mathscr H}.
\label{eq:sec8_horizon_vacuum_projector_def}
\end{align}
The exact sequential probability is
\begin{align}
&p(m,n;T_2,T_1)
:=
\bra{\eta(\Psi_{\rm H})}
M_n(T_1)^\dagger
U(T_2,T_1)^\dagger
\Pi_{\tilde b b}^{(m)}(T_2)
U(T_2,T_1)
M_n(T_1)
\ket{\eta(\Psi_{\rm H})}_{\rm phys},
\label{eq:sec8_joint_probability}
\end{align}
with \(M_n(T_1):=\Xi_{bE}^{(n)}(T_1)U(T_1,T_0)\).

The rest of this section calculates the correlator above in the order in which the experiment is performed. First we show that the topology changing Hamiltonian prepares the connected two leg branch with probability approaching one by the Page time. We then fix this connected branch and rewrite the early radiation purity measurement as a dressed projector on the first daughter index for the $\tilde b$ hard mode. Finally we pull the horizon vacuum projector back to the same dressed slice as that of the purity measurement, decompose its doubled connected kernel into direct and exchange contractions, and apply the horizon gluing map. The direct contraction carries a nonzero one sided boost discontinuity across the horizon and is therefore annihilated by the gluing map. The exchange contraction on the other hand is shown to be the smooth boost zero mode, and survives after gluing. We end by showing that the purity measurement and horizon vacuum measurement are actually equivalent as Dirac observables in the infalling observer algebra. 

\subsection{Connected branch dominance from time dependent perturbation theory\label{sec:dist}}

Between \(T_0\) and \(T_1\) there is nontrivial relational bulk evolution. The first question is whether, by the time the purity measurement is applied, support has been transferred from the one leg sector to the connected two leg sector. If \(P^{(2)}\) projects onto the connected two leg sector, then
\begin{align}
&P_{\rm conn}(T_1)
:=
\bra{\eta(\Psi_{\rm H})}
U(T_1,T_0)^\dagger P^{(2)}U(T_1,T_0)
\ket{\eta(\Psi_{\rm H})}_{\rm phys}.
\label{eq:sec8_Pconn_exact}
\end{align}
The Lorentzian representation of \eqref{eq:sec8_Pconn_exact} is a Schwinger--Keldysh contour.

After performing the construction in \cref{sec:RAQobserver}, the deparametrized constraint becomes
\begin{align}
&\hat C
=
\hat P_T+\hat K+\lambda\hat V_{\rm pants}
\approx 0.
\label{eq:sec8_extended_constraint}
\end{align}
Hence the clock conditioned extended state obeys
\begin{align}
&\ii\,\partial_T|\psi(T)\rangle
=
\bigl(\hat K+\lambda\hat V_{\rm pants}(T)\bigr)
|\psi(T)\rangle.
\label{eq:sec8_relational_schrodinger}
\end{align}
The branch growth problem is therefore ordinary time dependent perturbation theory in relational time. The zeroth order propagator is \(U_0(T_2,T_1)=\exp[-\ii(T_2-T_1)\hat K]\).
On a given timefold \(I_j=[T_{j-1},T_j]\), define the interaction picture vertex
\begin{align}
&\hat V_{{\rm pants},j}^{(I)}(t)
:=
U_0(t,T_{j-1})^\dagger\,\hat V_{\rm pants}(t)\,U_0(t,T_{j-1}).
\label{eq:sec8_Vpants_I}
\end{align}

If \(|\beta\rangle^{(1)}\) is a one leg state, the connected two leg final
state contains two independent kinds of labels. We denote the $\hat{K}$ spectrum by \(\alpha\), and an orthonormal basis of the
unobserved environment Hilbert space as
\(\{|a\rangle_{\tilde E}\}_{a=1}^k\):
\begin{align}
\mathcal H^{(2)}_{\rm conn,full}
\simeq
\mathcal H^{(2)}_{\rm conn}\otimes \mathcal H_{\tilde E},
\
|\alpha,a\rangle^{(2)}
:=
|\alpha\rangle^{(2)}_{\rm conn}\otimes |a\rangle_{\tilde E},
\
k:=\dim \mathcal H_{\tilde E}.
\label{eq:sec8_alpha_a_basis}
\end{align}
Here \(\mathcal H_{\tilde E}\) is the Hilbert space over which the
calculation is coarse grained. It contains the early radiation \(E\) and the
interior radiation bath labels \(\tilde R\) carried by the daughter interior
legs. The coarse graining is the trace over this Hilbert space.

Let \(p_\beta\) denote the weights of the initial one leg density matrix in the
\(\beta\) basis, with \(\sum_\beta p_\beta=1\). The leading contribution to the transition amplitude on the \(j\)th fold is
\begin{align}
&{}^{(2)}\!\langle \alpha,a|U_j|\beta;\chi_0\rangle^{(1)}
=
-\ii\lambda\int_{I_j}dt\,\mathcal A_{j;\alpha a,\beta}(t)
+
\mathcal O(\lambda^2),
\\
&\mathcal A_{j;\alpha a,\beta}(t)
:=
{}^{(2)}\!\langle \alpha,a|
\hat V_{{\rm pants},j}^{(I)}(t)
|\beta;\chi_0\rangle^{(1)}.
\label{eq:sec8_single_fold_amplitude}
\end{align}
Now, we denote the operator induced on the unobserved environment Hilbert space as
\begin{align}
\mathcal V_{j;\alpha\beta}(t)
:=
\langle \alpha|
\hat V_{{\rm pants},j}^{(I)}(t)
|\beta\rangle^{(1)}.
\end{align}
We then trace out the environment:
\begin{align}
\operatorname{Tr}_{\mathcal H_{\tilde E}}\!\left[
\mathcal V_{j;\alpha\beta}(t)\,
\rho_{\tilde E,0}\,
\mathcal V_{j';\alpha\beta}(t')^\dagger
\right]=\sum_{a=1}^k
\mathcal A_{j;\alpha a,\beta}(t)
\mathcal A^*_{j';\alpha a,\beta}(t')
,
\
\operatorname{Tr}\rho_{\tilde E,0}=1.
\label{eq:sec8_Etilde_trace}
\end{align}
Inserting this into \eqref{eq:sec8_Pconn_exact} gives
\begin{align}
&P_{\rm conn}(T_1)
=
|\lambda|^2
\sum_{\beta}p_\beta
\sum_{\alpha}
\sum_{a=1}^{k}
\sum_{j,j'=1}^{N_{\rm tf}}
\int_{I_j}dt\int_{I_{j'}}dt'\,
\langle t'|t\rangle\,
\mathcal A_{j;\alpha a,\beta}(t)
\mathcal A^*_{j';\alpha a,\beta}(t')
+
\mathcal O(|\lambda|^4).
\label{eq:sec8_Pconn_SK}
\end{align}

Group averaging supplies the clock propagator \(\langle t|t'\rangle = \mathcal O(\ee^{-S_0})\) for \(|t-t'|=\mathcal O(1)\) and \(t\neq t'\).
As a result, coincident timefold terms are order one, while interference between distinct timefolds separated by order one relational time is exponentially suppressed, that is, \(P^{\rm fold}_{\rm off}(T_1)\ll P^{\rm fold}_{\rm diag}(T_1)\).
For a single timefold, only the strip
\(|t-t'|\lesssim \Delta T_{\rm clock}\) near \(t=t'\) contributes to the inclusive probability:
\begin{align}
\delta P_j
:=
|\lambda|^2
\sum_{\beta}p_\beta
\sum_{\alpha}
\sum_{a=1}^{k}
\int_{I_j}dt\int_{I_j}dt'\,
\langle t'|t\rangle\,
\mathcal A_{j;\alpha a,\beta}(t)
\mathcal A^*_{j;\alpha a,\beta}(t') =
|\lambda|^2\,\sigma_j\,\delta t_j\,k\rho_{\rm eff}.
\label{eq:sec8_samefold_linear}
\end{align}
Here \(\langle t'|t\rangle\) is the clock propagator used to organize the timefold contour and $\sigma_j$ is an $\mathcal O(1)$ coefficient. By contrast, the horizon vacuum measurement uses the finite resolution infalling observer clock propagator \(\langle T'|T\rangle\) of \cref{sec:relational_time_vs_foliation}, for which \(d_{\rm clock}=\mathcal O(1)\) and order one separations remain unsuppressed.

Choose a basis of connected daughter states that are eigenstates of \(\hat K^{(2)}\) and the one sided boost generator \(\hat J^{(2)}\):
\begin{align}
&\hat K^{(2)}|K,J\rangle^{(2)}
=K\,|K,J\rangle^{(2)},
\\
&\hat J^{(2)}|K,J\rangle^{(2)}
=J\,|K,J\rangle^{(2)}.
\label{eq:sec8_PT_eigenbasis}
\end{align}
In our simplified setup there are no further continuous labels, so the density of states is
\begin{align}
\rho_{\rm eff}(K)
:=
\int dJ\,
{}^{(2)}\!\langle K,J|
\delta(K-\hat K^{(2)})|K,J\rangle^{(2)}.
\label{eq:sec8_rhoeff_def}
\end{align}
Since \((s,K)\) is a canonical chart on the same physical one boundary phase space as \((u_0, H_{\rm JT})\), the spectral measure inherits the usual scaling of the JT gravity density of states, up to the Jacobian of the canonical transformation \cite{GaoJafferisKolchmeyer2022, MaldacenaStanfordYang2016NearlyAdS2}:
\begin{align}
&\rho_{\rm eff}(K)=\ee^{S_0}\widehat\rho_{\rm eff}(K),
\;
\widehat\rho_{\rm eff}(K)=\mathcal O(1).
\label{eq:sec8_rhoeff_scaling}
\end{align}
In the old black hole JT gravity setting considered here, there is no actual dynamical semiclassical
evaporation during the interval \([T_0,T_1]\). We start with a static old black hole from the very start.
Thus the dimension $k=\dim\mathcal H_{\tilde E}$ of the unobserved environment Hilbert space traced out in the calculation is fixed throughout the time interval and scales as
\(k\sim \ee^{S_0}\).\footnote{The phrase ``Page time'' therefore refers solely to time evolution of the gravitational field. Matter states are not changing with \(T\) in our setup. The role of
the relational Hamiltonian evolution between \(T_0\) and \(T_1\) is instead to give the
nonperturbatively small pair of pants amplitude time to build up into a dominant
inclusive branch probability after summing over timefolds, radiation channels,
and connected daughter states.} 

The coincident timefold contribution of the \(j\)th timefold is therefore \(\delta P_j\approx \sigma_j|\lambda|^2k\rho_{\rm eff}\,\delta t_j\). At leading order in $\lambda$ the connected branch is dominated by the $1 \rightarrow 2$ transition,\footnote{The probability being computed is a $1 \rightarrow 1$ correlator with a two leg projector $P_2$ inserted at the turnaround in the Schwinger Keldysh contour. A \(2\to1\) vertex can therefore only contribute on the forward branch after a prior
\(1\to2\) transition, and, in order to survive the final \(P_2\) projection,
must be followed by another \(1\to2\) transition. Such terms are therefore
\(\mathcal{O}(\lambda^3)\) in the forward amplitude and only renormalize the leading
\(1\to2\) kernel at \(\mathcal{O}(\lambda^2)\).} hence its weight satisfies \(P_{j+1}=P_j+(1-P_j)\,\delta P_j\).
In the continuum limit,
\begin{align}
&\frac{dP_{\rm conn}}{dT}
=
\Lambda(T)\bigl(1-P_{\rm conn}(T)\bigr),
\
\Lambda(T)\sim |\lambda|^2\sigma(T)k\rho_{\rm eff},
\label{eq:sec8_continuum_rate}
\end{align}
whose solution is
\begin{align}
&P_{\rm conn}(T_1)=1-\exp\left[-\int_{T_0}^{T_1}dT \ \Lambda(T)\right]
\label{eq:sec8_Pconn_solution}
\end{align}
Using \(\lambda\sim \ee^{-S_0}\) and $k \sim e^{S_0}$ together with \eqref{eq:sec8_rhoeff_scaling}, each timefold contributes at $\mathcal{O}(1)$ to the exponent. Hence, since $T_1 - T_0 \sim S_0$ at the Page time, we have
\begin{align}
&1-P_{\rm conn}(T_1)\sim \ee^{-cS_0},
\;
P_{\rm conn}(T_1)\to 1,
\label{eq:sec8_Pconn_to_one}
\end{align}
for some positive order one constant \(c\).

Thus, preparation of the connected branch is the result of gravitational dynamics with topology change. It follows from the pair of pants interaction, the density of connected daughter states, the large number of radiation channels, and the suppression of long range timefold interference by the clock propagator.

\subsection{Purity measurement and the connected postselected branch\label{sec:postselection}}

We now analyze the early radiation purity measurement on the connected branch itself. Factor the early radiation Hilbert space as
\begin{align}
&\mathcal H_E\cong \mathcal H_{e_b}\otimes \mathcal H_{E'},
\;
\dim\mathcal H_{e_b}
=
\dim\mathcal H_{\tilde b}
=
d_b,
\label{eq:sec8_R_factorization}
\end{align}
and choose a basis \(|j,r'\rangle_E\) such that \(D|j,r'\rangle_E=|j\rangle_{e_b}\otimes |r'\rangle_{E'}\).
The index \(j\) labels the recovered subsystem \(e_b\), while \(r'\) labels the remaining early radiation.

Fix a connected two leg branch and work temporarily on the ordered cover \(\mathcal H_1\otimes\mathcal H_1\) of the symmetric two leg sector. We call the slot whose $\tilde b$ index is recovered by the distillation channel ``daughter 1''.\footnote{Strictly speaking, the physical two leg sector is
\(\mathrm{Sym}^2\mathcal H_1\), so the labels \(1,2\) used here are not physical
daughter labels. They are auxiliary labels on the ordered cover used to write kernels before
inserting the symmetrization operator. Equivalently, the recovery channel supplies a relational tag selecting
one of the two identical slots; summing over the two choices restores the symmetric description.
Because the pair of pants kernel is symmetric in the daughter slots and the horizon vacuum projection operator is
second quantized, exchanging the ordered cover labels merely exchanges the two Wick contractions
in \crefrange{eq:sec8_Kdir_from_topology}{eq:sec8_Kex_from_topology}. The matrix elements are therefore independent of this auxiliary
choice.} At fixed corner boost \(s\), the unobserved environment for the distillation channel is
\begin{align}
&\mathcal H_{\tilde E}
:=
\mathcal H_{E'}
\otimes
\mathcal H_{\tilde R,1}
\otimes
\mathcal H_{\tilde R,2}
\otimes
\mathcal H_{\tilde b,2},
\;
\rho_{\tilde E,0}
:=
|\chi_0\rangle\langle\chi_0|_{\tilde E}.
\label{eq:sec8_aux_state_def}
\end{align}
These are the same daughter labels introduced above, now with daughter indices. Let \(D_{\tilde E}\) denote the effective number of unobserved environment states in the coarse grained band used below. Recall that $D_{\tilde E} \sim e^{S_0}$.

Let \(U_{{\rm conn},s}(T_1,T_0)\) denote relational evolution restricted to the connected branch and let \(D\) denote the instantaneous distiller. Once the connected branch has been prepared, its matter state at \(T_1\) may be organized with the first daughter hard mode index left open. At fixed corner boost \(s\), we write
\begin{align}
&V_{s}(T_1)
:=
D
U_{{\rm conn},s}(T_1,T_0),
\label{eq:sec8_connected_isometry}
\\
&\mathcal N_{s}(X)
:=
\Tr_{\rm env}\!\left[
V_{s}(T_1)
\bigl(X\otimes \rho_{\tilde E,0}\bigr)
V_{s}(T_1)^\dagger
\right].
\label{eq:sec8_reduced_channel_def}
\end{align}
Here \(V_{s}(T_1):\mathcal H_{\tilde b}\otimes\mathcal H_{\tilde E}\to\mathcal H_{e_b}\otimes\mathcal H_{\rm env}\).
The corresponding connected branch state after distillation is
\begin{align}
&|\Psi_{s}(T_1)\rangle
:=
(\mathbf 1_b\otimes V_s(T_1))
\bigl(
|\Phi\rangle_{b\tilde b}\otimes |\chi_0\rangle_{\tilde E}
\bigr).
\label{eq:sec8_connected_branch_state_after_distillation}
\end{align}
Equation \eqref{eq:sec8_connected_branch_state_after_distillation} packages the already prepared connected branch together with the distiller at the time of the purity measurement. The only topology changing input is the growth of \(P_{\rm conn}(T_1)\) computed in the previous subsection.
Choosing an orthonormal basis \(|h\rangle_{\rm env}\),
\begin{align}
&V_{s}(T_1)
\bigl(|\tilde i\rangle_{\tilde b}\otimes |\chi_0\rangle_{\tilde E}\bigr)
=
\sum_{j,h}
A^{(s)}_{j h\mid \tilde i}(T_1)
|j\rangle_{e_b}\otimes |h\rangle_{\rm env},
\label{eq:sec8_connected_channel_amplitudes}
\end{align}
so
\begin{align}
&T^{(s)}_{j\tilde i;m\tilde n}(T_1)
:=
\sum_h
A^{(s)}_{j h\mid \tilde i}(T_1)
\bigl(A^{(s)}_{m h\mid \tilde n}(T_1)\bigr)^*,
\label{eq:sec8_reduced_channel_kernel}
\end{align}
with the trace preserving condition
\begin{align}
&\sum_{j=1}^{d_b}
T^{(s)}_{j\tilde i;j\tilde n}(T_1)
=
\delta_{\tilde i\tilde n}.
\label{eq:sec8_reduced_channel_trace_preserving}
\end{align}
The corresponding reduced \(b e_b\) state is
\begin{align}
&\rho^{(s)}_{b e_b}(T_1)
:=
(\mathbf 1_b\otimes \mathcal N_s)
\bigl(|\Phi\rangle\langle\Phi|_{b\tilde b}\bigr).
\label{eq:sec8_rho_be_def}
\end{align}
In the basis above,
\begin{align}
&\rho^{(s)}_{b e_b}(T_1)
=
\frac{1}{d_b}
\sum_{i,n=1}^{d_b}\sum_{j,m=1}^{d_b}
T^{(s)}_{j\tilde i;m\tilde n}(T_1)
|i\rangle_b\langle n|\otimes |j\rangle_{e_b}\langle m|.
\label{eq:sec8_rho_be_kernel}
\end{align}

The open matter indices admit the same two contraction patterns as in \cref{subsec:AMPS_warmup_contractions}. The exchange contraction identifies \(j\) with \(\tilde i\) and \(m\) with \(\tilde n\), while the direct contraction identifies \(j\) with \(m\) and \(\tilde i\) with \(\tilde n\). Define
\begin{align}
&(\mathbb S_{\times})_{j\tilde i;m\tilde n}
:=\delta_{j\tilde i}\delta_{m\tilde n},
\\
&(\mathbb S_{\parallel})_{j\tilde i;m\tilde n}
:=\frac{1}{d_b}\delta_{jm}\delta_{\tilde i\tilde n},
\label{eq:sec8_reduced_two_tensors}
\end{align}
together with the Hilbert--Schmidt inner product \(\langle A,B\rangle_{\rm HS}:=\sum_{j,\tilde i,m,\tilde n}A^*_{j\tilde i;m\tilde n}B_{j\tilde i;m\tilde n}\).
These are exactly the two index patterns isolated in \cref{subsec:AMPS_warmup_contractions}. A purely direct kernel gives \(\mathfrak M=d_b^{-2}\), while a purely exchange kernel gives \(\mathfrak M=1\).

After coarse graining over the unobserved environment band, write
\begin{align}
&T^{(s)}_{j\tilde i;m\tilde n}
=
\overline T^{(s)}_{j\tilde i;m\tilde n}
+
N^{(s)}_{j\tilde i;m\tilde n},
\;
\langle N^{(s)}\rangle_{\tilde E}=0,
\label{eq:sec8_random_kernel_extended}
\end{align}
with \(\bigl\langle \|N^{(s)}\|_{\rm HS}^2\bigr\rangle_{\tilde E}=\mathcal O(D_{\tilde E}^{-1})\).
The purity probability on the connected branch is
\begin{align}
&\mathfrak M(s;T_1)
:=
\langle \Phi|\rho^{(s)}_{b e_b}(T_1)|\Phi\rangle
=
\frac{1}{d_b^2}\langle T^{(s)},\mathbb S_{\times}\rangle_{\rm HS}.
\label{eq:sec8_maxent_functional}
\end{align}

The decoding assumption is most cleanly stated on the complementary channel. Define
\begin{align}
&\rho^{s}_{b\,{\rm env}}(T_1)
:=
\Tr_{e_b}\!\left[
(\mathbf 1_b\otimes V_{s})
\bigl(|\Phi\rangle\langle\Phi|_{b\tilde b}\otimes
\rho_{\tilde E,0}\bigr)
(\mathbf 1_b\otimes V_{s}^\dagger)
\right].
\label{eq:sec8_reduced_env_state}
\end{align}
and \(\rho^{s}_{\rm env}(T_1):=\Tr_b\,\rho^{s}_{b\,{\rm env}}(T_1)\).
We assume that the unobserved environment carries no appreciable information about the mode \(b\):
\begin{align}
&\Big\|
\rho^{s}_{b\,{\rm env}}(T_1)
-
\frac{\mathbf 1_b}{d_b}\otimes \rho^{s}_{\rm env}(T_1)
\Big\|_1
\le
\epsilon_{\rm dec}.
\label{eq:sec8_reduced_decoupling_statement}
\end{align}
This is Hayden Preskill recast in the present notation \cite{HaydenPreskill2007,SekinoSusskind2008,DupuisBertaWullschlegerRenner2014Decoupling,Szehr2013,BrownFawzi2015,RobertsYoshida2017,YoshidaKitaev2017}. The small hard mode factor is \(\mathcal H_{\tilde b}\). The unobserved environment is \(E'\) together with the connected daughter data not read by the horizon vacuum measurement. Since \(b\) began in the Hawking pair \(|\Phi\rangle_{b\tilde b}\), \eqref{eq:sec8_reduced_decoupling_statement} implies that, after the unobserved matter data are traced over, the purifier \(e_b\) carries the same small hard mode index that was originally labeled by \(\tilde b\), up to an error $\epsilon_{\rm rec} := \mathcal O(\epsilon_{\rm dec}^{1/2})$. Thus there exists a unitary \(U_{s}:\mathcal H_{\tilde b}\to\mathcal H_{e_b}\)
such that
\begin{align}
&\left\|
\rho^{(s)}_{b e_b}(T_1)
-
(\mathbf 1_b\otimes U_s)
|\Phi\rangle\langle\Phi|_{b\tilde b}
(\mathbf 1_b\otimes U_s^\dagger)
\right\|_1
\le
\epsilon_{\rm rec}.
\label{eq:sec8_recovered_maxent_state}
\end{align}
After rotating the recovered basis by \(U_s^\dagger\), the reduced kernel takes the exchange form
\begin{align}
&\overline T^{(s)}_{j\tilde i;m\tilde n}(T_1)
=
\delta_{j\tilde i}\delta_{m\tilde n}
+
\mathcal E^{(s)}_{j\tilde i;m\tilde n},
\;
\|\mathcal E^{(s)}\|_{\rm HS}
\le
d_b\,\epsilon_{\rm rec}.
\label{eq:sec8_exchange_kernel_after_basis}
\end{align}

In this recovered basis, the purity operator pulls back to the following projection operator:
\begin{align}
&(\mathbf 1_b\otimes U_s^\dagger)
\Pi^{(1)}_{b e_b}
(\mathbf 1_b\otimes U_s)
=
\frac{1}{d_b}
\sum_{i,j=1}^{d_b}
|i\rangle_b\langle j|
\otimes
|\tilde i\rangle\langle\tilde j|.
\label{eq:sec8_purity_projector_pullback}
\end{align}
Hence
\begin{align}
&p_{\rm pur}(T_1,s)
=
1
-
\mathcal O(\epsilon_{\rm rec})
-
\mathcal O(D_{\tilde E}^{-1/2}).
\label{eq:sec8_purity_conn_final}
\end{align}
Combining this with \eqref{eq:sec8_Pconn_to_one}, the unconditional success probability of the purity measurement is
\begin{align}
&p_{\rm pur}(T_1)
=
P_{\rm conn}(T_1)
\bigl[
1
-
\mathcal O(\epsilon_{\rm rec})
-
\mathcal O(D_{\tilde E}^{-1/2})
\bigr]
+
\bigl(1-P_{\rm conn}(T_1)\bigr)d_b^{-2},
\label{eq:sec8_ppur_final}
\end{align}
hence
\begin{align}
&p_{\rm pur}(T_1)
=
1
-
\mathcal O(\ee^{-cS_0})
-
\mathcal O(\epsilon_{\rm rec})
-
\mathcal O(D_{\tilde E}^{-1/2}).
\label{eq:sec8_ppur_to_one}
\end{align}

Using the recovered basis fixed above, the operator \(|\tilde i\rangle\langle\tilde j|\) in \eqref{eq:sec8_purity_projector_pullback} acts on the first daughter hard mode label:
\begin{align}
&\bigl(|\tilde i\rangle\langle\tilde j|\bigr)_{\rm conn}
|s;\tilde r_1,k;\tilde r_2,\tilde i_2\rangle
:=
\delta_{jk}|s;\tilde r_1,i;\tilde r_2,\tilde i_2\rangle .
\label{eq:sec8_interior_operator_conn_action}
\end{align}
Written in the same corner frame as the interior mode probed by the horizon vacuum measurement, the purity operator pulled back through the distiller on the connected branch is
\begin{align}
&\Pi_{b e_b}(T_1)
:=
\frac{1}{d_b}\sum_{i,j=1}^{d_b}|i\rangle_b\langle j|
\otimes
\Bigl(\ee^{-\ii k_{ij}\hat s}|\tilde i\rangle\langle\tilde j|\Bigr)_{\rm conn}.
\label{eq:sec8_purity_operator_branch}
\end{align}
The factor \(\ee^{-\ii k_{ij}\hat s}\) is the boost phase of the same dressed mode, now written directly on the connected branch. This is the operator to which the horizon vacuum measurement will be compared.

\subsection{Horizon vacuum measurement\label{sec:amps-operator-meaning}}

The previous subsection has rewritten the successful early radiation purity measurement as a dressed operator on the first daughter hard mode label. We now compute the horizon vacuum projector on the same connected branch and compare the two operators before gluing. By refined algebraic quantization, the horizon vacuum insertion can be pulled back to the same relational slice as the purity measurement:
\begin{align}
&U(T_2,T_1)^\dagger\Pi_{\tilde b b}(T_2)U(T_2,T_1)
\sim_{\rm phys}
\Pi_{\tilde b b}(T_1).
\label{eq:sec8_pullback}
\end{align}
The remaining question is therefore which contraction channel survives the gluing map on the connected two leg branch.

In the observer clock representation, the pulled back operator carries the usual pair of clock propagators associated with a group averaged relational insertion:
\begin{align}
&{}^{(2)}\!\langle T';1',2'|
\Pi_{\tilde b b}(T_2)|T'';1,2\rangle^{(2)}
=
\langle T' | T_2\rangle \langle T_2|T''\rangle\,\mathcal K_B(1',2';1,2).
\label{eq:sec8_late_kernel_with_clock}
\end{align}
Here the relevant kernel is the infalling observer clock propagator of \cref{sec:relational_time_vs_foliation}. Since the proper time from \(T_1\) to \(T_2\) is order one and \(d_{\rm clock}=\mathcal O(1)\), we have \(\langle T'|T_2\rangle=\mathcal O(1)\) and \(\langle T_2|T''\rangle=\mathcal O(1)\) for order one separations. The horizon vacuum measurement is thus not suppressed by the finite clock.\footnote{If the infalling observer's clock had resolution of order $\exp(-S_0)$ then for order one separated clock readings it yields an exponentially small overlap, namely $\langle T|T'\rangle \sim \exp(-S_0)$. The pulled back horizon vacuum projection operator carries two such clock propagators. Using that kernel for the infalling observer would therefore exponentially suppress the horizon vacuum success probability, even on the branch with the correct zero mode. Therefore, the fact that the infalling observer actually has a clock with $\mathcal{O}(1)$ resolution in $S_0$ counting is actually critical in resolving the AMPS paradox.}

Let
\begin{align}
&B^{(1),\varepsilon}_{T;ij}(\hat x',\hat x)
:=
{}^{(1),\varepsilon}\!\langle \hat x'|
\widetilde B^{(1),\varepsilon}_{ij}(T)
|\hat x\rangle^{(1),\varepsilon}.
\label{eq:sec8_Bkernel_decomposition}
\end{align}
Substituting this one leg matrix element into the two leg kernel gives
\begin{align}
&\mathcal K_B(1',2';1,2)
=
\frac{1}{d_b}\sum_{i,j=1}^{d_b}
\left[
\mathcal K^{\parallel,\varepsilon}_{ij}(1',2';1,2)
+
\mathcal K^{\times,\varepsilon}_{ij}(1',2';1,2)
\right]|i\rangle_b\langle j|,
\label{eq:sec8_KB_matrix_element_extended}
\end{align}
with direct terms
\begin{align}
&\mathcal K^{\parallel,\varepsilon}_{ij}(1',2';1,2)
:=
B^{(1),\varepsilon}_{T;ij}(1',1)G(2',2)
+
B^{(1),\varepsilon}_{T;ij}(2',2)G(1',1),
\label{eq:sec8_Kdir_from_topology}
\end{align}
and exchange terms
\begin{align}
&\mathcal K^{\times,\varepsilon}_{ij}(1',2';1,2)
:=
B^{(1),\varepsilon}_{T;ij}(2',1)G(1',2)
+
B^{(1),\varepsilon}_{T;ij}(1',2)G(2',1).
\label{eq:sec8_Kex_from_topology}
\end{align}
The direct contraction keeps the source insertion on the same daughter leg in bra and ket. The exchange contraction crosses the daughter labels. Exactly as in \cref{subsec:AMPS_warmup_contractions}, these are the gravitational versions of the direct and exchange contractions.

The horizon vacuum projector itself is defined from the smooth full slice operator. Therefore \eqref{eq:sec6_pi_zero_mode_on_cut_image} applies to each cut image state entering the doubled connected kernel:
\begin{align}
&\Pi_{\tilde b b}^{(1),\varepsilon}(T_1)\,
\mathcal C^{\varepsilon}_{\mathscr H}|\chi\rangle
=
\mathcal C^{\varepsilon}_{\mathscr H}\,
\hat O_0(T_1)\,|\chi\rangle
\label{eq:sec8_late_operator_zero_mode_on_connected_state}
\end{align}
for every state \(|\chi\rangle\) on the connected postselected branch. Thus the horizon vacuum projector itself is \(q\) independent on each smooth cut image. Any nonzero \(q\) dependence in the two leg calculation comes from the contraction of bra and ket daughter legs after the purity measurement.

The exchange channel pairs the leg carrying the dressed insertion with the leg selected by the purity measurement. The hard boost phase and the edge mode dressing therefore remain on the same contraction cycle, and the result depends only on the two sided boost \(s\):
\begin{align}
&\mathcal K^{\times,\varepsilon}_{ij}
=
\ee^{-\ii k_{ij}s}\,
\widetilde{\mathcal K}^{\times,\varepsilon}_{ij},
\;
\partial_q\widetilde{\mathcal K}^{\times,\varepsilon}_{ij}=0,
\label{eq:sec8_exchange_twosided_phase}
\\
&\bigl(\mathcal K^{\times,\varepsilon}_{ij}\bigr)_0
=
\mathcal K^{\times,\varepsilon}_{ij}.
\label{eq:sec8_exchange_zero_mode}
\end{align}
The direct channel pairs the leg carrying the dressed insertion with the same daughter leg in bra and ket. This is the semiclassical contraction after the purity measurement. The exterior frame and the exterior \(b,E\) factors are the same on the two sides of the matrix element. The residual phase therefore comes from a one sided boost of the interior daughter leg with the exterior frame held fixed. In the edge mode variables, the crossed product relation leaves
\begin{align}
&\mathcal K^{\parallel,\varepsilon}_{ij}
=
\ee^{-\ii k_{ij}q}\,
\widetilde{\mathcal K}^{\parallel,\varepsilon}_{ij},
\;
\partial_q\widetilde{\mathcal K}^{\parallel,\varepsilon}_{ij}=0.
\label{eq:sec8_direct_firewall_phase}
\end{align}
For \(i\neq j\), this is a nonzero one sided boost discontinuity, so the direct contraction is a firewall excitation of the extended Hilbert space. For \(i=j\), the phase is neutral, but the direct contraction is then the identity cycle already removed by the subtraction \eqref{eq:sec4_two_leg_subtraction}.\footnote{Here ``identity cycle'' refers to the cycle decomposition of the two slot index contraction. The direct \(i=j\) term has a \((1)(2)\) pattern: the dressed one leg insertion acts on one daughter slot, while the other slot is carried by an ordinary one leg propagator. It is therefore one of the disconnected products removed in \eqref{eq:sec4_two_leg_subtraction}. The exchange contraction is instead a \((12)\) cycle, which ties the two daughter slots into a single connected contraction.} Therefore
\begin{align}
&\bigl(\mathcal K^{\parallel,\varepsilon}_{ij}\bigr)_0=0.
\label{eq:sec8_direct_zero_mode_vanishes}
\end{align}
The exchange contraction instead obeys \eqref{eq:sec8_exchange_twosided_phase} and \eqref{eq:sec8_exchange_zero_mode}, so it is the smooth zero mode of the one sided boost. Gluing therefore removes the direct firewall channel because its connected part has no zero mode.

On the connected branch, the basis natural for comparison with the early measurement is \( |a;s,\tilde r_1,k;\tilde r_2,\tilde i_2\rangle:=|a\rangle_b\otimes |s;\tilde r_1,k;\tilde r_2,\tilde i_2\rangle\),
with inner product
\begin{align}
&\langle c;s',\tilde r_1',\ell;\tilde r_2',\tilde i_2'|a;s,\tilde r_1,k;\tilde r_2,\tilde i_2\rangle
=
\delta_{ca}\delta_{\tilde r_1'\tilde r_1}\delta_{\ell k}\delta_{\tilde r_2'\tilde r_2}\delta_{\tilde i_2'\tilde i_2}\delta(s'-s).
\label{eq:sec8_connected_matter_basis_inner_product}
\end{align}
Using \eqref{eq:sec8_interior_operator_conn_action}, the early radiation purity operator has matrix elements
\begin{align}
&\langle c;s',\tilde r_1',\ell;\tilde r_2',\tilde i_2'|
\Pi_{b e_b}(T_1)
|a;s,\tilde r_1,k;\tilde r_2,\tilde i_2\rangle
=
\frac{1}{d_b}
\ee^{-\ii k_{\ell k}s}
\delta_{ak}\delta_{c\ell}\delta_{\tilde r_1'\tilde r_1}\delta_{\tilde r_2'\tilde r_2}\delta_{\tilde i_2'\tilde i_2}\delta(s'-s).
\label{eq:sec8_purity_operator_matrix_element}
\end{align}

Write \(\Pi^{\times,\varepsilon}_{\tilde b b}\) and \(\Pi^{\parallel,\varepsilon}_{\tilde b b}\) for the two contributions obtained from \(\mathcal K^{\times,\varepsilon}\) and \(\mathcal K^{\parallel,\varepsilon}\). Since the exchange contraction is already equal to its zero mode, gluing keeps it:
\begin{align}
&\mathcal G^{\varepsilon}_{\mathscr H}
\Pi^{\times,\varepsilon}_{\tilde b b}(T_1)
\mathcal C^{\varepsilon}_{\mathscr H}
=
\bigl(\Pi^{\times,\varepsilon}_{\tilde b b}(T_1)\bigr)_0.
\label{eq:sec8_exchange_glues_to_zero_mode}
\end{align}
Its hard index structure is exactly the exchange contraction pattern of \cref{subsec:AMPS_warmup_contractions}. Using \eqref{eq:sec8_purity_projector_pullback}, its matrix elements reproduce \eqref{eq:sec8_purity_operator_matrix_element} up to $\mathcal O(\ee^{-cS_0}) + \mathcal O(\epsilon_{\rm rec}) + \mathcal O(D_{\tilde E}^{-1/2})$ corrections. On the other hand, the direct channel has no zero mode, so gluing annihilates it:
\begin{align}
&\mathcal G^{\varepsilon}_{\mathscr H}
\Pi^{\parallel,\varepsilon}_{\tilde b b}(T_1)
\mathcal C^{\varepsilon}_{\mathscr H}
=
0,
\label{eq:sec8_direct_removed_regulated_operator}
\end{align}
and hence
\begin{align}
&\langle c;s',\tilde r_1',\ell;\tilde r_2',\tilde i_2'|
\mathcal G^{\varepsilon}_{\mathscr H}
\Pi^{\parallel,\varepsilon}_{\tilde b b}(T_1)
\mathcal C^{\varepsilon}_{\mathscr H}
|a;s,\tilde r_1,k;\tilde r_2,\tilde i_2\rangle
=
0.
\label{eq:sec8_direct_removed_regulated}
\end{align}

Hence for arbitrary normalized smooth matter wavepackets \(|\psi\rangle\) and \(|\chi\rangle\) on the connected subspace,
\begin{align}
&\lim_{\varepsilon\to0}
\langle \psi|
\mathcal G^{\varepsilon}_{\mathscr H}
\Pi_{\tilde b b}^{\varepsilon}(T_1)
\mathcal C^{\varepsilon}_{\mathscr H}
|\chi\rangle
=
\langle \psi|\Pi_{b e_b}(T_1)|\chi\rangle
+
\mathcal O(\ee^{-cS_0})
+
\mathcal O(\epsilon_{\rm rec})
+
\mathcal O(D_{\tilde E}^{-1/2}).
\label{eq:sec8_matrix_element_operator_identity}
\end{align}
Hence, on the relevant connected states obtained after the purity measurement, the glued horizon vacuum projector and the purity operator pulled back through the distiller have the same matrix elements up to the displayed corrections. Before the purity measurement, the Hawking state is smooth by \eqref{eq:sec6_hawking_state_vacuum_projection}. After the purity measurement, the direct contraction is the firewall channel because it only carries nonzero gluing phase \(q\). Gluing keeps only the zero mode. The horizon vacuum projector already acts by that zero mode on each smooth cut image, because the dressing of \(\tilde b\) is tied to the corner frame. The exchange contraction lies in that same zero mode, while the direct firewall contraction does not.

Finally combine this with the relational pullback \eqref{eq:sec8_pullback}. The horizon vacuum projector obeys
\begin{align}
&U(T_2,T_1)^\dagger\Pi_{\tilde b b}(T_2)U(T_2,T_1)
\sim_{\rm phys}
\Pi_{b e_b}(T_1)
+
\mathcal O(\ee^{-cS_0})
+
\mathcal O(\epsilon_{\rm rec})
+
\mathcal O(D_{\tilde E}^{-1/2})
\label{eq:sec8_late_equals_early}
\end{align}
on the connected postselected branch, again understood through matrix elements on the relevant states.

This also yields an analogous statement about sequential measurement outcomes. Let \(Q_n(T_1)\) be the early measurement for outcome \(n\), and let \(L_m(T_2)\) be the horizon vacuum measurement for outcome \(m\), pulled back to \(T_1\). The same argument, applied in the rotated maximally entangled basis, gives
\begin{align}
&L_m(T_1)=Q_m(T_1)+E_m,
\\
&\|E_m\|
=
\mathcal O(\ee^{-cS_0})
+
\mathcal O(\epsilon_{\rm rec})
+
\mathcal O(D_{\tilde E}^{-1/2}),
\label{eq:sec8_late_bell_basis_error}
\end{align}
where the displayed norm is on the connected matter Hilbert space. Since \(Q_mQ_n=\delta_{mn}Q_n\),
\begin{align}
p(m|n)
=
\frac{\langle \Psi|Q_nL_mQ_n|\Psi\rangle}
{\langle \Psi|Q_n|\Psi\rangle}
=
\delta_{mn}
+
\mathcal O(\ee^{-cS_0})
+
\mathcal O(\epsilon_{\rm rec})
+
\mathcal O(D_{\tilde E}^{-1/2}),
\label{eq:sec8_conditional_late_given_early}
\end{align}
for any state with nonzero probability for the purity outcome. The horizon vacuum measurement evaluates the same Dirac observable as the purity measurement pulled back through the distiller after both are written in the extended Hilbert space and returned to the one boundary Hilbert space by gluing.

The mechanism is now explicit. Before the purity measurement, the Hawking state is smooth and the horizon vacuum projector succeeds. The mode \(\tilde b\) is identified first on the full slice \(\Sigma_T\). Cutting \(\Sigma_T\) across \(\mathscr{H}\) rewrites it as an observable gravitationally dressed to the corner boost frame. The same boost phase may be written on the left edge mode or on the interior hard mode \(\tilde b\). After the purity measurement, the direct contraction leaves this phase as the nonzero one sided boost discontinuity \(q\), while the exchange contraction keeps the hard phase and the edge mode dressing on the same contraction cycle. Gluing removes the direct firewall channel and keeps the exchange zero mode. The horizon vacuum measurement and the early radiation purity measurement therefore define the same Dirac observable on the connected branch, in the sense of equality of matrix elements on the relevant states.

\subsection{Emergent nonlocality from large diffeomorphisms\label{sec:emergent_nonlocality}}

The calculations thus far show that the horizon vacuum measurement and the early radiation purity measurement become the same Dirac observable on the connected postselected branch. In this final section we spell out the operator algebraic meaning of this statement. There are two distinct quotient structures involved. The first is the near horizon quotient described in \cref{sec:RAQobserver} that removes the one sided boost discontinuity \(q\) and kills the direct firewall contraction. The second is a global quotient that only emerges after Page time. It is the statement that the first daughter interior hard mode and the decoded early radiation subsystem are related by a large diffeomorphism on the connected branch. This second quotient is not the perturbative nonfactorization at the horizon. It is the dynamical nonlocality generated by the topology changing Hamiltonian.

For any \(T\in[T_0,T_1]\), let \(U_{{\rm conn},s}(T,T_0)\) denote the connected branch component of the relational time evolution at fixed smooth two sided boost parameter \(s\), and define
\begin{align}
&\widetilde V_s(T)
:=
DU_{{\rm conn},s}(T,T_0).
\label{eq:emergent_unnormalized_connected_map}
\end{align}
This is the same object as \eqref{eq:sec8_connected_isometry}, except that we have not yet set \(T=T_1\). The tilde reminds us that before conditioning on the connected branch its norm contains the branch probability. More explicitly, recall that
\begin{align}
&\hat U_{\rm grav}(T,T_0)
:=
{\cal T}\exp\left[
-i\int_{T_0}^{T}dt\,\hat H_{\rm grav}(t)
\right],
\\
&\hat H_{\rm grav}(t)
=
\hat K+\lambda \hat V_{\rm pants}(t).
\label{eq:emergent_Hgrav}
\end{align}
Let \(P_N\) denote the projector onto the \(N\) interior leg sector
\(H_N\subset H_\oplus\). Then
\begin{align}
&P_M\hat K P_N
=
\delta_{MN}\hat K_N,
\\
&P_2\hat V_{\rm pants}P_1
=
\hat V_{1\to 2},
\\
&P_1\hat V_{\rm pants}P_2
=
\hat V_{2\to 1}.
\label{eq:emergent_Hgrav_blocks}
\end{align}
The connected branch component at
fixed \(s\) is the restriction of $P_2\hat U_{\rm grav}(T,T_0)P_1$ to the associated \(s\) wavepacket. This restricted diffeomorphism is precisely what $U_{{\rm conn},s}(T,T_0)$ corresponds to. Thus all Page time dependence in the precursor below enters through the gravitational evolution
\(U_{{\rm conn},s}(T,T_0)\). Whenever the connected branch has nonzero weight, we define the normalized branch conditioned map
\begin{align}
&W_s(T)|\psi\rangle_{\tilde b_1}
:=
\frac{
\widetilde V_s(T)
\bigl(|\psi\rangle_{\tilde b_1}\otimes |\chi_0\rangle_{\tilde E}\bigr)
}{
\sqrt{
\bra{\psi,\chi_0}
\widetilde V_s(T)^\dagger \widetilde V_s(T)
\ket{\psi,\chi_0}
}
},
\label{eq:emergent_normalized_connected_map}
\end{align}
with the normalization understood in the coarse grained band.

Keeping \(O_{e_b}\) fixed as an operator on the decoded radiation subsystem, define its connected branch Heisenberg precursor by
\begin{align}
&\alpha_{s,T}(O_{e_b})
:=
W_s(T)^\dagger
\bigl(O_{e_b}\otimes \mathbf 1_{\rm env}\bigr)
W_s(T).
\label{eq:emergent_alpha_def}
\end{align}
At a generic time this is only a channel from the decoded radiation algebra into the connected branch input algebra. In particular, there is no reason for
\begin{align}
&\alpha_{s,T}(O_1O_2)
=
\alpha_{s,T}(O_1)\alpha_{s,T}(O_2)
\label{eq:emergent_alpha_not_hom}
\end{align}
to hold before the Page time. The statement that a nonlocal interior radiation algebra has emerged is precisely the statement that \(\alpha_{s,T}\) becomes an approximate map on the recovered hard mode algebra when the connected branch dominates and the complementary channel decouples.

Recall the definitions
\begin{align}
&B^0_{ij}:=|\tilde i\rangle_{\tilde b_1}\langle\tilde j|,
\
k_{ij}:=k_i-k_j,
\\
&B_{ij}(s)
:=
\ee^{-\ii k_{ij}\hat s}B^0_{ij}.
\label{eq:emergent_dressed_matrix_units}
\end{align}
The operators \(B_{ij}(s)\) are the matrix elements of the first daughter dressed hard mode algebra. After a Page time, decoupling of the complementary channel yields a branch dependent unitary \(U_s:\mathcal H_{\tilde b_1}\to\mathcal H_{e_b}\), as in \eqref{eq:sec8_recovered_maxent_state}. Define the recovered \(e_b\) matrix elements by
\begin{align}
&E_{ij}^{(s)}
:=
U_s B^0_{ij}U_s^\dagger
\in
{\rm End}(\mathcal H_{e_b}).
\label{eq:emergent_recovered_eb_matrix_units}
\end{align}
Then the pullback \eqref{eq:emergent_alpha_def} obeys
\begin{align}
&\alpha_{s,T_1}\bigl(E_{ij}^{(s)}\bigr)
=
B_{ij}(s)
+
\Delta_{ij}(s),
\label{eq:emergent_alpha_matrix_unit},
\end{align}
where $\|\Delta_{ij}(s)\|_{\rm code}=\mathcal O(\epsilon_{\rm rec})+\mathcal O(D_{\tilde E}^{-1/2})$. This is the operator algebra version of \eqref{eq:sec8_purity_operator_branch}.

One way to see the nonlocality emerging from time evolution under topology changing dynamics is to keep the \(e_b\) operator fixed and evolve the interior operator into the decoded radiation output. Let \(X_{\tilde b_1}\in{\rm End}(\mathcal H_{\tilde b_1})\). Then
\begin{align}
\left[
W_s(T)X_{\tilde b_1}W_s(T)^\dagger,
E_{ij}^{(s)}\otimes \mathbf 1_{\rm env}
\right]
=
W_s(T)
\left[
X_{\tilde b_1},
\alpha_{s,T}\bigl(E_{ij}^{(s)}\bigr)
\right]
W_s(T)^\dagger.
\label{eq:emergent_output_commutator_identity}
\end{align}
All of the time dependence is in \(W_s(T)\), hence in the topology changing relational time evolution. The operator \(E_{ij}^{(s)}\) is a fixed decoded radiation operator. At early times the connected branch either has negligible weight or the map \(\alpha_{s,T}\) is not an algebraic map. At the Page time, using \eqref{eq:emergent_alpha_matrix_unit},
\begin{align}
&\left[
X_{\tilde b_1},
\alpha_{s,T_1}\bigl(E_{ij}^{(s)}\bigr)
\right]
=
\ee^{-\ii k_{ij}\hat s}
\left[
X_{\tilde b_1},
B^0_{ij}
\right]
+
\mathcal O(\epsilon_{\rm rec})
+
\mathcal O(D_{\tilde E}^{-1/2}).
\label{eq:emergent_general_commutator}
\end{align}
For example, choosing \(X_{\tilde b_1}=B^0_{\ell m}\),
\begin{align}
&\left[
B^0_{\ell m},
\alpha_{s,T_1}\bigl(E_{ij}^{(s)}\bigr)
\right]
=
\ee^{-\ii k_{ij}\hat s}
\left(
\delta_{mi}B^0_{\ell j}
-
\delta_{j\ell}B^0_{im}
\right)
+
\mathcal O(\epsilon_{\rm rec})
+
\mathcal O(D_{\tilde E}^{-1/2}).
\label{eq:emergent_matrix_unit_commutator}
\end{align}
This is an order one commutator for generic indices. 

Intuitively, after a Page time, the decoded radiation operator has an interior representative on the very daughter leg that is split and rejoined by the pair of pants Hamiltonian. This is the precise sense in which the topology changing Hamiltonian builds a nonlocality between the interior and exterior. The size of this effect is controlled by the connected branch probability. The nonlocality is therefore nonperturbative in the following sense. A single pair of pants transition is suppressed by \(\lambda\sim \ee^{-S_0}\), but the inclusive sum over timefolds, radiation channels, and connected daughter states makes the connected branch dominate by Page time. Once this happens, the recovered \(e_b\) algebra and the first daughter hard mode algebra become two representatives of the same logical algebra.

We now package this statement as a crossed product. Let
\begin{align}
&\mathcal A_{\tilde b_1}(s)
:=
{\rm span}\{B_{ij}(s)\},
\\
&\mathcal A_{e_b}(s)
:=
{\rm span}\{E_{ij}^{(s)}\}.
\label{eq:emergent_two_algebras}
\end{align}
Define a large diffeomorphism in terms of the automorphism
\begin{align}
&\Theta_s:
\mathcal A_{\tilde b_1}(s)
\oplus
\mathcal A_{e_b}(s)
\to
\mathcal A_{\tilde b_1}(s)
\oplus
\mathcal A_{e_b}(s),
\label{eq:emergent_theta_domain}
\\
&\Theta_s\bigl(B_{ij}(s)\oplus 0\bigr)
=
0\oplus E_{ij}^{(s)},
\label{eq:emergent_theta_first}
\\
&\Theta_s\bigl(0\oplus E_{ij}^{(s)}\bigr)
=
B_{ij}(s)\oplus 0.
\label{eq:emergent_theta_second}
\end{align}
Since the dressed operators obey
\begin{align}
&B_{ij}(s)B_{mn}(s)
=
\delta_{jm}B_{in}(s),
\
E_{ij}^{(s)}E_{mn}^{(s)}
=
\delta_{jm}E_{in}^{(s)},
\label{eq:emergent_matrix_unit_product}
\end{align}
\(\Theta_s\) extends to a \(*\) automorphism and \(\Theta_s^2=1\) in the ideal recovered limit. At finite \(S_0\), this statement holds in the code subspace after a Page time up to the same \(\mathcal O(\epsilon_{\rm rec})+\mathcal O(D_{\tilde E}^{-1/2})\) errors. The corresponding Page time algebra is
\begin{align}
&\mathfrak A_{\rm Page}(s)
:=
\left(
\mathcal A_{\tilde b_1}(s)
\oplus
\mathcal A_{e_b}(s)
\right)
\rtimes_{\Theta_s}
\mathbb Z_2.
\label{eq:emergent_page_crossed_product}
\end{align}
It is generated by \(\mathcal A_{\tilde b_1}(s)\oplus\mathcal A_{e_b}(s)\) and an implementing unitary \(\mathcal W_s\), subject to $\mathcal W_s^2=1$ and $\mathcal W_s A\mathcal W_s^{-1}=\Theta_s(A)$. Equivalently, for \(a,b\in\mathbb Z_2\),
\begin{align}
&(A\mathcal W_s^a)(B\mathcal W_s^b)
=
A\,\Theta_s^a(B)\,\mathcal W_s^{a+b}.
\label{eq:emergent_page_crossed_product_multiplication}
\end{align}
This is the large diffeomorphism analogue of the crossed product relation \eqref{eq:sec6_quantum_interior_crossed_product}. The difference is that the automorphism here is not the small diffeomorphism generated by the horizon boost generator. It is instead generated dynamically by the gravitational Hamiltonian and made sharp by decoupling of the complementary channel. It exchanges the interior and radiation representatives of the same hard mode algebra.\footnote{This is closely related to the mechanism proposed by Maxfield in \cite{Maxfield:2022sio}.}

The physical meaning is that there is no canonical split
\begin{align}
&\mathcal H_{\rm code}
\neq
\mathcal H_{\tilde b_1}\otimes \mathcal H_{e_b}\otimes \mathcal H_{\rm rest}
\label{eq:emergent_no_split}
\end{align}
on the connected branch after a Page time. The labels \(\tilde b_1\) and \(e_b\) are two sections of the same quotient algebra. In terms of physical matrix elements,
\begin{align}
&B_{ij}(s)
\sim_{\rm Page}
E_{ij}^{(s)},
\label{eq:emergent_page_equivalence_matrix_units}
\end{align}
where \(\sim_{\rm Page}\) means equality on the Page time connected code subspace up to
\(\mathcal O(\epsilon_{\rm rec})+\mathcal O(D_{\tilde E}^{-1/2})\) corrections. Thus
\begin{align}
&\Pi_{b e_b}(T_1)
=
\frac{1}{d_b}
\sum_{i,j=1}^{d_b}
|i\rangle_b\langle j|
\otimes
E_{ij}^{(s)}
\sim_{\rm Page}
\frac{1}{d_b}
\sum_{i,j=1}^{d_b}
|i\rangle_b\langle j|
\otimes
B_{ij}(s).
\label{eq:emergent_purity_page_equivalence}
\end{align}
This is exactly the operator that appeared in \eqref{eq:sec8_purity_operator_branch}. The large diffeomorphism supplies the global identification \(e_b\simeq \tilde b_1\). What follows, after applying the gluing map, is the identification \eqref{eq:sec8_late_equals_early}.

\section*{Acknowledgements}
For helpful discussions and/or comments on the draft we thank Ahmed Almheiri, Stefano Antonini, Raphael Bousso, Adam Levine, Guanda Lin, Don Marolf, Henry Maxfield, Geoff Penington, Eva Silverstein, Pratik Rath, and Misha Usatyuk.

\appendix

\section{Higher order terms}

\subsection{Closed universe sectors}
The main text keeps the part of the third quantized Hilbert space built from interior legs. Here an interior leg means the black hole interior portion of the observer dressed slice, with one endpoint at the left horizon corner and the other on the end of the world brane. These are the sectors counted by \(N\). They carry the horizon corner frame, the interior hard mode labels, and the end of the world brane labels, and they are precisely the sectors that can be returned to the one boundary Hilbert space by the horizon gluing map. A more complete third quantized Hilbert space would also include compact components disconnected from the exterior, namely closed universe sectors. Schematically,
\begin{align}
&\mathcal H_{\rm full}^{\varepsilon}
=
\bigoplus_{N,M\ge 0}
\Bigl(
\widehat{\mathcal H}_{\rm kin}^{(N),\varepsilon}
\otimes
\mathcal H_{\rm c}^{(M)}
\Bigr),
\label{eq:app_horizon_compact_sector_decomp}
\end{align}
where \(\mathcal H_{\rm c}^{(0)}=\mathbb C\), \(N\) counts interior legs, and \(M\) counts closed universe components.

Let \(\mathbf P_0\) project onto the \(M=0\) block and let \(\mathbf Q_{\rm c}:=\mathbf 1-\mathbf P_0\). The operators in the AMPS experiment are pulled back from the observer dressed slice and the extended Hilbert space. They act on the interior leg factors and do not change \(M\). In matrix elements between states with no closed universe components, only the restricted operators
\begin{align}
&\widehat O_{\rm AMPS}^{(0)}
=
\mathbf P_0\widehat O_{\rm AMPS}\mathbf P_0,
\\
&\mathbf P_0\widehat O_{\rm AMPS}\mathbf Q_{\rm c}
=
\mathbf Q_{\rm c}\widehat O_{\rm AMPS}\mathbf P_0
=
0
\label{eq:app_AMPS_operator_compact_projection}
\end{align}
appear.
Thus the explicit AMPS calculation is an amplitude between states with no disconnected compact components. The interior legs may split and rejoin through the pair of pants Hamiltonian and are later glued back to the exterior. A compact component that disconnects from the exterior is not glued back and is not measured by the AMPS operators.

The full gravitational Hamiltonian may be written schematically as
\begin{align}
&\widehat H_{\rm grav}
=
\widehat H_{\mathscr H}
+
\widehat H_{\rm disc}
+
\widehat H_{\rm mix},
\;
\widehat H_{\rm mix}=\mathcal{O}(\lambda),
\;
\lambda\sim e^{-S_0},
\label{eq:app_H_horizon_compact_mix}
\end{align}
where \(\widehat H_{\mathscr H}\) acts on the interior leg sectors and contains the pair of pants interaction. The part relevant for branch preparation is
\begin{align}
&\mathbf P_2\widehat H_{\mathscr H}\mathbf P_1
=
\lambda\,\mathbf P_2\widehat V_{\rm pants}\mathbf P_1,
\label{eq:app_interior_one_to_two}
\end{align}
where \(\mathbf P_N\) projects onto the \(N\) interior leg sector. This gives the leading connected branch used in \cref{sec:dist}: an amplitude of order \(\lambda\) from one interior leg to two interior legs.

Disconnected compact sectors can still affect this amplitude after they are integrated out. The effective Hamiltonian on the \(M=0\) block is the Feshbach operator
\begin{align}
&\widehat H_{\rm eff}(E)
=
\mathbf P_0\widehat H_{\rm grav}\mathbf P_0
+
\mathbf P_0\widehat H_{\rm grav}\mathbf Q_{\rm c}
\frac{1}{E-\mathbf Q_{\rm c}\widehat H_{\rm grav}\mathbf Q_{\rm c}}
\mathbf Q_{\rm c}\widehat H_{\rm grav}\mathbf P_0.
\label{eq:app_Feshbach_horizon_effective_H}
\end{align}
Consequently
\begin{align}
&\mathbf P_2\widehat H_{\rm eff}(E)\mathbf P_1
=
\lambda\,\mathbf P_2\widehat V_{\rm pants}\mathbf P_1
+
\mathcal O(\lambda^2),
\label{eq:app_effective_horizon_one_to_two}
\end{align}
up to order one matrix elements and possible density of states factors. Thus, absent a compensating exponentially large degeneracy of closed universe sectors, their leading effect is down by an additional factor of \(e^{-S_0}\) in the amplitude relative to the pair of pants transition between interior leg sectors. In the regime studied in the main text they renormalize the effective Hamiltonian acting on the interior leg sectors and do not change the AMPS calculation.

\subsection{Higher topologies}
In \cref{sec:cubicpants} we kept only the cubic pair of pants interaction. This is the minimal topology changing term that transfers support from the one leg sector to the connected two leg sector. The AMPS observables are insensitive to the detailed history by which that connected branch was prepared. They only require that the branch exist by the time the purity measurement is applied, and that the horizon vacuum measurement then act on that branch.

Let the interaction in the interior leg sectors be written as
\begin{align}
&\widehat H_{\rm int}
=
\lambda\widehat V_{\rm pants}
+
\sum_a\lambda_a\widehat W_a.
\label{eq:app_general_topology_interaction}
\end{align}
Only the projected matrix element \(\mathbf P_2\widehat H_{\rm int}\mathbf P_1\) enters the leading branch preparation rate. If a higher topology operator has \(\mathbf P_2\widehat W_a\mathbf P_1\neq0\), then it contributes to the same measured one leg to two leg channel and simply changes the effective transition kernel. In that case the kernel used in the main text is replaced by
\begin{align}
&\lambda\,\mathbf P_2\widehat V_{\rm pants}\mathbf P_1
+
\sum_a\lambda_a\,\mathbf P_2\widehat W_a\mathbf P_1.
\label{eq:app_transition_kernel_renormalization}
\end{align}
If instead \(\mathbf P_2\widehat W_a\mathbf P_1=0\), its first contribution to the measured channel has intermediate sectors and is of the form
\begin{align}
&\mathbf P_2\widehat W_a\,
\frac{1}{E-\widehat H_X}\,
\widehat W_b\mathbf P_1,
\label{eq:app_virtual_topology_correction}
\end{align}
where \(X\) denotes the intermediate interior leg sectors or closed universe sectors. Such terms renormalize the effective one leg to two leg kernel and the effective branch preparation rate \(\Lambda(T)\). They can change coefficients and the detailed edge mode distribution on the postselected connected branch, but they do not change the basic mechanism by which AMPS is resolved.

\subsection{Dyson series}
We justify the use of the leading term in the Dyson series in \cref{sec:dist}. The estimate is local on a single timefold \(I_j\). The long time contour has already been accounted for by summing over folds and passing to the rate equation \eqref{eq:sec8_continuum_rate}.

Consider a term in the Dyson series on \(I_j\) that contributes to the one leg to two leg transition. Since each insertion of \(\widehat V_{\rm pants}\) changes the number of interior legs by one, the number of insertions must be odd, \(n=2m+1\). For \(m\ge1\), the history leaves the measured two leg sector and returns to it. These terms are therefore corrections to the effective one leg to two leg transition kernel. On a single timefold with \(\delta T_j=\mathcal O(1)\), the ordered integral scales as
\begin{align}
&A_{1\to2}^{(2m+1)}(I_j)
\sim
\lambda^{2m+1}
\int_{I_j}\Dd t_1\cdots \Dd t_{2m+1}\,
\theta(t_1>\cdots > t_{2m+1})
\sim
\lambda^{2m+1}\frac{(\delta T_j)^{2m+1}}{(2m+1)!},
\label{eq:app_higher_dyson_scaling_single_fold}
\end{align}
up to order one matrix elements. Therefore
\begin{align}
&\frac{A_{1\to2}^{(2m+1)}(I_j)}{A_{1\to2}^{(1)}(I_j)}
=
\mathcal{O}\!\bigl(e^{-2mS_0}\bigr),
\;
\delta T_j=\mathcal{O}(1).
\label{eq:app_higher_dyson_suppression_ratio}
\end{align}
Thus the single insertion term dominates on each fold. Including the higher rank odd insertions changes the rate in \eqref{eq:sec8_continuum_rate} only by
\begin{align}
&\Lambda(T)
=
|\lambda|^2 k\rho_{\rm eff}
\left[
\sigma_0(T)+\mathcal O(\ee^{-2S_0})
\right],
\label{eq:app_rate_renormalization}
\end{align}
with \(\sigma_0(T)=\mathcal O(1)\). The parametrically long distillation time is therefore already included in the exponent of \eqref{eq:sec8_Pconn_solution}; it does not promote higher Dyson insertions on a single fold to new leading AMPS branches.

\bibliographystyle{JHEP}

\bibliography{canonical_qg_intro_refs, firewall_intro_additional_refs, main_body_suggested_refs, section_revision_extra_refs}

\end{document}